\documentclass[11pt]{article}

\usepackage{amsthm}
\usepackage{graphicx} 
\usepackage{array,makecell} 

\usepackage{amsmath, amssymb, amsfonts, verbatim}
\usepackage{bbm}
\usepackage{hyphenat,epsfig,subfigure,multirow}
\usepackage{nicefrac}
\usepackage{paralist}

\usepackage{booktabs,tabularx}

\usepackage[usenames,dvipsnames]{xcolor}
\usepackage[ruled]{algorithm2e}

\DeclareFontFamily{U}{mathx}{\hyphenchar\font45}
\DeclareFontShape{U}{mathx}{m}{n}{
      <5> <6> <7> <8> <9> <10>
      <10.95> <12> <14.4> <17.28> <20.74> <24.88>
      mathx10
      }{}
\DeclareSymbolFont{mathx}{U}{mathx}{m}{n}
\DeclareMathSymbol{\bigtimes}{1}{mathx}{"91}

\usepackage{tcolorbox}
\tcbuselibrary{skins,breakable}
\tcbset{enhanced jigsaw}

\usepackage[normalem]{ulem}
\usepackage[compact]{titlesec}

\definecolor{DarkRed}{rgb}{0.5,0.1,0.1}
\definecolor{DarkBlue}{rgb}{0.1,0.1,0.5}

\usepackage{nameref}
\definecolor{ForestGreen}{rgb}{0.1333,0.5451,0.1333}
\definecolor{Red}{rgb}{0.9,0,0}
\usepackage[linktocpage=true,
	pagebackref=true,colorlinks,
	linkcolor=DarkRed,citecolor=ForestGreen,
	bookmarks,bookmarksopen,bookmarksnumbered]
	{hyperref}
\usepackage[noabbrev,nameinlink]{cleveref}
\crefname{property}{property}{Property}
\creflabelformat{property}{(#1)#2#3}
\crefname{equation}{eq}{Eq}
\creflabelformat{equation}{(#1)#2#3}

\usepackage{bm}
\usepackage{url}
\usepackage{xspace}
\usepackage[mathscr]{euscript}

\usepackage{tikz}
\usetikzlibrary{arrows}
\usetikzlibrary{arrows.meta}
\usetikzlibrary{shapes}
\usetikzlibrary{backgrounds}
\usetikzlibrary{positioning}
\usetikzlibrary{decorations.markings}
\usetikzlibrary{patterns}
\usetikzlibrary{calc}
\usetikzlibrary{fit}
\usetikzlibrary{decorations}

\usepackage{mdframed}

\usepackage[noend]{algpseudocode}
\makeatletter
\def\BState{\State\hskip-\ALG@thistlm}
\makeatother

\usepackage{cite}
\usepackage{enumitem}

\usepackage[margin=1in]{geometry}

\newtheorem{lemma}{Lemma}[section]
\newtheorem{proposition}[lemma]{Proposition}
\newtheorem{corollary}[lemma]{Corollary}
\newtheorem{claim}[lemma]{Claim}
\newtheorem{fact}[lemma]{Fact}
\newtheorem{assumption}[lemma]{Assumption}

\newtheorem{definition}[lemma]{Definition}

\newtheorem*{claim*}{Claim}
\newtheorem*{assumption*}{Assumption}
\newtheorem*{proposition*}{Proposition}
\newtheorem*{lemma*}{Lemma}
\newtheorem*{problem5*}{Problem}

\crefname{lemma}{Lemma}{Lemmas}
\crefname{claim}{claim}{claims}

\newtheorem{mdresult}{Result}
\newenvironment{result}{\begin{mdframed}[backgroundcolor=lightgray!40,topline=false,rightline=false,leftline=false,bottomline=false,innertopmargin=2pt]\begin{mdresult}}{\end{mdresult}\end{mdframed}}

\newtheorem{observation}[lemma]{Observation}

\newtheoremstyle{restate}{}{}{\itshape}{}{\bfseries}{~(restated).}{.5em}{\thmnote{#3}}
\theoremstyle{restate}

\allowdisplaybreaks

\renewcommand{\qed}{\nobreak \ifvmode \relax \else
      \ifdim\lastskip<1.5em \hskip-\lastskip
      \hskip1.5em plus0em minus0.5em \fi \nobreak
      \vrule height0.75em width0.5em depth0.25em\fi}

\newcommand{\tvd}[2]{\ensuremath{\norm{#1 - #2}_{\mathrm{tvd}}}}
\newcommand{\Ot}{\ensuremath{\widetilde{O}}}
\newcommand{\eps}{\ensuremath{\varepsilon}}

\newcommand{\bracket}[1]{\left[#1\right]}
\newcommand{\paren}[1]{\ensuremath{\left(#1\right)}\xspace}
\newcommand{\card}[1]{\left\vert{#1}\right\vert}
\newcommand{\Omgt}{\ensuremath{\widetilde{\Omega}}}

\newcommand{\norm}[1]{\ensuremath{\|#1\|}}

\newcommand{\ceil}[1]{{\left\lceil{#1}\right\rceil}}
\newcommand{\floor}[1]{{\left\lfloor{#1}\right\rfloor}}

\newcommand{\expect}[1]{\Exp\bracket{#1}}

\newcommand{\set}[1]{\ensuremath{\left\{ #1 \right\}}}
\newcommand{\poly}{\mbox{\rm poly}}

\newcommand{\alg}{\ensuremath{\mathcal{A}}\xspace}

\DeclareMathOperator*{\Exp}{\ensuremath{{\mathbb{E}}}}
\DeclareMathOperator*{\Prob}{\ensuremath{\textnormal{Pr}}}
\renewcommand{\Pr}{\Prob}

\newenvironment{tbox}{\begin{tcolorbox}[
		enlarge top by=5pt,
		enlarge bottom by=5pt,
		 breakable,
		 boxsep=0pt,
                  left=4pt,
                  right=4pt,
                  top=10pt,
                  arc=0pt,
                  boxrule=1pt,toprule=1pt,
                  colback=white
                  ]
	}
{\end{tcolorbox}}

\newcommand{\rv}[1]{\ensuremath{{\mathsf{#1}}}\xspace}
\newcommand{\rA}{\rv{A}}
\newcommand{\rB}{\rv{B}}

\newcommand{\kl}[2]{\ensuremath{\mathbb{D}(#1~||~#2)}}
\newcommand{\js}[2]{\ensuremath{\operatorname{JS}(#1,#2)}}

\newcommand{\ind}{\ensuremath{\mathbbm{1}}}
\newcommand{\HH}{\ensuremath{\mathbb{H}}}
\newcommand{\II}{\ensuremath{\mathbb{I}}}
\newcommand{\mi}[2]{\ensuremath{\def\mione{#1}\def\mitwo{#2}\mireal}}
\newcommand{\mireal}[1][]{
  \ifx\relax#1\relax%
    \II(\mione \,; \mitwo)%
  \else%
    \II(\mione \,; \mitwo\mid #1)%
  \fi
}
\newcommand{\en}[1]{\ensuremath{\HH(#1)}}

\newcommand{\itfacts}[1]{\Cref{fact:it-facts}-\ref{part:#1}\xspace}

\newcommand{\jstar}{\ensuremath{j^{\star}}}
\newcommand{\rjstar}{\ensuremath{\rv{j}^{\star}}}
\newcommand{\nprime}{\ensuremath{n'}}

\newcommand{\MEproblem}{\ensuremath{\textnormal{\textsc{ME}}}}

\newcommand{\protME}{\ensuremath{\pi^{\textnormal{ME}}}}
\newcommand{\prot}{\pi}

\newcommand{\sens}{\ensuremath{sens}}
\newcommand{\dgn}{\ensuremath{dgn}}
\renewcommand{\inf}{\ensuremath{inf}}

\newcommand{\sleft}{\ensuremath{s_{\ell}}\xspace}
\newcommand{\sright}{\ensuremath{s_r}\xspace}

\newcommand{\cD}{\mathcal{D}}

\newcommand{\tvdfacts}[1]{\Cref{fact:tvd-facts}-\ref{fact:#1}}

\newcommand{\Inp}{\ensuremath{\textnormal{\textsc{Inp}}}}

\newcommand{\err}{\ensuremath{\textnormal{\textsc{Err}}}}

\newcommand{\cDalg}{\ensuremath{\cD^{\alg}}}
\newcommand{\eventbad}{\ensuremath{\mathcal{E}_{\textnormal{bad}}}}

\newcommand{\rb}{\ensuremath{\rv{b}}}

\newcommand{\xstar}{\ensuremath{x^{\star}}}

\newcommand{\foplus}{\ensuremath{f \oplus}}

\newcommand{\Lall}{\ensuremath{L}}
\newcommand{\rLall}{\ensuremath{\rv{L}}}

\newcommand{\Rpublic}{\ensuremath{\rv{R}}_{\text{pub}}}

\newcommand{\rP}{\rv{P}}
\newcommand{\ry}{\rv{y}}

\newcommand{\rX}{\rv{X}}
\newcommand{\rY}{\rv{Y}}

\newcommand{\dist}{\ensuremath{\textnormal{dist}}}

\newcommand{\rZ}{\rv{Z}}

\newcommand{\ic}{\textnormal{\textsc{IC}}}

\newcommand{\player}{Q}

\newcommand{\Prot}{\ensuremath{\Pi}}

\newcommand{\ralpha}{\ensuremath{\bm{\alpha}}}
\newcommand{\rbeta}{\ensuremath{\bm{\beta}}}
\newcommand{\rrho}{\ensuremath{\bm{\rho}}}
\newcommand{\rtheta}{\ensuremath{\bm{\theta}}}

\renewcommand{\js}[3]{\ensuremath{\mathbb{J}_{#1}(#2~||~#3)}}

\newcommand{\charalpha}{\ensuremath{'\!\!\alpha'}}
\newcommand{\charbeta}{\ensuremath{'\!\!\beta'}}
\newcommand{\rProt}{\ensuremath{\bm{\Prot}}}

\newcommand{\Rpublicn}{R_{\text{pub}}}

\newcommand{\eventfail}{\mathcal{E}_{\mathrm{fail}}}

\newcommand{\protgood}{\prot^{\textsc{bad}}}
\newcommand{\rProtgood}{\rProt^{\textsc{bad}}}
\newcommand{\eventgood}{\mathcal{E}^{\textsc{bad}}}
\newcommand{\pgood}{p^{\textsc{bad}}}
\newcommand{\Protgood}{\Prot^{\textsc{bad}}}

\newcommand{\Rpublictwo}{\Rpublic^{\textnormal{two}}}

\newcommand{\two}{\textnormal{two}}

\newcommand{\specialtime}[1]{\ensuremath{\widehat{t}_{#1}}}
\newcommand{\specialbit}[1]{\ensuremath{\widehat{b}_{#1}}}
\newcommand{\specialstate}[1]{\ensuremath{\widehat{\rP}_{#1}}}

\newcommand{\rM}{\ensuremath{\rv{M}}}
\newcommand{\rh}{\ensuremath{\rv{h}}}

\newcommand{\rPrott}{\ensuremath{\rProt^{\textnormal{two}}}}

\newcommand{\fbar}{\ensuremath{\overline{f}}}
\newcommand{\gbar}{\ensuremath{\overline{g}}}
\newcommand{\Pbar}{\ensuremath{\overline{P}}}

\newcommand{\tribes}{\textsc{tribes}}

\newcommand{\commented}[1]{}

\newcommand{\istar}{\ensuremath{i^{\star}}}

\newcommand{\cH}{\mathcal H}
\newcommand{\cG}{\mathcal G}

\newcommand{\eventeq}{\ensuremath{\mathcal{E}_{\mathrm{eq}}}}

\newcommand{\EQ}{\ensuremath{\textsc{EQ}}}

\newcommand{\rz}{\rv{z}}

\def\DEBUG{true}

\ifdefined\DEBUG
\newcommand{\snote}[1]{\textcolor{red}{[SV: #1]}}
\fi

\ifdefined\DEBUG
\newcommand{\sg}[1]{\textcolor{blue}{[SG: #1]}}
\fi

\title{XOR-Functions and Beyond in the Random Query Model}

\date{}

\author{
Sumegha Garg\footnote{  (sumegha.garg@rutgers.edu) Department of Computer Science, Rutgers University.  Supported by NSF Grant CCF-2542741 and Rutgers Individual Fulcrum Award.\smallskip} 
	\\ {\small Rutgers University} \and
	Janani Sundaresan\footnote{(js16503@nyu.edu) Courant Institute of Mathematical Sciences, New York University. \smallskip} \\ {\small New York University} \and Santhoshini Velusamy \footnote{(santhoshini.velusamy@uwaterloo.ca) David R. Cheriton School of Computer Science, University of Waterloo. Supported by NSERC Discovery grant RGPIN-2026-04689 and ECR supplement DGECR-2026-00421. Supported in part by NSF award CCF 2348475 when the author was affiliated with Toyota Technological Institute at Chicago.}
	\\ {\small University of Waterloo} 
}
\begin{document}
\maketitle

\pagenumbering{roman}

\begin{abstract}

We study time--space tradeoffs for computing Boolean functions $f:\set{0,1}^n \to \set{0,1}$ in the random query model. On input $x$, a branching program receives independent samples $(i,x(i))$, where $i$ is uniform in $[n]$. We want to evaluate $f(x)$ ($w.h.p.$) while minimizing the width $2^S$ and the length $T$ of the branching programs. There is a simple upper bound of $T = \tilde{O}(n), \;S = O(n)$: coupon-collector shows that $\tilde{O}(n)$ samples suffice, and they can be stored in $O(n)$ space. 

Raz and Zhan (ITCS 2020) introduced this model and studied the zero-error setting. In the bounded-error setting, Dinur (SODA 2024) proved $TS = \Omgt(n \cdot \inf[f])$, where $\inf[f]$ is the total influence of $f$. For many functions with low influence such as Equality and Tribes, this lower bound can still leave a substantial gap from the best known upper bounds. Our main result is a stronger lower bound of $T(S+\log T) = \Omega(n \cdot \sens[g])$ for any XOR-function $g(x,y)= f(x \oplus y)$, where $\sens[g]$ is the sensitivity of $g$. 
Through simple corollaries, we get tight nearly-quadratic time--space lower bounds for Tribes and Equality, despite their low influence.
Our proof introduces a new notion of information complexity tailored to the random query model along with a corresponding direct-sum argument which may be of independent interest.

Dinur also proved a stronger bound of $TS = \Omega(n^2)$ for Majority, whose influence is only $\Theta(\sqrt{n})$, and asked whether an alternative characterization of Boolean functions could give a unified proof of tight results for a broader class of functions, including Majority.
We answer this open question by proving $TS = \Omgt(n \cdot \dgn[f])$, where $\dgn[f]$ is the degeneracy of the sensitivity graph of $f$ (for Majority, this value is $\Theta(n)$).  We also prove that this bound is tight for all symmetric functions up to logarithmic factors.
\end{abstract}

\clearpage

\setcounter{tocdepth}{3}
\tableofcontents

\clearpage

\pagenumbering{arabic}
\setcounter{page}{1}

\section{Introduction}

Time-space tradeoffs have been studied extensively in various models of computation. 
The random query model was introduced by \cite{RazZ20} as a new form of space-bounded computation for computing Boolean functions. Given a function $f:\set{0,1}^n \rightarrow \set{0,1}$, the aim is to design a branching program to find $f(x)$ on an unknown input $x$. The bits of the input cannot be queried directly, but indices $i_1, i_2, \ldots, i_T$ are chosen independently from the uniform distribution over $[n]$ for a total of $T$ time steps. The branching program receives a pair $(i_j, x(i_j))$, corresponding to index $i_j$ and the value $x(i_j)$ at index $i_j$ as input in each time step $j$.
The branching program has time $T$ (or length $T$) corresponding to the number of samples, and bounded memory of $S$ bits (or width $2^S$) in which to process the input. 
 We want to study  tradeoffs between the number of required samples $T$ and the width $2^S$ of the branching programs.
 
\cite{RazZ20} provides strong motivation for studying the random-query model. Time–space tradeoffs for standard branching programs, in which the algorithm can adaptively choose the index to query at each vertex, have been extensively studied (see, e.g., \cite{BabaiNS89,ajtai1999non,beame2001time}).  Nevertheless, proving strongly superlinear time–space tradeoffs, of the form $TS=\Omega(n^{1+\epsilon})$ for any constant $\epsilon>0$, for explicit single-output Boolean functions remains a longstanding open problem, even for oblivious branching programs. The random-query model offers a promising approach to this challenge, as it can simulate any oblivious branching program with only a $\tilde{O}(n)$ overhead in time. Moreover, \cite{RazZ20} showed that extending lower bounds in the random-query model to recurring distributions, where certain query indices are known to be equal, would already imply nontrivial lower bounds for oblivious branching programs.

Beyond its connections to standard branching programs, the random-query model is a natural model of computation to study in its own right. Computing a function $f$ in this model can be viewed as a memory-bounded testing problem: can an algorithm distinguish between the sets $\set{x \mid f(x) = 0}$ and $\set{x \mid f(x) = 1}$ when, at each time step, it observes a random index (sample) together with the value of $x$ at that index? This perspective connects the model to a long line of work establishing memory-sample tradeoffs for learning \cite{shamir2014fundamental,steinhardt2016memory,raz17,KolRT17,GargRT18,sharan2019memory,GargRT19,Raz19,beame18,moshkovitz2018entropy,gonen20,lyu2023tight,liu2023memory}. More recently, there has also been growing interest in massive graph processing under random edge sampling \cite{kapralov2020space,GargHP26}. A key challenge that distinguishes the random-query model from these settings is that, in the sample regimes of interest, the algorithm observes the same input coordinate multiple times (up to $O(n)$ times). Such repetitions create dependencies across samples that require new techniques to analyze.

As their main result, \cite{RazZ20} proves that any branching program with zero-error\footnote{Here, for every input, the branching program is allowed to not have an output with probability at most $1/2$. However, if an output is produced, it must be correct.} that evaluates functions $f$ with sensitivity $\sens[f]$ \footnote{Sensitivity of $f$ at input $x$ is the number of bits in $x$ that, when flipped, the output changes from $f(x)$. The sensitivity of $f$ is the maximum number of sensitive bits  taken over all inputs $x$. } using $T$ time and $S$ space must satisfy: $T \cdot (S + \log n) = \Omega(n \cdot \sens[f])$. \commented{\snote{I think we can keep the footnote that defines sensitivity.}} For many functions like parity and majority, this gives a tight quadratic time-space lower bound up to logarithmic factors. However, it is easy to see that the sensitivity based lower bound of \cite{RazZ20} does not extend to the bounded-error regime. The AND function only requires $T \cdot S = O(n)$ for outputting the answer with any constant advantage. Using a coupon-collector argument, it is not hard to show that with high probability, all the $n$ indices will appear in $O(n \log n)$ independent samples. Then using just constant space, one can easily check if all their values are $1$, or not. \commented{\snote{I am a bit puzzled here. Don't we have only $T \cdot S = \tilde{O}(n)$? I would perhaps write it like the following.}}  And yet, the sensitivity of the AND function is $n$.  

Later, \cite{Dinur24} proved that in bounded-error computation, any branching program computing function $f$ with total influence\footnote{Total influence is the average number of sensitive bits where the average is taken uniformly over all inputs.} $\ell$ must satisfy $T \cdot S = \Omgt(n \cdot \ell)$.\footnote{The $\Tilde{\Omega}$ hides logarithmic factors in $n$.} This bound is tight (up to logarithmic factors) for the parity function. However, for functions with $O(\log n)$ influence, it does not give better than the trivial $T \cdot S= \Omega(n)$ lower bound applicable for any non-constant function. 

Moreover, for the majority function, with total influence $\Theta(\sqrt{n})$,  \cite{Dinur24} gives a separate argument to show that the majority function indeed requires $T \cdot S = \Omega(n^2)$. Hence, total influence does not provide a complete picture, and functions can have much larger time-space lower bounds.

The gap between total influence and the best-known upper bounds is much more profound for the well-known TRIBES function, which is an OR of disjoint tribes of ANDs on inputs $(n/2)$-bit strings $x, y \in \set{0,1}^{n/2}$:
\[
\tribes_2(x,y) = \bigvee_{i \in [n/2]} \paren{x(i)\wedge y(i)}.
\]

This function, introduced in \cite{Ben-OrL89}, has since then been studied in many contexts like the two-party communication complexity \cite{KalyanasundaramS92,Razborov92,BarYossefJKS04} and learning theory \cite{Dachman-SoledFTWW15,BlancLT20}. It is known to have an influence of $O(1)$, and yet the best upper bound is the trivial $T \cdot S = \Ot(n^2)$.

This gap is the question that our paper addresses:
\begin{quote}
    \textit{Can we prove $\omega(n)$ lower bounds for a large class of Boolean functions with low influence, like TRIBES?}
\end{quote}

 We focus on XOR functions $g:\set{0,1}^{2n} \rightarrow \set{0,1}$, which have the form
$g = f \circ \oplus_2$, i.e., a base function
$f:\set{0,1}^n\to\set{0,1}$ is composed with two-bit parity.
This class includes equality and Hamming-distance, and it
has been studied extensively in communication complexity
\cite{TsangWXZ13,Zhang14,HatamiHL16,GirishSTW23}.
 XOR composition preserves total influence up to a
factor of two, with $\inf[g]=2\inf[f]$, making this a natural
class in which to study time-space lower bounds for
low-influence functions.

This brings us to our main result:
\begin{result}\label{thm:equal-lb}
For any XOR-function $g:\set{0,1}^{2n}\rightarrow\set{0,1}$ with  sensitivity $k$, any branching program that computes $g$  in the random query model with probability of success at least $2/3$ with width $2^S$ and time $T$ must satisfy:
\[
	T \cdot (S + \log T) = \Omega(n \cdot k).
\]
\end{result}

Through a simple reduction, we get almost tight lower bounds for the TRIBES function with any inner tribe size. 
Our result also gives nearly-tight lower bounds for the Element Distinctness problem via another simple reduction, which is a fundamental problem studied in classical time-space tradeoffs \cite{Ajtai05,BeameCM13,LyuZ23,ChenJWW22}. 
These reductions are presented in \Cref{sec:extend-other-fns}. This lower bound evidences the newness in our techniques. 


To prove \Cref{thm:equal-lb}, we introduce a \textbf{new notion of information complexity} tailored to the random query model, where indices $i_j$ (at the $j$-th time-step) are sampled independently but the input bits $x(i_j)$ are correlated across time-steps. With appropriate conditioning, this notion admits a direct-sum decomposition across matching pairs of input bits. We develop an embedding argument that exploits this decomposition to reduce the global information cost to that of a carefully chosen local problem. We believe this approach to proving information lower bounds in the random query model may find applications beyond \Cref{thm:equal-lb}, particularly to distributed computation settings in which multiple players share common data.

In addition, we also answer the following question explicitly asked by \cite{Dinur24}: can the influence-based lower bound be strengthened with another complexity measure to recover the quadratic lower bound for majority?
We show that the degeneracy of the sensitivity graph is a measure that unifies these results. 

For any function $f$, let $G_f$ be the graph with the vertex set $\set{0,1}^n$, and an edge between two vertices $x, y$ if they differ in exactly one co-ordinate and $f(x) \neq f(y)$. The combinatorial and spectral properties of this graph have been studied before \cite{GopalanSW16,AaronsonBKRT21,MittalNP24}.
The total influence $\inf[f]$ and  sensitivity $\sens[f]$ of $f$ are the average
degree and maximum degree of $G_f$ respectively. The degeneracy of
$G_f$ is the largest minimum degree among its subgraphs:
\[
\dgn[f]
=
\max_{H \subseteq G_f} (\text{Minimum degree in the subgraph $H$}).
\]

It is a standard fact that degeneracy is at least half of the average degree of any graph, and hence, $\dgn[f] = \Omega(\inf[f])$ for any function $f$. For the majority function, the degeneracy is $\Theta(n)$.

We show that $\Omgt(n \cdot \dgn[f])$ is a lower bound on $T \cdot S$ for \emph{all} functions. We also prove that this lower bound is \emph{almost tight} for all symmetric functions with a new upper bound result. 

\begin{result}\label{thm:dgn-lb}
	 Any branching program that computes a Boolean function $f$ with probability of success at least $2/3$ in the random query model, with width $2^S$ and $T$ samples must have
		\[T \cdot S = \Omgt(n \cdot \dgn[f]).\]
    For symmetric functions, this bound is tight up to logarithmic factors for any $T = \Omega(n \log n)$. 
    \commented{\snote{What is this condition that $T = \Omega(n \log n)$? Why do we need it? - Because only for $\Omega(n \log n)$ do we get a matching algorithm at this point in the tradeoff. The algorithm does not transfer if $T = O(n)$ directly, although I have not thought too much about this}}
\end{result}

To place \Cref{thm:dgn-lb} and \Cref{thm:equal-lb} in context, \Cref{tab:results} summarizes the relevant complexity measures and compares our bounds with prior work for several representative functions:

\begin{table}[htbp]
    \centering
    \small
    \setlength{\tabcolsep}{3pt}
    \renewcommand{\arraystretch}{1.35}
    \begin{tabular}{
        @{}>{\centering\arraybackslash}m{3cm}
        cc@{\hspace{12pt}}c@{\hspace{12pt}}c@{\hspace{12pt}}c@{}
    }
        \toprule
        \textbf{Function}
        & \textbf{Influence}
        & \textbf{Degeneracy}
        & \makecell[c]{\textbf{Prior}\\\textbf{Bound}}
        & \makecell[c]{\textbf{Our Lower}\\\textbf{Bounds}}
        & \makecell[c]{\textbf{Upper}\\\textbf{Bounds}} \\
        \midrule

        Threshold $k$,\newline
        $1\leq k< n/2$ 
        & $n2^{1-n}\binom{n-1}{k-1}$
        & $k$
        & $\Omega(n\max\{1,\inf[f]\})$
        & \makecell[c]{$\Omgt(nk)$\\\Cref{thm:dgn-lb}}
        & \makecell[c]
        {$\widetilde{O}(nk)$\\\Cref{thm:dgn-lb}}\\
        \addlinespace[5pt]

        Equality of\newline
        $n$-bit strings
        & $O(n2^{-n})$
        & $2$
        & $\Omega(n)$
        & \makecell[c]{$\Omgt(n^2)$\\\Cref{thm:equal-lb}}
        & \makecell[c]
        {$\Ot(n^2)$\\Trivial}\\
        \addlinespace[5pt]

        TRIBES with\newline
        tribe size 2
        & $O(n\cdot (3/4)^{n/2})$
        & $2$
        & $\Omega(n)$
        & \makecell[c]{$\Omgt(n^2)$\\Corollary of \Cref{thm:equal-lb}}
        & \makecell[c]
        {$\Ot(n^2)$\\Trivial}\\ 

        TRIBES with\newline
        tribe size $w$
        & $O(n2^{-w})$
        &  $\min\set{w, n/w}$
        & $\Omega(n)$
        & \makecell[c]{$\Omgt(n^2/w)$\\Corollary of \Cref{thm:equal-lb}}
        & \makecell[c]
        {$\Ot(n^2/w)$\\Trivial}\\
        \bottomrule
    \end{tabular}
    \caption{The prior bounds are from \cite{Dinur24}, or the
    trivial $\Omega(n)$ lower bound. We ignore logarithmic factors.}
    \label{tab:results}
\end{table}

For symmetric functions, \Cref{thm:dgn-lb} provides a complete characterization via degeneracy with new upper and lower bounds. However, even the degeneracy-based lower bound falls short for functions like Equality, TRIBES, and other XOR-functions, and \Cref{thm:equal-lb} introduces new information-theoretic techniques to get tight lower bounds via sensitivity.

\commented{
\snote{I think the table above should also include TRIBES with arbitrary tribe size and element distinctness, and have an additional column for the upper bounds. I also believe the upper bound for general symmetric functions was not there in prior works and so we should mention it somewhere among results and also refer to it in the table. The explicit symmetric function characterization is interesting and has connections to quantum query complexity which I also think is worth mentioning.
For element distinctness, I would suggest spending a few sentences to explain the reduction from equality (inequality?) as it is not entirely trivial.}
}

\subsection{Our Techniques}\label{sec:techniques}

We only give the high level overview of \Cref{thm:equal-lb} in this subsection. As the proof of \Cref{thm:dgn-lb} is comparatively shorter, we refer the reader to \Cref{sec:dgn-lb}.

Our approach works with an information cost for branching programs. Prior work in this model used reductions from the Missing Element problem, where the branching program was used to detect whether some randomly chosen index $i \in [n]$ was \emph{entirely} missing from the list of queries, or the queries covered $[n]$. To do so, the branching program was given access to \emph{all} of $x(j)$ for $j \in [n], j \neq i$, and asked to answer $f(x)$ for a randomly chosen $x$. The program was incorrect with a probability proportional to the influence of $f$, leading to the lower bound of \cite{Dinur24}.

The lower bound gives the program strictly more information than it has access to: any program with $S = o(n)$ cannot store all the values of $x(j)$ for $j \neq i$ in memory. In contrast, we directly analyze how low space affects the success probability without a reduction to Missing Element. We show that any successful branching program still stores sufficient information \emph{about the random indices}, while processing input $x$.

Our information complexity notion is adapted from \cite{BravermanGW20} for streaming algorithms, and subsequently used in other works \cite{BrownBS22,Aden-AliHNY24,BravermanGLWWZ24,LiWZ25}. Rather than measuring information about the input bits directly, which is challenging because the same input bit may be queried multiple times, we measure information about the independently sampled query indices \emph{conditioned on the entire input $x$}. 
This information complexity can be analyzed to give the lower bounds in \Cref{thm:equal-lb} using the following main ideas: 

\begin{enumerate}
\item \emph{A multi-party communication lower bound in the blackboard model for a two-bit problem.} We show that any protocol deciding whether two bits, given to an arbitrary \emph{subset} of players, are equal must reveal information not only about the values of the bits, but also about which players received them.
\item \emph{A direct sum argument over the sensitive bits of the outer function.} We develop a new embedding protocol that embeds the two-bit problem described above into the input $x$. Our information complexity measure allows conditioning on the entire input $x$, which is crucial for both the embedding and the subsequent standard direct-sum argument. This gives the $\Omega(\sens[f])$ factor in the lower bound on $T \cdot S$.
\item \emph{Exploiting the gaps between repeated queries.} To obtain the additional $\Omega(n)$ factor in the lower bound on $T\cdot S$, we show that information associated with an occurrence of an index $i$ (and its corresponding input bit $x(i)$) must be retained until the next occurrence of its matching index. While a simple argument gives a gap of $\Omega(\sqrt{n})$ between repeated occurrences, the expected gap is $\Omega(n)$. To handle the varying gaps between repetitions, we adapt the argument from \cite{BrownBS22}, originally developed to handle different random subpopulations in memory lower bounds for learning tasks.
\end{enumerate} 

\commented{
Direct sum arguments typically involve embedding a smaller problem into the sensitive bits of a larger problem. In our case, the smaller problem is not straight-forward to analyze.
There are two uniformly random bits $\alpha, \beta$ and many players where each player receive one of the two bits randomly \emph{only if} some privately held fair random coin lands on one, and they want to output the parity of $\alpha, \beta$. Communication complexity problems with randomness in allocation have been studied before \cite{ChakrabartiCM08,ChakrabartiGMV20,AssadiS23} in other contexts. 

The direct sum argument requires a very specific information lower bound on the smaller problem: we prove an $\Omega(1)$ information lower bound for the two-bit parity problem with random allocations, even after conditioning on both inputs. This lower bound is then lifted to  the sensitive coordinates of the outer function using a \textbf{new embedding protocol}, giving an $\Omega(\sens[f])$ lower bound in the information complexity. This lower bound uses the Jensen-Shannon distance  to move between mutual information and total variation distance \cite{EndresS03}.  

}

\subsection{Related Work}

\paragraph{Memory-Sample Lower Bounds for Learning Problems.}

As discussed earlier, a long line of work studies memory-sample tradeoffs for learning; here we focus on the results most closely related to our techniques. To establish tight memory bounds for mean estimation from $i.i.d.$ samples, \cite{BravermanGW20} introduced an averaged information-complexity measure for one-pass streaming.  This measure has since been used to prove tight space bounds for approximate counting in insertion streams~\cite{Aden-AliHNY24} and lower bounds for learning problems, including sparse linear classifiers and $k$-term 2-DNFs~\cite{BrownBS22}. However, this measure does not handle correlations arising from repeated observations of the same samples. To address this challenge in multi-pass streaming, where the learner processes the same input in multiple passes, \cite{BravermanGLWWZ24} introduced a new information-complexity measure, which has subsequently yielded strong memory-sample tradeoffs for various learning and estimation problems~\cite{LiWZ25,garg2026unified}. The random-query and multi-pass streaming models are incomparable, and both offer promising avenues toward unconditional lower bounds for log-space Turing machines. Crucially, the information-complexity measure for multi-pass streaming exploits the fact that repeated observations occur in a known, fixed order, allowing for carefully chosen conditioning. This structure is absent in the random-query model, where repeated observations occur at random. In this work, we introduce a new information-complexity measure tailored to the random-query model, which we hope will prove useful for future time-space lower bounds, as previous information-complexity measures have been in other settings.


\paragraph{Notions of Complexity for Boolean Functions.}
There are many notions of complexity for analyzing Boolean functions like randomized decision tree complexity, certificate complexity, sensitivity, block sensitivity and approximate degree and numerous results study their connections to each other \cite{BealsBCMW01,AmbainisBBLSS17,BunKT18,AaronsonBKRT21}, see \cite{AaronsonBKRT21} for more details. 

It is easy to see that for general functions, there is a trivial upper bound of $\tilde{O}(n \cdot R(f)^2)$ on the product of time and space ($T \cdot S$)  where $R(f)$ is the randomized decision tree complexity of $f$\footnote{We spend $T = \Ot(n)$ samples at each node of the randomized decision tree, for a total of $\Ot(n \cdot R(f))$ queries. The tree can be directly seen as a branching program of width $2^{R(f)}$.}. Since sensitivity and randomized decision tree complexity are polynomially related for all total Boolean functions~\cite{huang2019induced}, our lower bound in \Cref{thm:equal-lb} is tight for XOR-functions up to constant factors in the exponent of $\sens$.

Viewing $T\cdot S/n$ as another way of studying Boolean function complexity, our \Cref{thm:dgn-lb} gives, up to logarithmic factors,
\[
\dgn[f] \leq \frac{T \cdot S}{n}(f) \leq R(f)^2. 
\]

To the best of our knowledge, degeneracy of Boolean functions has received little attention in the literature. \cite{GuLX26} uses this notion to study the noisy query complexity, and lower bounds the error of an estimator after specific partial revelations of the input. In our work, we use the same distribution for a different application: we lower bound the error of an estimator where all of the input barring one uniformly random bit are revealed, and combine this with the Missing Element reduction of \cite{Dinur24} to prove \Cref{thm:dgn-lb}.



\commented{
\subsection{Open Directions}
We conclude our introduction with some discussion on further directions. 
\begin{enumerate}[label=$(\roman*)$]
    \item Placing $T\cdot S/n$ among other notions of Boolean complexity: It is not clear if the trivial upper bound of $R(f)^2$ is tight for any function $f$. Moreover, we showed that in some cases, $T \cdot S/n$ is equal to the sensitivity, and it is an open question as to whether this lower bound can be unified with degeneracy. 
    \item For partial functions: Our argument based on embedding into the sensitive bits does not apply for partial functions. 
    \item More general compositions: The randomized decision tree complexity of composed functions has been well-studied, and  
\end{enumerate}
}
\paragraph{AI Methodology.} For \Cref{thm:equal-lb}, AI was not used except for proof reading. For the upper bound for symmetric functions in \Cref{thm:dgn-lb}, AI was not used.

AI was used in identifying the degeneracy notion for \Cref{thm:dgn-lb}. More specifically,
the authors had the lower bound argument in \Cref{thm:dgn-lb} for symmetric functions without AI use with the characterization by \cite{Paturi92} (see \Cref{def:sided-sens}). Further connections to degeneracy for all Boolean functions were given by ChatGPT 5.6 Sol. 
The authors have verified the results and take full responsibility for the correctness of this paper.

\paragraph{Human Acknowledgements.}
We thank Sepehr Assadi, Eric Blais, Mark Bun, and Madhur Tulsiani for useful discussions on this paper. 


\section{Technical Overview}\label{sec:ov-rehaul}

The prior works in this model, \cite{RazZ20} and \cite{Dinur24}, employ
reductions from the Missing Element (ME) problem. This is a multiplayer communication complexity problem.  Each player has a randomly chosen subset of $[n]$, and the
players have to distinguish whether their union covers $[n]$ or whether one uniformly random element is missing from all the sets.
\cite{Dinur24} proves that $\Omega(n)$ communication is required over
a blackboard for this task.

Given an algorithm $P$ computing $f$, the players of ME execute
the algorithm $P$ on a publicly
chosen uniformly random input from $x \in \set{0,1}^n$.  Each player runs the algorithm on a random subset of the indices that corresponds to their input in the ME problem. \cite{Dinur24} analyses the error probability of the algorithm for solving $f(x)$:
\begin{itemize}
    \item If there is no missing element, the queries are uniformly random, and the simulation of the algorithm computing $f(x)$ is exact. There is no additional error. 
    \item If there is a missing element $\jstar \in [n]$, the value of $x(\jstar)$ is never seen by the algorithm. Even if values of $x(1), x(2), \ldots, x(\jstar-1), x(\jstar+1), \ldots, x(n)$ are all fixed and given to the algorithm, there is still an error in outputting $f(x)$ proportional to the total influence of $f$.
\end{itemize}
\vspace{-2mm}
 This gives the $\Omega(n \cdot \inf[f])$ lower bound from \cite{Dinur24}. 
\\
 
\noindent
\textbf{Brief Ideas behind \Cref{thm:dgn-lb}:}
We show that the same ME problem can be analyzed with a different hard
distribution  based on degeneracy in place of a uniformly random input from $\set{0,1}^n$ to get a stronger lower bound of $\Omgt(n\cdot\dgn[f])$ for all
functions.

We give a new algorithm for all symmetric functions as well, showing that this bound is tight. This algorithm goes via an alternate characterization of degeneracy for symmetric functions given in \Cref{def:sided-sens}. We do not cover the details of this result in the overview, as both the analysis of the hard distribution, and the new algorithm for symmetric functions in \Cref{sec:dgn-lb} are not too long. Instead, we focus on conveying the intuition behind \Cref{thm:equal-lb}.

Notice that in the analysis of the reduction from ME problem, the algorithm was given  more information than it has access to: \emph{all of the values of $x(i)$} for $i \neq \jstar$ were given for free. The error probability was analyzed only when $x(\jstar)$ is missing, but any algorithm with $o(n)$ space does not see these elements together. Nonetheless, this framework is powerful enough to  give tight bounds for all symmetric functions.\footnote{For symmetric functions, the specific values of $x(i)$ for $i \neq \jstar$ are irrelevant, as long as the number of ones in $\set{x(i): i \in [n], i \neq \jstar}$ is fixed. This additional power given to the algorithm is not useful in computing $f(x)$.}

\noindent
\textbf{Detailed Overview for \Cref{thm:equal-lb}:} The reduction from ME problem does not capture the additional difficulty arising from
the pairing of input coordinates when dealing with XOR-functions. 
Recall the Equality function on $2n$ bits:
\[
    \EQ(x)=1 \quad\Longleftrightarrow\quad
    x(2i-1)=x(2i)\text{ for every }i\in[n].
\]
For any uniformly random index $\jstar \in [2n]$, sampled independently of the input $x$, if $x(i)$ is given to the algorithm for all $i \neq \jstar$, the answer to $\EQ$ is incorrect only with probability $O(1/n)$. To see this, consider a mismatch $x(2i) \neq x(2i-1)$. If $\jstar \notin \set{2i-1, 2i}$, the answer is already known to any algorithm that sees $x(2i-1)$ and $x(2i)$. The uniformly random missing index $\jstar$ hides the answer to $\EQ(x)$ only with probability $O(1/n)$. 

 The reduction in \cite{Dinur24}, and a more direct analysis about missing elements in \cite{RazZ20}, do not capture how long the algorithm has to retain information about any bit till its matching counterpart has been seen through the queries. This is necessary for \Cref{thm:equal-lb} and our technique addresses this issue.

To focus on the main steps of our proof,
we only consider the Equality function and explain how we obtain a nearly
quadratic time-space lower bound. 
 We want to prove $T \cdot S = \Omega(n^2)$ for any algorithm that solves this function.

\subsection{A standard direct sum}

A natural starting point is an information-complexity direct sum, in the
spirit of~\cite{BarYossefJKS04}. 
The Equality function can be viewed as $n$-many XOR functions on two bits.

For one such XOR, after seeing one bit, we must wait $\Omega(n)$ time
steps in expectation to see its matching bit. It is tempting to argue
that the algorithm must remember the first bit throughout this wait,
contributing $\Omega(n)$ bits of information when summed over time.
Lifting this argument to $\Omega(n)$ such XORs would give a total of
$\Omega(n^2)$ bits of information measured across all time steps.
Since the state at each step contains at most $S$ bits, this would give
the desired bound $TS=\Omega(n^2)$.

This intuition is wrong: a protocol can check whether Equality is $1$
without recording the individual bits explicitly. To see why, consider
the following algorithm:
\begin{enumerate}[label=$(\roman*)$]
    \item Sample a uniformly random $r \in \set{0,1}^n$. 
    \item Compute the two $1$-bit parities: \[\ell_1 =  \bigoplus_{i \in [n]} r_i \cdot x(2i-1) \qquad \text{and} \qquad \ell_2 = \bigoplus_{i \in [n]} r_i \cdot x(2i). \] 
    \item Output that the answer is $1$ if $\ell_1 = \ell_2$. 
\end{enumerate}
This simple algorithm succeeds with constant advantage, but it does not explicitly store any bit. When a random query $(j, x(j))$ arrives at time step $t$, and this is the first time step that $x(j)$ was seen, the value of the sketches can be updated. Otherwise, the algorithm moves to the next random query.\footnote{This algorithm is the same as the well-known $O(1)$ randomized communication cost protocol for two-party Equality \cite{Yao79}.}

The final sketches contain very little information about the individual
input bits. In fact, even conditioned on $r$, the pair of $(\ell_1,\ell_2)$ contains at
most two bits of information about the entire input. 
The algorithm can compress information about the input $x$ into $O(1)$ bits using linear sketches. However, implementing the above algorithm in the random query model still takes $\Ot(n)$ space with $\Ot(n)$ samples, because it records whether any index $j \in [2n]$ was seen before or not. 

 Our approach builds on the intuition underlying previous works: the algorithm must keep track of which indices have appeared among the random queries. Even when a direct reduction from ME cannot be used, as in~\cite{Dinur24}, for functions like Equality that have both low influence and degeneracy,
we can directly measure information about the randomness of the
queries. This helps us get around the fact that not enough information may be stored about the input $x$ itself. 
 We first illustrate this approach with a weaker bound of $\Omega(n\sqrt{n})$.

\subsection{An $\Omega(n\sqrt{n})$ lower bound via direct sum on random queries}
\label{sec:ov-rehaul-blocks}

We first try a direct sum over the $n$ pairs, measuring how much
information the algorithm keeps about \emph{random queries} to each pair. We formulate this direct sum via communication complexity, which is a standard tool in proving lower bounds on space complexity.

Let $P_t$ denote the state of the algorithm after the first
$t$ queries, with $P_0$ being its initial state. 
We choose a block length $1\leq L\leq n$ and let $q=(T/L)$ be the total number of blocks. We have $q$ players, and each player receives $L$ independent samples of the form $(j, x(j))$ for randomly chosen $j \in [2n]$. They want to output the answer $\EQ(x)$ by communicating on a shared blackboard. 

More concretely, player $a$ gets:
\[
(j_{(a-1)L + 1}, x(j_{(a-1)L + 1}), \qquad (j_{(a-1)L + 2}, x(j_{(a-1)L + 2}),\qquad \ldots, \qquad (j_{aL}, x(j_{aL})),
\]
where each $j_t$ is chosen uniformly at random and independently from $[2n]$. The last player is responsible for outputting $\EQ(x)$ on the blackboard. 

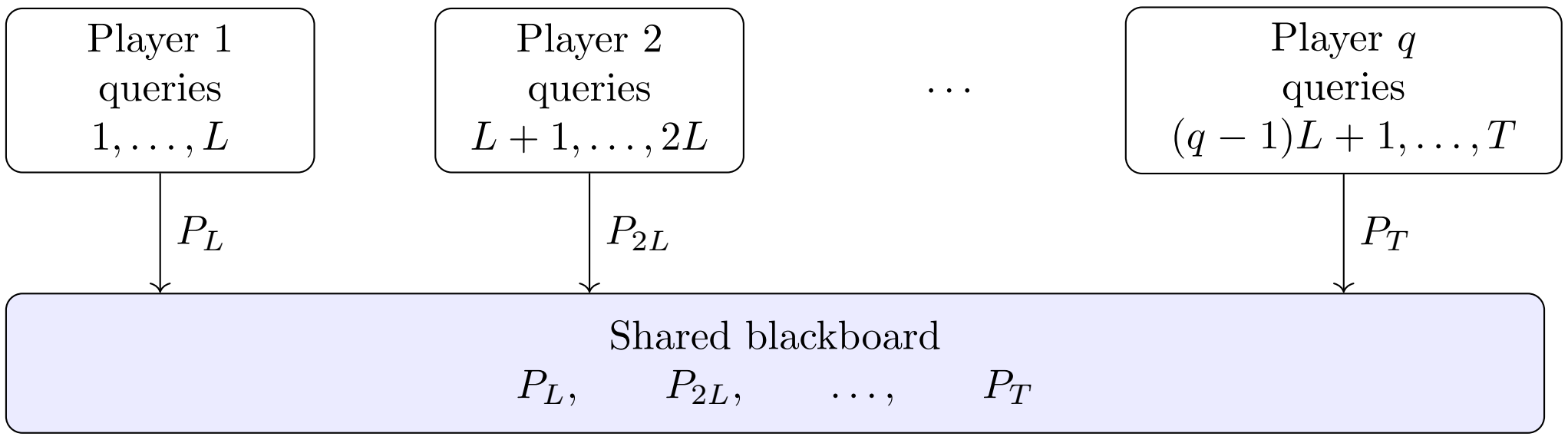
\begin{figure}[ht]
\centering
\begin{tikzpicture}[font=\small,
    block/.style={draw,rounded corners,text width=2.4cm,minimum height=1.2cm,align=center}]
    \node[block] (a) at (0,0) {Player $1$\\queries\\$1,\ldots,L$};
    \node[block] (b) at (3.7,0) {Player $2$\\queries\\$L+1,\ldots,2L$};
    \node[inner sep=1pt] (dots) at (6.8,0) {$\ldots$};
    \node[block,text width=3.5cm] (c) at (10.2,0)
        {Player $q$\\queries\\$(q-1)L+1,\ldots,T$};
    \node[draw,rounded corners,fill=blue!8,text width=13cm,
        minimum height=1.2cm,align=center] (board) at (5.3,-2.35)
        {Shared blackboard\\
         $P_L,\;P_{2L},\;\ldots,\;P_T$};
    \draw[->] (a.south)--node[right] {$P_L$} (a.south |- board.north);
    \draw[->] (b.south)--node[right] {$P_{2L}$} (b.south |- board.north);
    \draw[->] (c.south)--node[right] {$P_T$} (c.south |- board.north);
\end{tikzpicture}
\caption{For $a \in [q]$, each player $a$ receives its own block of $L$ samples. They can simulate any $S$-space algorithm with a total communication of $q \cdot S$ bits by sending the intermediate memory states. }
\label{fig:ov-rehaul-blocks}
\end{figure}

It is easy to see that if an algorithm uses $S$-bits of space to solve $\EQ(x)$, the $q$ players in the communication problem can solve for $\EQ(x)$ with a total communication of $q \cdot S$, by sending the memory state of the algorithm and simulating it on their samples. See \Cref{fig:ov-rehaul-blocks} for an illustration. Hence, proving a lower bound on the communication required to solve $\EQ(x)$ by the $q$-players also proves a lower bound on $S$.

To obtain a lower bound for the communication problem via direct sum, we want to measure the
information on the blackboard about the random queries to
each pair. If each of the $n$ pairs $\set{2i-1, 2i}$ for $i \in [n]$ forces a constant amount of
information, adding these costs would give an $\Omega(n)$ lower
bound on the total communication. 

We will see later why an $\Omega(n)$ communication lower bound gives
$TS=\Omega(n\sqrt n)$. First, we formulate the small problem that
arises from one pair when $L=\Theta(\sqrt n)$, i.e., when focusing on just one pair $\set{2i-1, 2i}$, what the $q$ players solve for.

\paragraph{The information bound from a two-bit problem.}
There are two independent uniform bits $\alpha,\beta$ and $q$ players.
Each player independently receives $\alpha$ with probability
$L/2n$, receives $\beta$ with probability $L/2n$, and receives
neither bit otherwise. Each player also knows which bit they receive. The
players write on a shared blackboard in order and want to compute
$\alpha\oplus\beta$. 

Observe that this is exactly similar to the $q$-player communication complexity problem we defined to analyze the branching program:
\begin{itemize}
    \item The probability that a random index $j \in [2n]$ is received by a player is $L/2n$, as they receive $L$ uniform samples. 
    \item They communicate to the blackboard to solve for whether there exists a pair with $x(2j-1) \neq x(2j)$, which corresponds to finding $\alpha \oplus \beta$.
\end{itemize}

\begin{figure}[ht]
\centering
\begin{tikzpicture}[font=\small,
    block/.style={draw,rounded corners,text width=2.8cm,minimum height=1.5cm,align=center}]
    \node[block] (a) at (0,0)
        {Player $1$ gets\\
         $\alpha$ w.p. $L/2n$\\$\beta$ w.p. $L/2n$\\
         independently};
    \node[block] (b) at (3.7,0)
        {Player $2$ gets\\
         $\alpha$ w.p. $L/2n$\\$\beta$ w.p. $L/2n$\\
         independently};
    \node[inner sep=1pt] (dots) at (6.8,0) {$\ldots$};
    \node[block] (c) at (10.2,0)
        {Player $q$ gets\\
         $\alpha$ w.p. $L/2n$\\$\beta$ w.p. $L/2n$\\
         independently};
    \node[draw,rounded corners,fill=blue!8,text width=13.4cm,
        minimum height=1.2cm,align=center] (board) at (5.1,-2.45)
        {Shared blackboard\\
         $\Pi=(\pi_1,\pi_2,\ldots,\pi_q)$};
    \draw[->] (a.south)--node[right] {$\pi_1$} (a.south |- board.north);
    \draw[->] (b.south)--node[right] {$\pi_2$} (b.south |- board.north);
    \draw[->] (c.south)--node[right] {$\pi_q$} (c.south |- board.north);
\end{tikzpicture}
\caption{The two-bit problem focused on only one pair $\set{2j-1, 2j}$. Each player
receives $\alpha$ with probability $L/2n$ and $\beta$ with probability $L/2n$ independently. The players write messages in order on the
shared blackboard, and player $q$ outputs $\alpha\oplus\beta$.}
\label{fig:ov-rehaul-sparse-game}
\end{figure}
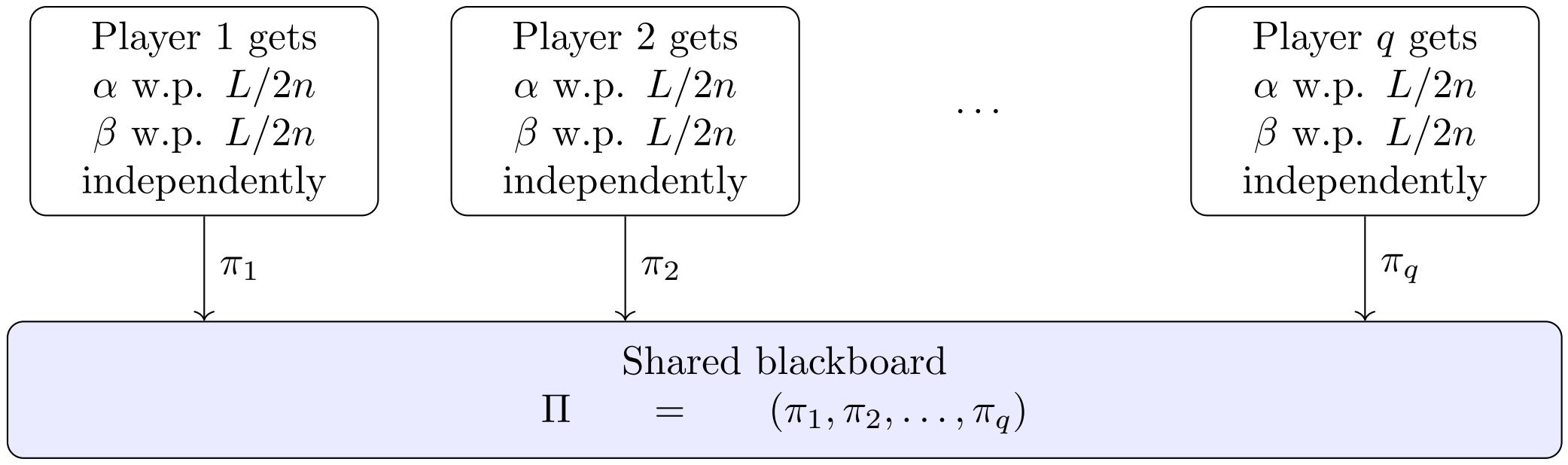

Refer to \Cref{fig:ov-rehaul-sparse-game} for an illustration. 

A player can receive both bits and compute their parity locally.
This happens with probability $L^2/4n^2$, so the probability that any
player receives both is at most $qL^2/4n^2$. If this is $o(1)$, or even
a sufficiently small constant, conditioning on no player receiving
both bits preserves constant advantage.

Let $\rz_a \in \set{0,1}^2$ denote the two indicator random variables for whether player $a$ receives $\alpha$ and $\beta$ respectively.
We can prove
an $\Omega(1)$ external information lower bound: the messages must reveal a
constant amount about which players received which bits, even after
conditioning on equal input values and the public randomness. More concretely, we show that there exists some $b \in \set{0,1}$ with:
\begin{equation}\label{eq:ov1}
\mi{\Prot}{\rz_1, \ldots, \rz_q \mid \alpha = \beta = b, \text{Public Randomness}} = \Omega(1),
\end{equation}
for any protocol transcript $\Prot$ that outputs $\alpha \oplus \beta$ correctly with high enough constant advantage. 
Notice that we only have guarantees that the protocol reveals information about $\alpha \oplus \beta$; we are still able to use it to argue that information about $\rz_a$ will also be revealed. Intuitively, any message from any player that changes the distribution of $\alpha, \beta$ significantly will also reveal whether this player got any of these two bits. We crucially use that with high enough constant probability, no player receives \emph{both} bits $\alpha, \beta$. Otherwise, they can communicate only when $\alpha \neq \beta$, and the bound in \Cref{eq:ov1} would be false.

Now, we can use the original protocol for the $q$-player game that solves $\EQ$ to solve this two-bit parity problem. 
We embed $\alpha,\beta$ in a randomly chosen pair $\set{2\istar-1, 2\istar}$ for $\istar \in [n]$. For every other pair $\set{2j-1, 2j}$ for $j \neq \istar$, we set the two bits $x(2j-1)= x(2j)$ with a uniformly random bit chosen with \emph{public randomness}.
 Observe that input $x$ (barring $x(2\istar-1), x(2\istar)$) being chosen with public randomness is necessary for the players to run the protocol for $\EQ$: the samples that each player sends are correlated through $x$. After input $x$ is fixed, the protocol can be run using standard ideas in embedding arguments.
 
 \Cref{eq:ov1} measures information conditioned on $\alpha = \beta$, and hence, all pairs $\set{2j-1,2j}$ for $j \in [n]$ look identical to the protocol for EQ. An $\Omega(n)$ lower bound on the following information cost can be proven:

\begin{equation}\label{eq:ov2}
    \mi{\Prot_{eq}}{j_1,\ldots,j_T\mid x,\text{public randomness}}
    =\Omega(n),
\end{equation}
where $\Prot_{eq}=(P_L,P_{2L},\ldots,P_T)$ is the transcript of the
protocol for $\EQ$, and $j_1,\ldots,j_T$ are the queried indices at all $T$ time steps.
The input $x$ has an independent uniform common bit in each pair.
The bound holds when $qL^2/n^2$ is sufficiently small, as required
for the two-bit parity problem lower bound in \Cref{eq:ov1}.

\paragraph{Balancing collisions and block length.}
We now see how \Cref{eq:ov2} gives a time-space lower bound. The
transcript $\Prot_{eq}$ contains $q$ states of at most $S$ bits each,
so its information cost is at most $qS$. Hence,
\begin{equation}\label{eq:ov3}
    \Omega(n)\leq qS=\paren{\frac{T}{L}}S,
    \qquad\text{giving}\qquad TS=\Omega(nL).
\end{equation}
We would like to take $L$ as large as possible. However, increasing
$L$ also makes it more likely that a player receives both bits of
the hidden pair.  The
probability of a collision is $O(qL^2/n^2)=O(TL/n^2)$, and \Cref{eq:ov1} requires this to be some small constant. 
If so, we directly get $TS=\Omega(nL)$ from \Cref{eq:ov3}. Otherwise, $T=\Omega(n^2/L)$, in either case, we have:
\[
TS = \Omega(\min\set{nL, n^2/L}).
\]
Choosing
$L=\Theta(\sqrt n)$ balances these two bounds and gives
$TS=\Omega(n\sqrt n)$.

To get a stronger lower bound, we need to also look at the information retained by the memory states \emph{inside} each block. The blackboard model charges only the boundary states $P_L$, $P_{2L}$, $\ldots$, $P_{T}$, and we need a notion of information cost that considers all time steps $t \in [T]$. 

\subsection{Our Notion of Information Cost}

Recall that $P_t$ denotes the state of the algorithm after the first $t$ queries.
The information cost introduced in \cite{BravermanGW20} gives a way to measure what the memory states have retained across all time steps. We adapt their approach with our insights: we measure the information about
the queried indices $j_1,\ldots,j_T$ across all the time steps, after conditioning on the entire input $x$. 
\begin{equation*}
\ic_1:= \sum_{t=1}^{T}\sum_{s=1}^{t}
    I(P_t;j_s\mid P_{s-1},x,\text{public randomness})\leq T \cdot S.
\end{equation*}
These indices remain independent even
after conditioning on input $x$ and the public randomness, and thus, an upper bound of $T \cdot S$ can be proven on this cost similar to \cite{BravermanGW20}.

Each term with $s \leq t$ measures how much information the state $P_t$ has about the
index $j_s$, given the state $P_{s-1}$. For each time step $t \in [T]$, all earlier queries $s\leq t$ are accounted for in the information cost. \commented{\snote{This sentence sounds weird and I am not sure what you intended to say here.}.}
For example, if index $j_t$ is retained for $d$ steps after $t$, this adds a value of $\Omega(d)$ to $\ic_1$. Now, it is sufficient to prove a lower bound of $\Omega(n^2)$ on $\ic_1$.

We make the following small change
to the query model, as it makes the proof more convenient.

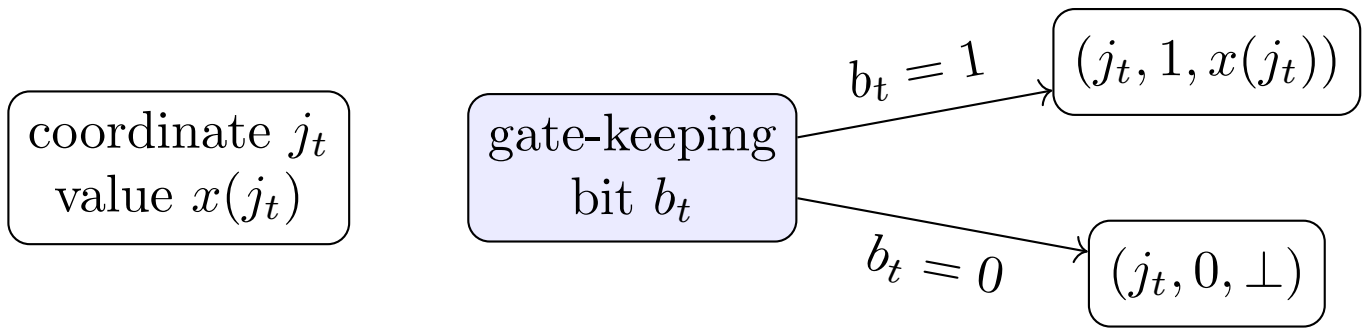
\begin{figure}[ht]
\centering
\begin{tikzpicture}[font=\small,
    box/.style={draw,rounded corners,align=center,minimum height=.7cm}]
    \node[box] (sample) at (0,0) {coordinate $j_t$\\value $x(j_t)$};
    \node[box,fill=blue!8] (gate) at (3,0) {gate-keeping\\bit $b_t$};
    \node[box] (yes) at (6.8,.7) {$(j_t,1,x(j_t))$};
    \node[box] (no) at (6.8,-.7) {$(j_t,0,\perp)$};
    \draw[->] (gate)--node[above,sloped] {$b_t=1$} (yes);
    \draw[->] (gate)--node[below,sloped] {$b_t=0$} (no);
\end{tikzpicture}
\caption{The gate-keeping bit $b_t$ determines whether the queried
value $x(j_t)$ is given. The index $j_t$ is always given. }
\label{fig:ov-rehaul-reveal}
\end{figure}

\paragraph{Gate-keeping bits.}
At time step $t$, the sample contains an independent uniform bit $b_t$ called the gate-keeper. 
 The algorithm always receives the index $j_t$, but receives $x(j_t)$
value only when $b_t=1$. It is easy to see that any algorithm in the original model can also be used to solve for $\EQ(x)$ in this model, by ignoring samples with $b_t = 0$ and feeding it only the samples with $b_t = 1$. This needs $O(T)$ samples, and $O(\log T)$ extra bits of space to count how many such samples have been fed so far, i.e., which layer of the original branching program the simulation is in. \commented{ \snote{I don't understand why you end up with a logarithmic overhead}.} An illustration of what the new sample contains is given in \Cref{fig:ov-rehaul-reveal}.

We can condition on the input $x$, all the queried indices, and
the public randomness. The gate-keeping bits remain independent,
so there is still randomness in whether each value was received.
This is our final notion of information cost:
\begin{equation}\label{eq:ov-rehaul-cost}
    \ic_2:=\sum_{t=1}^{T}\sum_{j=1}^{t}
    I(P_t;b_j\mid P_{j-1},x,\text{indices, public randomness})
    \leq TS.
\end{equation}
The upper bound of $T \cdot S$ holds because the gate-keeping bits are independent across all the samples. 
Our aim is to prove an $\Omega(n^2)$ lower bound on $\ic_2$.

\paragraph{The two-bit information bound.}
To lower bound $\ic_2$ using a similar direct sum as in \Cref{sec:ov-rehaul-blocks}, we use a version of the two-bit problem where the input contains gate-keeping bits also. 

\begin{figure}[ht]
\centering
\begin{tikzpicture}[font=\small,
    block/.style={draw,rounded corners,text width=2.4cm,minimum height=1.2cm,align=center}]
    \node[block] (a) at (0,0)
        {Player $1$\\private input\\$(\rho_1,\theta_1,v_1)$};
    \node[block] (b) at (3.7,0)
        {Player $2$\\private input\\$(\rho_2,\theta_2,v_2)$};
    \node[inner sep=1pt] (dots) at (6.8,0) {$\ldots$};
    \node[block] (c) at (10.2,0)
        {Player $k$\\private input\\$(\rho_k,\theta_k,v_k)$};
    \node[draw,rounded corners,fill=blue!8,text width=13cm,
        minimum height=1.2cm,align=center] (board) at (5.3,-2.35)
        {Shared blackboard\\
         $\Pi=(\pi_1,\pi_2,\ldots,\pi_k)$};
    \draw[->] (a.south)--node[right] {$\pi_1$} (a.south |- board.north);
    \draw[->] (b.south)--node[right] {$\pi_2$} (b.south |- board.north);
    \draw[->] (c.south)--node[right] {$\pi_k$} (c.south |- board.north);
\end{tikzpicture}
\caption{Player $a$ receives $(\rho_a,\theta_a,v_a)$, where $\rho_a$ points to one of $\alpha, \beta$ at random, $\theta_a$ plays the role of gate-keeper and is uniform over $\set{0,1}$. Lastly, $v_a$
is the value that $\rho_a$ points to if $\theta_a=1$, and $\perp$ otherwise.
The players communicate on the blackboard to compute $\alpha \oplus \beta$.}
\label{fig:ov-rehaul-game}
\end{figure}
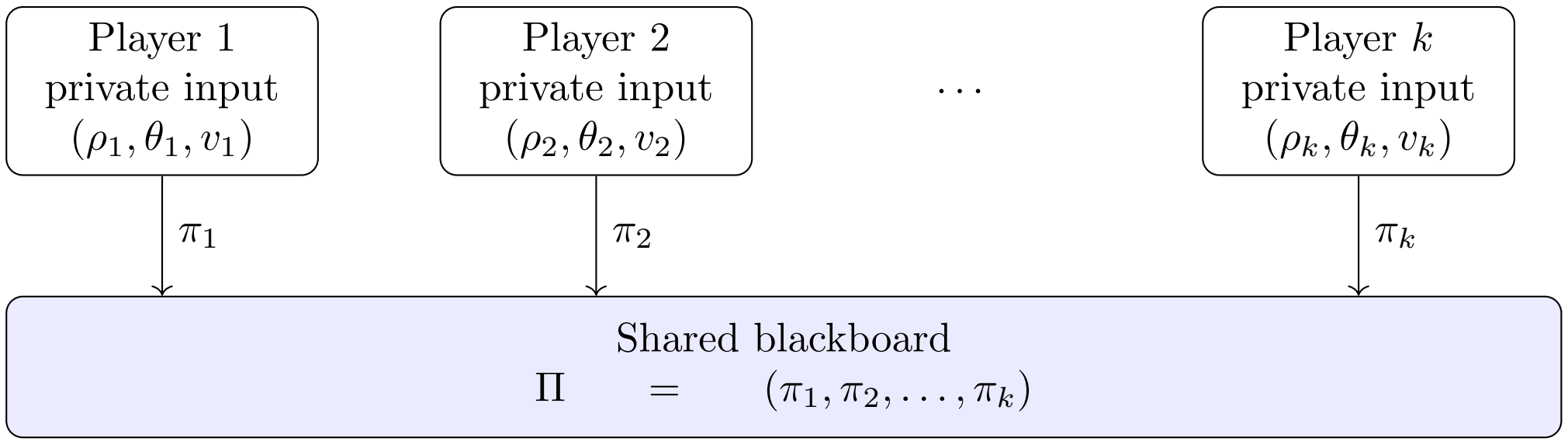

There are $k$ players and two independent
uniform bits $\alpha,\beta$. Each player $a$ independently gets a
uniform pointer $\rho_a$ to one of $\alpha$ or $\beta$ and an independent fair bit
$\theta_a$. It knows both, but receives the value that pointer $\rho_a$ points to only
if $\theta_a=1$. The players communicate on a blackboard to compute
$\alpha\oplus\beta$. An illustration is provided in \Cref{fig:ov-rehaul-game}.

We can prove that any protocol with constant advantage reveals
$\Omega(1)$ information about $\theta_1,\ldots,\theta_k$ in its
transcript $\Prot$. As in \Cref{eq:ov1}, for some bit $u\in\set{0,1}$,
\begin{equation}\label{eq:lb-theta-ov}
    \mi{\Prot}{\theta_1,\ldots,\theta_k\mid\rho_1,\ldots,\rho_k,\alpha=\beta=u,\text{public randomness}}
    =\Omega(1).
\end{equation}
This statement appears as \Cref{lem:two-bit-game}.
Again, \Cref{eq:lb-theta-ov} holds only because each player gets at most one of $\set{\alpha, \beta}$, as we have conditioning on $\alpha = \beta = u$. We use Jensen-Shannon distance \cite{EndresS03} to analyze a successful protocol of the two-bit parity problem, and the details can be found in \Cref{sec:2bitgame}.

\subsection{Simulation of the Algorithm}

We now explain how the two-bit lemma from \Cref{eq:lb-theta-ov} can be used to get the desired time-space
bound for $\EQ$. We need to relate the information revealed by
the protocol to the information retained by the algorithm for $\EQ$ over time.

The framework is similar to \Cref{sec:ov-rehaul-blocks}, except for one important change: \emph{the number of blocks that each player simulates is a random variable} based on the sequence of indices. 
 Similar to \Cref{sec:ov-rehaul-blocks}, we choose a uniform pair $\set{2\istar-1, 2\istar}$ and embed
$\alpha,\beta$ in these bits. We fill all other pairs $\set{2j-1, 2j}$ with independent
uniform bits $x(2j-1) = x(2j)$ chosen publicly.

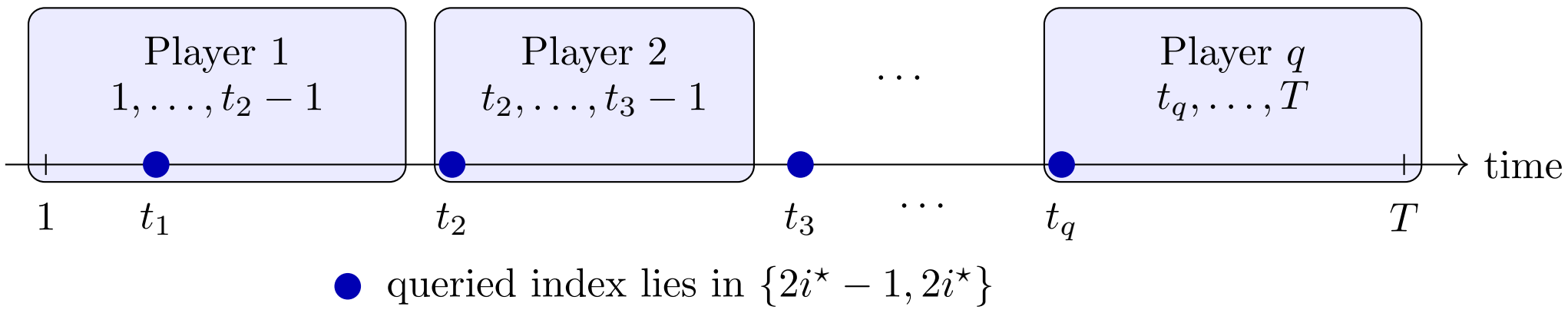
\begin{figure}[ht]
\centering
\begin{tikzpicture}[font=\small,
    interval/.style={draw,rounded corners,fill=blue!8},
    hit/.style={circle,fill=blue!70!black,inner sep=2.3pt}]
    \draw[interval] (0,-.15) rectangle (3.25,1.35);
    \draw[interval] (3.5,-.15) rectangle (6.25,1.35);
    \draw[interval] (8.75,-.15) rectangle (12,1.35);

    \node at (1.625,.95) {Player $1$};
    \node at (1.625,.55) {$1,\ldots,t_2-1$};
    \node at (4.875,.95) {Player $2$};
    \node at (4.875,.55) {$t_2,\ldots,t_3-1$};
    \node at (7.5,.75) {$\ldots$};
    \node at (10.375,.95) {Player $q$};
    \node at (10.375,.55) {$t_q,\ldots,T$};

    \draw[->] (-.2,0)--(12.4,0) node[right] {time};
    \draw (.15,-.09)--(.15,.09);
    \draw (11.85,-.09)--(11.85,.09);
    \node[below] at (.15,-.2) {$1$};
    \node[below] at (11.85,-.2) {$T$};
    \foreach \x/\lab in {1.1/1,3.65/2,6.65/3,8.9/q}{
        \node[hit] at (\x,0) {};
        \node[below] at (\x,-.2) {$t_{\lab}$};
    }
    \node[below] at (7.7,-.2) {$\ldots$};

    \node[hit] at (2.75,-1.05) {};
    \node[anchor=west] at (2.95,-1.05)
        {queried index lies in $\set{2\istar-1,2\istar}$};
\end{tikzpicture}
\caption{The queries at $t_1,t_2,\ldots,t_q$ determine the simulation
blocks. Player $1$ also handles the queries before $t_1$.}
\label{fig:ov-rehaul-intervals}
\end{figure}

In addition, we also publicly sample all the queried indices independently and uniformly.\footnote{There is a small technicality where the pointer $\rho_a$ from the players of the two-bit parity problem are also used in sampling the sequence of indices, but we ignore it in the overview.}
Once the sequence is fixed, we define $t_a$ as the time step of the $a$-th query to one of
$\set{2\istar-1, 2\istar}$. 
We use $q\approx T/n$ players to simulate successive blocks, as shown in
\Cref{fig:ov-rehaul-intervals}. Player $1$ starts at time $1$, and
each later player $a$ takes over at time $t_a$.

Each player runs at most one query to $\set{2\istar-1,2\istar}$.
Player $a$ can supply this query at $t_a$ using its input
$(\rho_a,\theta_a,v_a)$ from the two-bit game: $\rho_a$ gives the
coordinate, and $\theta_a$ determines value of gate-keeping bit $b_{t_a}$. The corresponding input value $x(2\istar-1)$ or $x(2\istar)$ can also be provided by player $a$ if $\theta_a = b_{t_a} = 1$. 
All other $x(j)$ for $j \notin \set{2\istar-1, 2\istar}$ are public, and so are the indices. Therefore, each player can generate the other
queries using privately chosen gate-keeping bits.

Player $1$ starts from $P_0$. Each player simulates its block and
writes the resulting state on the blackboard, so the next player
can continue from this state.

\begin{figure}[ht]
\centering
\begin{tikzpicture}[font=\small,
    block/.style={draw,rounded corners,text width=2.4cm,minimum height=1.2cm,align=center}]
    \node[block] (a) at (0,0) {Player $1$\\queries\\$1,\ldots,t_2-1$};
    \node[block] (b) at (3.7,0) {Player $2$\\queries\\$t_2,\ldots,t_3-1$};
    \node[inner sep=1pt] (dots) at (6.8,0) {$\ldots$};
    \node[block,text width=3.5cm] (c) at (10.2,0)
        {Player $q$\\queries\\$t_q,\ldots,T$};
    \node[draw,rounded corners,fill=blue!8,text width=13cm,
        minimum height=1.2cm,align=center] (board) at (5.3,-2.35)
        {Shared blackboard\\
         $P_{t_2-1},\;P_{t_3-1},\;\ldots,\;P_T$};
    \draw[->] (a.south)--node[right] {$P_{t_2-1}$} (a.south |- board.north);
    \draw[->] (b.south)--node[right] {$P_{t_3-1}$} (b.south |- board.north);
    \draw[->] (c.south)--node[right] {$P_T$} (c.south |- board.north);
\end{tikzpicture}
\caption{Each player writes the state at the end of its block on
the shared blackboard for the next player.}
\label{fig:ov-rehaul-simulation}
\end{figure}
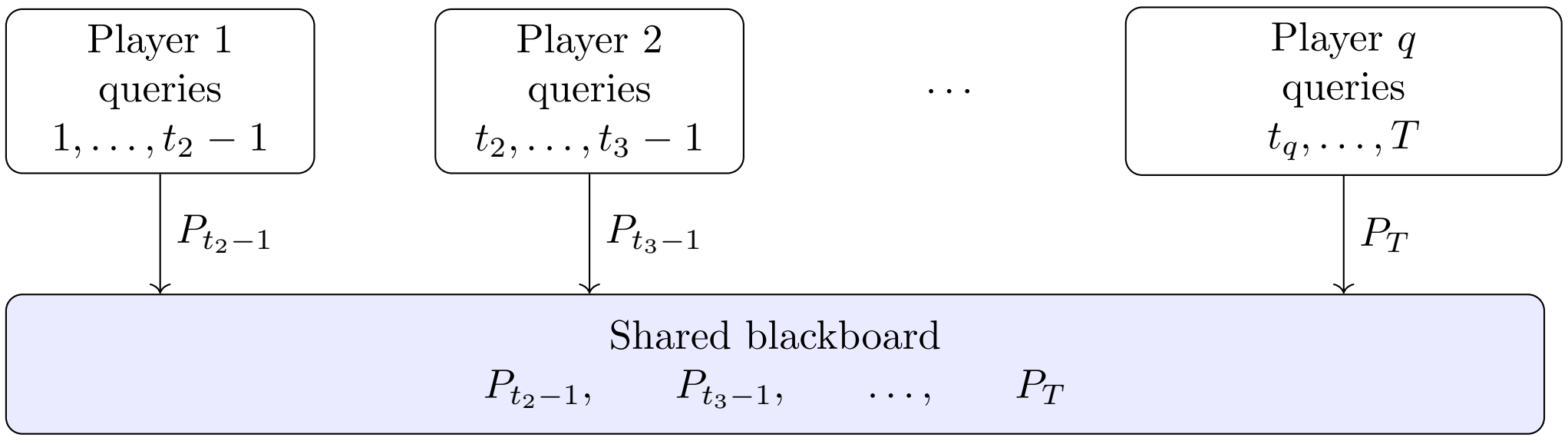

We can prove that the information cost of this protocol for the two-bit problem from \Cref{eq:lb-theta-ov} is at most $O(1/n^2) \cdot \ic_2$. There are two distinct sources of $(1/n)$ factors:

\begin{description}
    \item[Direct Sum] This part is largely similar to the $\Omega(n)$ factor with fixed block length in \Cref{sec:ov-rehaul-blocks}. There are $n$ pairs $\set{2j-1, 2j}$, and the value of $\istar$ is uniform over $[n]$. Each memory state of the algorithm only stores $(1/n)$-fraction of information about the specific $\set{2\istar-1, 2\istar}$ on average. 
    \item[Random Block Lengths] The other $n$-factor arises from the unpredictability of block lengths. After the query at time $t_1$, any state
$P_t$ with $t<T$ is written just when the next query belongs to
$\istar$. This happens with probability $1/n$, independently of
everything the algorithm has seen till time step $t$. 
Any state that contains information about the gate-keeping bits of $\set{2\istar-1, 2\istar}$ only appears on the blackboard with probability $1/n$.  A similar use of unpredictable arrivals appears in \cite{BrownBS22}.
\end{description}

The two-bit lower bound from \Cref{eq:lb-theta-ov} and these two factors give:
\begin{equation*}
    TS\geq
    \underbrace{\Omega(1)}_{\text{two-bit information bound}}
    \cdot\underbrace{n}_{\text{direct sum}}
    \cdot\underbrace{n}_{\text{random block lengths}}
    =\Omega(n^2),
\end{equation*}
for the $\EQ$ function. 

For general XOR-functions $f\circ\oplus_2$, the same argument over
$\Omega(\sens[f])$ sensitive pairs gives
$T(S+\log T)=\Omega(n\cdot\sens[f])$, and this concludes the overview.

\section{Preliminaries}

\paragraph{Notation.} We use sans-serif font $(\rA)$ to denote random variables. For any $x \in \{0,1\}^n$, and index $i \in [n]$,  we use $x(i)$ to denote the value of $x$ at index $i$. We use $x(-i)$ to denote the value of $x$ at the $(n-1)$ indices excluding $i$. More generally for any set $S \subseteq [n]$, we use $x(S)$ to denote the sequence of $\card{S}$ many values of $x$ at the indices in $S$. For an event $E$, we use $\ind[E]$ to denote its indicator.

We use $e_i$ for $i \in [n]$ to denote the element of $\set{0,1}^n$ which has value $1$ at index $i$ and $0$ everywhere else.

We use the following standard form of Chernoff bound and its extension to negatively correlated variables in~\cite{PanconesiS97}. 

\begin{proposition}[Chernoff Bound; cf.~\cite{DubhashiP09}]\label{prop:chernoff}
	Let $X_1,\ldots,X_n$ be $n$ independent random variables in $[0,1]$ and $X := \sum_{i=1}^{n} X_i$. For any $\delta > 0$ and $\mu_{min} \leq \expect{X} \leq \mu_{max}$, 
	\begin{align*}
		&\Pr\paren{X \geq (1+\delta) \cdot \mu_{max}} \leq \exp\paren{-\frac{\delta^2 \cdot \mu_{max}}{2+\delta}} \quad , \quad \Pr\paren{X \leq (1-\delta) \cdot \mu_{min}} \leq \exp\paren{-\frac{\delta^2 \cdot \mu_{min}}{2}}.
	\end{align*} 	
	Moreover, the upper tail bound continues to hold as long as $X_1,\ldots,X_n$ are negatively correlated, i.e., for every $S \subseteq [n]$, 
	\[
		\expect{\prod_{i \in S} X_i} \leq \prod_{i \in S} \expect{X_i}. 
	\]
\end{proposition}

We use standard definitions and facts from information theory throughout this paper; more details can be found in \Cref{sec:information-theory}.
We refer the reader to \cite{CoverT03} for a comprehensive introduction to the topic.

\subsection{Model of Computation}\label{subsec:model}

We use the random query model introduced in \cite{RazZ20} and further studied in \cite{Dinur24}. There is a Boolean function $f:\set{0,1}^n \rightarrow \set{0,1}$, and some input $x$ that we want to compute. 

The input is accessed via random samples of the form $(i, x(i))$ where $i \in [n]$ is chosen randomly according to some distribution. We study only the uniform distribution  over $[n]$.

The computation can be modeled with a randomized branching program. More specifically, for each $n \geq 1$, there is a distribution $\cDalg_n$ which is supported over branching programs with only private randomness with the following properties:
\begin{enumerate}[label=$(\roman*)$]
    \item There are $T+1$ layers, each with $2^S$ vertices. In each layer, every node is assigned a distinct label from $\set{0,1}^S$. Thus, a node is uniquely specified by its layer $t \in [T] \cup \set{0}$ together with its $S$-bit label.
    \item The first layer has a starting node from which computation proceeds. 
    \item Each node in the first $T$ layers has a distribution over $2n$ outgoing edges, corresponding to the $2n$ possible inputs of $(i, x(i))$. One such set of outgoing edges is sampled, and computation moves to the next layer based on the input $(i, x(i))$. The sampled edges are not revealed to the later nodes.  
    \item Each node in the final layer has a label from $\set{0,1}$ which corresponds to the output of the branching program on input $x$.
\end{enumerate}

For any input $x \in \set{0,1}^n$, a branching program from distribution $\cDalg_n$ is picked, and this is used to output the final answer. We use $\alg(x)$ to denote this answer. 

The probability of success of the branching program is defined as:
\[
\text{Probability of success on input $x$ is}  \Pr[\alg(x) = f(x)],
\]
where the randomness is over the choice of branching program from distribution $\cDalg_n$, the choice of sampled indices at each node, and the randomness of the indices $i \in [n]$. 
The final success probability is the minimum success probability over all input $x$.


\subsection{Boolean Function Complexity Measures}\label{subsec:boolean-complexity}
For any Boolean function $f:\set{0,1}^n \rightarrow \set{0,1}$, we frequently refer to some notions of complexity, and we define them here. We refer the readers to \cite{AaronsonBKRT21} for more details and other complexity measures.

\begin{definition}[Sensitivity Graph]\label{def:sens-graph}
    Given a Boolean function $f:\{0,1\}^n \rightarrow \{0,1\}$, we define a bipartite graph $G_f = (V_1 \cup V_0, E_f)$ as follows:
    \begin{align*}
    V_0 &= \set{x \in \set{0,1}^n \mid f(x) = 0} \\
    V_1 &= \set{x \in \set{0,1}^n \mid f(x) = 1} \\
    E_f &= \set{(x,y) \mid x \in V_0, y \in V_1, \text{$x$ and $y$ differ in exactly one bit}}.
    \end{align*}
\end{definition}

Any vertex $x \in \set{0,1}^n$ must belong to either $V_1$ or $V_0$ as $f$ is a total function. The degree of a vertex $x$, denoted by $\sens[f,x]$ is the total number of bits on which $x$ is sensitive, i.e.,
\[
    \sens[f,x] = \card{ i \in [n] \mid f(x) \neq f(x \oplus e_i)}.
\]
The maximum degree in $G_f$ is called the sensitivity of $f$ and we use $\sens[f]$ to denote it. Moreover, the bits or positions $i \in [n]$ at which $f(x) \neq f(x \oplus e_i)$ are referred to as the sensitive bits of $x$.
We use $\sens_1[f, x]$ and $\sens_0[f,x]$ to denote the number of sensitive bits for $x$ where the value in $x$ is $1$ and $0$ respectively. Trivially, we have $\sens_1[f,x]+\sens_0[f,x] = \sens[f,x]$.

The average degree in $G_f$, denoted by $\inf[f]$ is called the influence of $f$.
\[
    \inf[f] = \frac1{2^n} \sum_{x \in \set{0,1}^n} \sens[f,x].
\]

We use $\dgn[f]$ to denote the degeneracy of the graph $G_f$. Alternately, $\dgn[f]$ is the minimum value such that any vertex-induced subgraph of $G_f$ contains a vertex of degree at most $\dgn[f]$.

\section{Composition with Parity}

We begin this section with the set up for the proof. Then we formalize the small XOR on two bit problem that we prove a lower bound for in the blackboard model of communication. Finally, we show that the lower bound for our two-bit XOR problem also implies a lower bound for the $g_{\foplus}$ function via a direct sum. 

\subsection{Our Notion of Information Cost}

We give a different type of input to the algorithm, and argue that this is the same up to constant factors in the length and width of the branching program. This modification makes the rest of the analysis simpler.

\noindent
\paragraph{New form of Input given to Branching Programs.}
In \Cref{subsec:model}, the branching programs receive as input $(i, x(i))$ where $i \in [n]$ is chosen uniformly at random. Instead, the branching program is given as input $(i, b, w(i))$ where:
\begin{itemize}
    \item The index $i \in [n]$ is chosen uniformly at random from $[n]$.
    \item The bit $b \in \set{0,1}$ is chosen uniformly at random independent of $i$.
    \item The value $w(i)$ is equal to $x(i)$ if $b = 1$ and $\perp$ otherwise. 
\end{itemize}

If we fix $b = 1$ for all the samples, the input distribution is exactly the same as in \Cref{subsec:model}. It is easy to see that any branching program that works with inputs of the form $(i, x(i))$ can also be made to work with $(i, b, w(i))$.

\begin{claim}\label{clm:change-bp-input}
Let
$P$ be a branching program of length $T$ and width $2^S$ whose input
consists of independent samples
\[
(i_1,x(i_1)),\ldots,(i_T,x(i_T)),
\qquad
i_1,\ldots,i_T \text{ are uniform over $[n]$}.
\]

Consider instead the distribution in which each sample is of the form
$
(i,b,w(i))$ where $(i,b)$ is uniform over $[n] \times \set{0,1}$ independently and 
\[
w(i)=
\begin{cases}
x(i), & \text{ if }b=1,\\
\perp, & \text{ if }b=0.
\end{cases}
\]

Then, for every $\delta>0$, there is a branching program $P'$ of
length $T' = O(T + \log(1/\delta))$ and width at most $O(T \cdot 2^S)$
such that $P'$ exactly simulates $P$
 with probability at least $1-\delta$. 
\end{claim}

\begin{proof}
Branching program $P'$ maintains a state $(j,v)$, where
$j\in\{0,\ldots,T\}$ denotes the number of samples with $b = 1$ that have 
been received so far in the simulation, and $v$ is the state of $P$ after these
$j$ samples. In other words, $P'$ explicitly maintains the current node of $P$ while processing, by using $j$ to denote the layer and $v$ to point to the vertex in the specific layer.

Upon receiving a sample $(i,b,w(i))$, if $b=0$, then $P'$ ignores the
sample and leaves $(j,v)$ unchanged.  If $b=1$, then
$w(i)=x(i)$, and $P'$ feeds the sample $(i,x(i))$ to the $(j+1)^{\textnormal{th}}$ layer of
$P$. The state is updated from $(j,v)$ to $(j+1,v')$ if program $P$ moves from state $v$ to $v'$ on receiving $(i, x(i))$.

It is easy to see that this simulation is exact, as the indices $i$ are generated uniformly at random even in the inputs given to $P'$. For sufficiently large $T'  = 4T + 16 \ln(1/\delta)$, we know that in expectation, $P'$ receives $T'/2$ samples with $b = 1$ and getting the actual value of the input at random index $i$.
By the Chernoff bound for independent random variables from \Cref{prop:chernoff}, we get that the total number of samples with $b = 1$ is at least $T$ with probability at least $1-\delta$.

Finally, the state of $P'$ is one of the $(T+1) \cdot 2^S$ many nodes of branching program $P$. Hence the width of $P'$ is $O(T \cdot 2^S)$.
\end{proof}

Given this model, we are ready to define the notion of information complexity that we use in our proofs. 
We start by describing the distribution of inputs $x$ to the branching program that we analyze. 

We start with an assumption about the function $f$ that simplifies our proof. 
\begin{assumption}\label{ass:sens-0-bit}
    Given a function $f$, let $\xstar_f \in \set{0,1}^n$ be an input such that $\sens[f] = \sens[f, \xstar_f]$. That is, $\xstar$ is one of the inputs with the most number of sensitive bits in $f$.
    We assume that $\sens_0[f, \xstar_f]\geq \sens_1[f, \xstar_f]$. Under our assumption, $\sens_0[f, \xstar_f] = \Omega(\sens[f])$.
\end{assumption}

We only prove \Cref{thm:equal-lb} for functions obeying \Cref{ass:sens-0-bit} for now, and we will see how this proves a lower bound for other functions as well in \Cref{sec:extend-other-fns}.

\begin{tbox}
    \textbf{Distribution $\cD_{\foplus}$ over $\set{0,1}^{2n}$ which are inputs to $g_{\foplus}$:}

\vspace{2mm}

    Sample $y \in \set{0,1}^{2n}$ as follows:
    \begin{enumerate}[label=$(\roman*)$]
        \item  For $i \in [n]$, if $i$ is one of the sensitive bits for $f$ in $\xstar_f$ and $\xstar_f(i) = 0$, we sample $y(2i-1) \in \set{0,1}$ uniformly at random and fix $y(2i) =  y(2i-1)$.
        \item For all other $i \in [n]$, we fix $y(2i-1) =0$ and $y(2i) = \xstar_f(i)$.
    \end{enumerate}
\end{tbox}

It is easy to see that the output of $g_{\foplus}$ is fixed on all inputs sampled from $\cD_{\foplus}$.

\begin{observation}\label{obs:output-cDg}
    For any $y \sim \cD_{\foplus}$, $g_{\foplus}(y) = f(\xstar_f)$. Further, if $i \in [n]$ is sensitive for $\xstar_f$ with $\xstar_f(i) = 0$, then $g(y \oplus e_{2i}) = g(y \oplus e_{2i-1})  = \neg f(\xstar_f)$.
\end{observation}
\begin{proof}
    If $i$ is not sensitive for $\xstar_f$, or $\xstar_f(i) = 1$, then by definition, we have $y(2i-1) \oplus y(2i) = \xstar_f(i)$. Otherwise, we know $y(2i-1) \oplus y(2i) = 0$ by construction. If these $y(2i)$ or $y(2i-1)$ are flipped, the output changes, as $i$ is sensitive for $\xstar_f$.
\end{proof}

Our notion of information complexity will be measured with respect to inputs sampled from $\cD_{\foplus}$ on protocols which are correct for all inputs.

\paragraph{Notation.}
For $t \in [T]$, we use $\rP_t$ to denote the random variable corresponding to the $S$-bit label of the node reached after processing the first $t$ samples. This randomness is both over the samples of the input and the choice of which branching program to run from the distribution $\cDalg_n$ of the algorithm.

Let $(\ell_t, b_t, w_{\ell}(t))$ be the samples given at time $t \in [T]$ to the branching program. Let $\Lall$ denote the entire sequence of all the indices $\ell_1, \ldots, \ell_T$. Let $\Rpublic$ denote the random variable that corresponds to the choice of randomness in sampling the branching program.

\begin{tbox}
For any branching program $P$, distribution $\mu$ over inputs which solves $g_{\foplus}(y)$, we define the \textbf{information cost} as: \commented{\sg{We should define the notion with distribution ($\mu$) as argument instead of accepting input. So $\mathsf y$ is sampled from $\mu$}} \[
\ic_\mu(P) = \sum_{i \in[T]} \sum_{j  \leq i}\mi{\rP_i}{\rb_j \mid \rP_{j-1}, \ry, \rLall, \Rpublic},
\]
where $\ry$ is sampled from the input distribution $\mu$.
\end{tbox}

\commented{This information cost is the same as the one defined in \cite{BravermanGW20}, except we only focus on the mutual information between the memory states, and whether an actual input from $y$ was sent during the corresponding sample. \sg{I would delete the previous line as we already acknowledged BGW'20 in the introduction.} Using the ideas of \cite{BravermanGW20},} We show that it is sufficient to give a lower bound on the information cost to prove \Cref{thm:equal-lb}.

\begin{claim}\label{clm:information-cost-upper-bound}
For any branching program $P$ of length $T$ and width $2^S$, for any input distribution $\mu$, we have
\[
    \ic_{\mu}(P) \leq TS.
\]
\end{claim}

\begin{proof}
For the proof of this claim, we set $\rZ=(\ry,\rLall,\Rpublic)$. Recall that
$\rP_{<j}=(\rP_0,\ldots,\rP_{j-1})$, all the prior states of the branching program.

Conditioned on $\rP_{j-1}$ and $\rZ$, the next steps of the branching program depends only on the bits $\rb_j,\ldots,\rb_i$
and fresh private randomness. Thus,
$\rP_i\perp\rP_{<j-1},\rb_{<j}\mid\rP_{j-1},\rZ$, and using \itfacts{both-light-more},
\begin{align*}
\mi{\rP_i}{\rb_j\mid\rP_{j-1},\rZ}
&\leq \mi{\rP_i}{\rb_j\mid\rP_{j-1},\rP_{<j-1},\rb_{<j},\rZ}.
\end{align*}

Continuing, by chain rule of mutual information \itfacts{mi-chain}, we get,
\begin{align*}
    \mi{\rP_i}{\rb_j\mid\rP_{j-1},\rP_{<j-1},\rb_{<j},\rZ} &\leq \mi{\rP_i}{\rP_{j-1},\rb_j\mid\rP_{<j-1},\rb_{<j},\rZ}.
\end{align*}

We can add the terms in the information complexity directly now.
\begin{align*}
\sum_{j \leq i}\mi{\rP_i}{\rb_j\mid\rP_{j-1},\rZ} &\leq \sum_{j \leq i}
\mi{\rP_i}{\rP_{j-1},\rb_j\mid\rP_{<j-1},\rb_{<j},\rZ}\tag{using the two inequalities above}\\
&=\mi{\rP_i}{\rP_{< i},\rb_{\leq i}\mid\rZ}
  \tag{by chain rule of mutual information \itfacts{mi-chain}}\\
&\leq \en{\rP_i \mid \rZ} \tag{by definition of conditional mutual information} \\
&\leq S,
\end{align*}
where in the last inequality, we have used that each state of the branching program can be encoded in $S$ bits. 
Summing over all time steps $i \in [T]$, we get the proof. 
\end{proof}

Our proof strategy captures the underlying reason why solving $g_{\foplus}$ is hard in the random query model. There are two bits $2i-1, 2i$ for all $i \in [n]$ that need to be seen together to find their parity. In expectation, there is a gap of $\Omega(n)$ samples between the branching program observing these two bits. 
\begin{itemize}
\item We formulate a simpler problem called the two-bit parity game. The players  want to find the parity of two independent uniformly random bits, but the bits are allocated to them similar to the random query model. We show that at least some information on whether the players \emph{have seen these bits} is transferred through the nodes of the branching program. 
    \item In the next step, we show that roughly $1/(n \cdot \sens_0[f, \xstar_f])$ fraction of the information cost of $g_{\foplus}$ is an upper bound on the information revealed by the smaller problem. This step uses an information complexity direct sum, and measuring the information retained about the smaller problem over all time steps. 
\end{itemize}

The intermediate problem we call the two-bit parity game simulates finding only an XOR of two bits in the random query model. 
\begin{tbox}
\begin{definition}[Two-Bit Parity Game]
There are $k$ players $\player_1,\ldots,\player_k$. The inputs are generated
and the communication proceeds as follows.
\begin{enumerate}
    \item There are two bits $\ralpha$ and $\rbeta$ chosen independently and uniformly
    at random from $\set{0,1}$.

    \item For every $a\in[k]$, independently, $\rrho_a$ is chosen uniformly from
    $\set{\charalpha, \charbeta}$ and $\rtheta_a$ is chosen uniformly from
    $\set{0,1}$. 
    \item Player $\player_a$ receives $\rrho_a$ and $\rtheta_a$. In addition, if
    $\rtheta_a=1$, the player also receives $\ralpha$ when
    $\rrho_a=\charalpha$, and $\rbeta$ when $\rrho_a=\charbeta$.

    \item The players communicate in the blackboard model in the order
    $\player_1,\player_2,\ldots,\player_k$. Each player writes one message
    on the blackboard after observing its input and all preceding messages.

    \item The last player $\player_k$ must write $\ralpha\oplus\rbeta$ on the
    blackboard.
\end{enumerate}
\end{definition}
\end{tbox}

 We use bolded $\ralpha$, $\rbeta$, and for $a \in [k]$, $\rrho_a$, and $\rtheta_a$ to denote the random
variables.  We also use:
\[
\rrho:=(\rrho_1,\ldots,\rrho_k)
\qquad\text{and}\qquad
\rtheta:=(\rtheta_1,\ldots,\rtheta_k).
\]
We present a summary of these variables in the following table:

\begin{table}[h]
\centering
\begin{tabular}{|c|p{0.72\textwidth}|}

\hline
\textbf{Random Variable} & \textbf{Meaning} \\
\hline
$\ralpha,\rbeta$ & The two input bits chosen independently and uniformly at random  \\
$\rtheta_a$ & Indicates whether player $\player_a$ receives any information about the input bits.\\
$\rrho_a$ & Indicates whether player $\player_a$ selects the left bit $\ralpha$ or the right bit $\rbeta$. \\
\hline
\end{tabular}
\caption{Variables in the Two-Bit Parity Game.}
\label{tab:two-bit-parity-variables}
\end{table}

We show that a successful protocol in the two bit parity game must reveal some information about whether the inputs $\alpha, \beta$ are present for at least few of the $k$ players. 
We use $\Prot$ to denote the transcript of a protocol $\prot$. We use $\rProt$ to denote the random variable corresponding to $\Prot$, and $\rProt_a$ for the random variable corresponding to the message sent by $\player_a$, for $a\in[k]$.
This is the main lemma of this subsection.

Let $\eventeq := \{\ralpha=\rbeta\}$ denote the event that the two input bits are equal.

\begin{lemma}[External Information of the two-bit parity game]\label{lem:two-bit-game}
Let $\prot$ be a protocol for the two-bit
parity game that outputs $\ralpha\oplus\rbeta$ with probability of success at least $1/2 + \eps$.
Then,
\[
\mi{\rProt}{\rtheta
\mid \rrho,\Rpublic,\ralpha,\eventeq}
=\Omega_{\eps}(1).
\]
\end{lemma}

The proof of \Cref{lem:two-bit-game} is deferred to the next section. In the remainder of this section, we show that the lower bound in \Cref{lem:two-bit-game} gives an $\Omega(n \cdot \sens[f])$ on the information cost of branching programs that solve $g_{\foplus}$.

\subsection{Protocol for Two-Bit Game}\label{sec:dirsum}

We give a protocol for the two bit parity game using a branching program that solves $g_{\foplus}$ with high probability. 

To disambiguate between the public randomness of the two-bit game from the public randomness of the branching program, we use $\Rpublictwo$ to denote the public randomness of the protocol for the two-bit game that we construct. 
Throughout this section, we assume that we are given a randomized branching program $P$ that solves $g_{\foplus}$ of length $T$ and width $2^S$ with probability of success at least $2/3$.

We give the protocol in multiple steps for readability.

\noindent
\textbf{Constructing the index sequences $\rLall$.}
We know $g_{\foplus}$ has $\sens_0[f, \xstar_f]$ many pairs of indices where we can hope to embed a small two-bit parity problem with the hope of solving it. First, we pick a special one of these indices to embed the problem into uniformly at random. Then we show how the randomness of $\rrho$ in the two-bit game can be used alongside public randomness to construct a valid sequence $\rLall$. 

The two-bit game has $k$ players, and the lower bound in \Cref{lem:two-bit-game} holds for any value of $k$. However, $k$ corresponds to the number of samples or inputs the players get to try to find $\ralpha \oplus \rbeta$. We work with $k \geq (4T \log n)/n$. 

\begin{tbox}
\textbf{Protocol $\prot_{\rLall}$: Constructing $\rLall$ with public randomness and inputs:}
\begin{enumerate}[label=$(\roman*)$]
\item Pick a special index $h \in [n]$ uniformly at random from
    \[
        \set{i \in [n] \mid \xstar_f(i)=0
        \text{ and } f(\xstar_f) \neq f(\xstar_f \oplus e_i)},
    \]
    i.e., from the set of indices which are sensitive for $\xstar_f$ with value $0$.
    \item For each $t \in [T]$, sample $m_t$ uniform over $[n]$ with public randomness. 
    \item For each $t \in [T]$:
    \begin{itemize}
        \item If $m_t \neq h$, set $\ell_t$ as one of $2m_t-1$ or $2m_t$ at random with public randomness.
        \item If $m_t = h$, then let this be the $a^{\textnormal{th}}$ occurrence of $h$ in the sequence of $m_t$ values. If $a \leq k$, we set $\ell_t = 2m_t - 1$ if $\rho_a = \charalpha$ and $\ell_t = 2m_t$ if $\rho_a = \charbeta$. Otherwise, we set $\ell_t$ to be one of $2m_t-1$ or $2m_t$ at random using public randomness.
    \end{itemize}
\end{enumerate}
\end{tbox}

We can prove that we indeed generate a uniformly random sequence of indices from $[2n]$ with high probability. 

\begin{claim}\label{clm:sequence-construct}
Protocol $\prot_{\rLall}$ generates a uniformly random sequence of $T$ integers. Moreover, the probability that this sequence contains more than $k$ values $m_t$ for $t \in [T]$ with $m_t = h$ is at most $1/n^2$.
\end{claim}

\begin{proof}
First, we argue that the sequence is uniformly random. Each $m_t$ is chosen from $[n]$ uniformly random. If $m_t \neq h$, or if $m_t =h$ is the $a^{\textnormal{th}}$ occurrence of $h$ with $a > k$, then one of $2m_t-1$ or $2m_t$ is also chosen uniformly. Thus, each $2j$ or $2j-1$ is selected with probability $1/2n$ as required, for $j \neq h$. For $m_t = h$, if this is the $a^{\textnormal{th}}$ occurrence of $h$ for $a \leq k$, index $\rrho_a$ is uniform over $\set{\charalpha, \charbeta}$ for $a \in [k]$ also. Hence, $2h-1$ or $2h$ are also sampled with probability $1/2n$.

    Each value $m_t$ is chosen from $[n]$ uniformly random and independently. Thus, in expectation, the number of $t \in [T]$ with $m_t = h$ is $T/n \leq k/(4\log n)$, by our assumption $k \geq (4T \log n)/n$. Using \Cref{prop:chernoff} for independent random variables, we get that the number of $t \in [T]$ with $m_t = h$ is at most $k$ with probability at least $1-1/n^2$.
\end{proof}

For $a\in[k]$, let $\specialtime{a}$ be the time step $t \in [T]$ such that $m_t$ is the $a^{\textnormal{th}}$ occurrence of $h$ in the sequence $(m_t)_{t\in[T]}$. $\specialtime{a}=T+1$ if there are fewer than $a$ occurrences. We also set $\specialtime{k+1}=T+1$. We call a player $\player_a$ for $a\in [k]$ an \textbf{active} player if  $\specialtime{a}\leq T$. These are the players who send messages on the board that depend on the input. Observe that which players are active depends solely on the sequence $(m_t)_{t \in [T]}$.

We explicitly state among the $k$ players who can see what part of the sequence $\ell_t$ for $t \in [T]$.

\begin{observation}\label{obs:rL-public}
    For each $t \in [T]$, conditioned on $\ell_t \notin \set{2h-1, 2h}$, the value of $\ell_t$ is known to all the players of the two-bit parity game. Moreover, for $a \in [k]$, if this is the $a^{\textnormal{th}}$ time that $m_t = h$ in the sequence of $(m_t)_{t \in [T]}$, then $\player_a$ knows the value of $\ell_t$.
\end{observation}
\begin{proof}
    The values of $m_t$ are chosen with public randomness, and if $m_t \neq h$, the value of $\ell_t$ is also chosen with public randomness.
    For $a \in [k]$, when $m_t = h$ for the $a^{\textnormal{th}}$ time in the sequence $(m_t)_{t \in [T]}$, the value of $\ell_t$ is chosen based on the input $\rho_a$ given to $\player_a$.
\end{proof}

Next, we look at how the players generate the actual input to branching program $P$.

\noindent
\textbf{Constructing the input $\ry$.}
We defined distribution $\cD_{\foplus}$ from which to sample $\ry$. We embed the input $\alpha, \beta$ into $2h-1$ and $2h$, and the other bits will be fixed by some $y \sim \cD_{\foplus}$.

\begin{tbox}
\textbf{Protocol $\prot_{\ry}$: Constructing $\ry$ with public randomness and inputs:}
\begin{enumerate}[label=$(\roman*)$]
\item Sample some $y'$ from distribution $\cD_{\foplus}$ using public randomness. 
\item For $j \notin \set{2h-1, 2h}$, set $y(j) = y'(j)$.
\item We set $y(2h-1) = \alpha$ and $y(2h) = \beta$.
\end{enumerate}
\end{tbox}

We can argue that the answer from the input we construct reveals $\ralpha \oplus \beta$.

\begin{observation}\label{obs:newprot-output}
    For any $y \in \set{0,1}^{2n}$ constructed by $\prot_{\ry}$, the value of $g_{\foplus}(y)$ is $f(\xstar_f) \oplus (\ralpha \oplus \rbeta)$.
\end{observation}
\begin{proof}
    We know $h$ is sensitive for $\xstar_f$ and $\xstar_f(h)=0$. For any $i \neq h$, the value of $y(2i-1) \oplus y(2i)$ is unchanged from $y' \sim \cD_{\foplus}$. Otherwise, $y(2h-1) \oplus y(2h)$ is set as $\alpha \oplus \beta$. Therefore:
    \begin{align*}
        g_{\foplus}(y) &= g_{\foplus}(y') \oplus (\alpha \oplus \beta) \tag{as $h$ is sensitive for $\xstar_f$ with $\xstar_f(h) = 0$} \\
    &= f(\xstar_f) \oplus (\alpha \oplus \beta),
    \end{align*}
    where the last equality follows by \Cref{obs:output-cDg}.
\end{proof}

We also analyze which bits of $y$ are known publicly to all the players. 

\begin{observation}\label{obs:newprot-yinfo}
    For any $y $ generated by $\prot_{\ry}$, the value of $y(2i)$ or $y(2i-1)$ is known to all the $k$ players if $i \neq h$. Moreover, value of $y(2i-1)$ is known by every player who holds $\alpha$ and $y(2i)$ is known by every player who knows $\beta$.
\end{observation}
\begin{proof}
    In the first step of $\prot_{\ry}$, $y' \sim \cD_{\foplus}$ is sampled publicly. Only bits $y(2h-1)$ and $y(2h)$ are changed, so all the players know the remaining values. The values of $y(2h-1)$ and $y(2h)$ are set to $\alpha, \beta$ respectively, so the second part of the observation is also true.
\end{proof}

 We have constructed the input and the index sequences that will be given to $P$. We are ready to describe the final protocol to solve the two-bit parity game. 

\noindent
\textbf{Final Protocol.}
We need some terminology to describe the protocol.
 For every active player, we define
$\specialbit{a}:=b_{\specialtime{a}}$ for ease of notation.
For each player $\player_a$ for $a \in [k]$, we define a \textbf{simulation block} as:
\begin{itemize}
    \item If $\player_a$ is active, the $a^{\textnormal{th}}$ simulation block is from time step $\specialtime{a}$ to $\min\set{\specialtime{a+1}-1, T}$.
    \item If $\player_a$ is not active, they have no simulation block. 
\end{itemize}
These are the time steps that $\player_a$ is in charge of executing in the branching program $P$.
For $a\in\set{0,\ldots,k}$, we define the state sent at the end of the
$a^{\textnormal{th}}$ simulation block by $  \specialstate{a}:=
    \rP_{\min\set{\specialtime{a+1}-1,T}}$.
These are the states that will be sent to the blackboard via $S$-bit labels. 
To start with, $\specialstate{0}$ is the state immediately before the first
occurrence of $h$ in sequence $(m_t)_{t \in [T]}$. If $\player_a$ is the last active player, then
$\specialstate{a}=\rP_T$, which is the last time block.

\begin{tbox}
   \textbf{Protocol $\prot_{\two}$: Solving the two-bit parity game.}
\begin{enumerate}[label=$(\roman*)$]
\item Sample $\rLall$ using protocol $\prot_{\rLall}$ and sample $\ry$ using protocol $\prot_{\ry}$. If $h$ occurs more than $k$ times in $(m_t)_{t\in[T]}$, declare FAIL and output $0$.
\item Use public randomness to sample a deterministic branching program from the distribution associated with the randomized branching program $P$. 
\item 
The players simulate $P$ as follows: (see \Cref{clm:run-prog} for more details)
\begin{enumerate}[label=$(\alph*)$]
    \item Sample the $S$-bit label $\specialstate{0}$ using public randomness conditioned on $\ry$ and $\rLall$.
    \item For every active player $\player_a$ with $a \in [k]$, the value of $\specialbit{a} = b_{\specialtime{a}}$ is set as $\theta_a$. For every other time step $t$ in the simulation block of $a$, where $\specialtime{a} < t \leq \specialtime{a+1}-1$, $b_t$ is sampled from $\set{0,1}$ uniformly at random with private randomness of $\player_a$.
    \item Active player $\player_a$ then writes the $S$-bit label corresponding $\specialstate{a}$ on the blackboard. 
    \item Players who are not active, with $\specialtime{a}=T+1$ write $\perp$ on the board.
\end{enumerate}
\item The last player $\player_k$ uses the last non-null state, which is $\rP_T$, to get the output $P(\ry)$, and fixes $P(\ry)\oplus f(\xstar_f)$ as the answer.
\end{enumerate}
\end{tbox}

We show that the simulation of $P$ is possible in $\prot_{\two}$.
\begin{claim}\label{clm:run-prog}
Protocol $\prot_{\two}$ simulates branching program $P$ on the input $y$ and the sequence of indices generated by $\prot_{\rLall}$. 
\end{claim}
\begin{proof}
First, we argue that the players have sufficient information to simulate $P$. By \Cref{obs:rL-public}, for every active player $\player_a$, if $t = \specialtime{a}$, then $\ell_t$ is known to $\player_a$, or otherwise, if $t < \specialtime{a+1}-1$, the value of $\ell_t$ is fixed by public randomness. 

The value of all indices of $\ry$ except for $2h-1$ and $2h$ are fixed by public randomness. For the time step $\specialtime{a}$, if $\theta_a = 1$, the value of the corresponding bit is given to $\player_a$ as input, by \Cref{obs:newprot-yinfo}. Therefore, each active player knows all the input samples to $P$ in their simulation blocks. 

The $S$-bit label of $P$ just before the active block is also known. $\specialstate{0}$ is sampled with public randomness, and $\specialstate{a-1} = \rP_{\specialtime{a}-1}$ is written on the board by $\player_{a-1}$ for $a > 1$. Hence, simulation of $P$ proceeds faithfully. 

The distribution that the players sample $y$ from is the same as $\cD_{f \oplus}$ when $\eventeq$ occurs, directly by construction. 
\end{proof}

We prove that $\prot_{\two}$ has high probability of success. 

\begin{claim}\label{clm:newprot-success}
    The probability of success of $\prot_{\two}$ is at least $2/3-1/n^2$.
\end{claim}

\begin{proof}
By \Cref{clm:sequence-construct}, the probability that the number of $(m_t)_{t \in [T]}$ with $m_t = h$ is more than $k$ is at most $1/n^2$. Let $\mathcal{E}_{\textnormal{F}}$ be the event that protocol $\prot_{\two}$ outputs FAIL. 

Conditioned on $\neg \mathcal{E}_{\textnormal{F}}$, the output of $\prot_{\two}$ is the same as that of branching program $P$. 
There are two sources of failure for protocol $\prot_{\two}$:
\begin{enumerate}[label=$(\roman*)$]
    \item Event $\mathcal{E}_{\textnormal{F}}$ happens, in which case, the protocol outputs fail. 
    \item Event $\mathcal{E}_{\textnormal{F}}$ does not happen and $P$ is incorrect. 
\end{enumerate}

We can upper bound the probability of failure of $\prot_{\two}$ as:
\begin{align*}
    \Pr[\prot_{\two}\textnormal{ is incorrect}] &= \Pr[\mathcal{E}_{\textnormal{F}}] + Pr[P\textnormal{ is incorrect} \cap \neg \mathcal{E}_{\textnormal{F}}] \tag{by union bound}\\
    &\leq \Pr[\mathcal{E}_{\textnormal{F}}] + Pr[P\textnormal{ is incorrect}], \\
    &\leq \frac13 + \frac1{n^2},
\end{align*}
where in the last line, we have measured probability of failure of $P$ on a truly random sequence of indices, for the input given to $\prot_{\two}$. 
\end{proof}

It remains to compare the external information of this protocol $\prot_{\two}$ with the
information cost of $P$. The rest of this section covers this comparison, and proves the final lower bound. 

\subsection{External Information of the Protocol}

Let $d:=\sens_0[f,\xstar_f]$, for ease of notation in the rest of this section. We analyse the external information of $\prot_{\two}$ first. 
The variables sampled with public randomness in $\Rpublictwo$ are:
\begin{itemize}
    \item The special index $\rh \in [n]$ from among $d$ many indices chosen uniformly at random. 
    \item The sequence $(m_t)_{t \in [T]}$, and further $\ell_t$ for all $t$ with $m_t \neq h$. Let the sequence of $m_t$ be $\rM$, and the sequence of $\ell_t$ we have already denoted using $\rLall$.
    \item The value of $\ry$ on indices which are not $2h-1$ or $2h$.
    \item The $S$-bit label $\specialstate{0}$, and the public randomness $\Rpublic$ used to sample a deterministic branching program from $P$.
\end{itemize}

Some small observations are in order. 
\begin{observation}\label{obs:var-in-conditioning}
    For the random variables in $\Rpublictwo$ and in the input of the two-bit parity game,
    \begin{enumerate}[label=$(\roman*)$]
        \item The random variables $\rM$, $\rrho$ and the other choices of $2m_t-1$ or $2m_t$ given value of $m_t$ from $\prot_{\rLall}$ fix $\rLall$.
        \item The random variables $\ralpha, \rbeta$ and the values of $\ry$ at indices which are not $2 \rh$ and $2\rh-1$ fix the value of input $\ry$. Conditioned on $\eventeq$, this input is sampled from distribution $\cD_{\foplus}$. 
        \item The distributions of $\rLall$, $\ry, \rb$, $\Rpublic$ and all $\rP_i$ for $i \in [T]$ are independent of $\rh$, conditioned on $\eventeq$.
    \end{enumerate}
\end{observation}

\begin{proof}
    Parts $(i)$ and $(ii)$ follow directly from our construction. We elaborate on part $(iii)$. 

    The distribution of $\rLall$ is uniform over $[2n]^{T}$ irrespective of the value of $\rh$. This is because $\rh$ is a uniformly random value from $[n]$ that simply chooses some indices in $\rLall$ to be sampled via public randomness, and some to come from $\rrho$ input given to the players. 

    Similarly, $\ry$ is always sampled from $\cD_{\foplus}$ irrespective of what the value of $\rh$ is, when conditioned on $\eventeq$. All the $d$ many bits of $\xstar_f$ which are sensitive and have value $0$ look identical. Random variable $\rh$ points to one of them uniformly at random. 

    The values of $\rb_i$ are uniform over $\set{0,1}$ irrespective of value of $\rh$, and the public randomness $\Rpublic$ does not depend on $\rh$ either. Hence, as all the input random variables are independent of $\rh$, so are the $S$-bit labels $\rP_i$ for $i \in [T]$.
\end{proof}

We expand the external information of protocol $\prot_{\two}$ to prove the main lemma. We use $\rPrott_a$ to denote the message sent by $\player_a$ in $\prot_{\two}$.
We simplify this mutual information as a first step. We can focus only on the information revealed about earlier values of $\rtheta_c$ for any message $\rPrott_a$ sent by $\player_a$ with $a \geq c$.
\begin{claim}\label{clm:msg-cost-2bit-simple}
For any protocol with transcript $\rPrott_a$ by $\player_a$,
\[
\mi{\rPrott}{\rtheta \mid \rrho ,\Rpublictwo, \ralpha, \eventeq} \leq \sum_{a =1}^k \sum_{c=1}^a \mi{\rPrott_a}{\rtheta_c \mid \rPrott_{<c}, \rrho ,\Rpublictwo, \ralpha, \eventeq}.
\]
\end{claim}
\begin{proof}
    First, we use chain rule of mutual information \itfacts{mi-chain} to get,
\begin{align*}
&\mi{\rPrott}{\rtheta
\mid \rrho,\Rpublictwo,\ralpha,\eventeq} \\
&=
\sum_{a=1}^{k}\sum_{c=1}^{k}
\mi{\rPrott_a}{\rtheta_c
\mid \rPrott_{<a},\rtheta_{<c},
\rrho,\Rpublictwo,\ralpha,\eventeq}.
\end{align*}

For $c > a$, we know $\rPrott_a$ is fixed by $\rPrott_{<a}$, $\rrho, \rtheta_a$, $\ralpha, \rbeta$, $\Rpublictwo$ and the private randomness of $\player_a$. Conditioned on $\rPrott_{<a}$, $\rtheta_{<c}$, $\rrho$, $\ralpha, \rbeta$ and $\Rpublictwo$, $\rPrott_a$ depends only on the private randomness of $\player_a$, which is independent of $\rtheta_c$. By \itfacts{mi-nonnegative}, for $c > a$,
\begin{equation}\label{eq:inter-cgeqa}
    \mi{\rPrott_a}{\rtheta_c
\mid \rPrott_{<a},\rtheta_{<c},
\rrho,\Rpublictwo,\ralpha,\eventeq} = 0.
\end{equation}

For $c < a$, again, conditioned on $\rPrott_{<a}$, $\rtheta_{<c}$, $\rrho, \ralpha, \rbeta$ and $\Rpublictwo$, $\rPrott_a$ depends only on $\rtheta_a$ and the private randomness of $\player_a$. These random variables are  independent of all $\rtheta_c$ for $c < a$ after conditioning on $\rPrott_{<a}$, $\rtheta_{<c}$, $\rrho, \ralpha, \rbeta$ and $\Rpublictwo$. By \itfacts{mi-nonnegative}, for $c < a$,
\begin{equation}\label{eq:inter-cleqa}
    \mi{\rPrott_a}{\rtheta_c
\mid \rPrott_{<a},\rtheta_{<c},
\rrho,\Rpublictwo,\ralpha,\eventeq} = 0.
\end{equation}

For $c=a$, we know $\rtheta_{<a}$ is independent of $\rPrott_a$ when conditioned on $\rtheta_a, \rPrott_{<a}, \rrho, \Rpublictwo, \ralpha, \rbeta$. Once its input and the preceding messages are fixed, the message of $\player_a$ depends only on its private randomness. Using \itfacts{both-heavy-less}, for $c = a$,
\begin{equation}\label{eq:inter-ceqa}
    \mi{\rPrott_a}{\rtheta_c
\mid \rPrott_{<a},\rtheta_{<c},
\rrho,\Rpublictwo,\ralpha,\eventeq} \leq \mi{\rPrott_a}{\rtheta_c
\mid \rPrott_{<a},
\rrho,\Rpublictwo,\ralpha,\eventeq}
\end{equation}

Using \Cref{eq:inter-ceqa}, \Cref{eq:inter-cgeqa} and \Cref{eq:inter-cleqa}, we get,
\begin{align*}
\mi{\rProt}{\rtheta
\mid \rrho,\Rpublictwo,\ralpha,\eventeq} 
&= \sum_{a=1}^{k}
\mi{\rPrott_a}{\rtheta_a
\mid \rPrott_{<a},
\rrho,\Rpublictwo,\ralpha,\eventeq} \\
&\leq \sum_{a=1}^{k} \sum_{c=1}^a
\mi{\rPrott_a}{\rtheta_c
\mid \rPrott_{<c},
\rrho,\Rpublictwo,\ralpha,\eventeq},
\end{align*}
where in the last inequality, we have used that mutual information is nonnegative from \itfacts{mi-nonnegative}.
\end{proof}

\Cref{clm:msg-cost-2bit-simple} moves us one step towards the information cost of $P$. Let $\eventfail$ be the event that $\prot_{\two}$ declares FAIL in step $(i)$. We have to account for the probability of $\eventfail$, as it affects the distribution of the transcript.

Let $\rProt'$ be the auxiliary transcript which always simulates the
first $k$ blocks without declaring FAIL. If there is a $(k+1)^{\textnormal{th}}$
occurrence of $h$, the last block ends just before this occurrence, and the last player outputs $\specialstate{k} = \rP_{\specialtime{k+1}}$. We use
$\rP'_a$ to denote its $a^{\textnormal{th}}$ message.

\begin{claim}\label{clm:aux-transcript}
    For the auxiliary transcript, we get,
    \[
   \mi{\rPrott}{\rtheta
\mid \rrho,\Rpublictwo,\ralpha,\eventeq} \leq H_2(1/n^2)+
\mi{\rProt'}{\rtheta
\mid \rrho,\Rpublictwo,\ralpha,\eventeq}, 
    \]
    where $H_2()$ is the binary entropy function.
\end{claim}
\begin{proof}
The transcript $\rPrott$ is fixed by $\rProt'$ and the indicator of $\eventfail$. This is because, when $\eventfail$ does not happen, $\rPrott$ replicates $\rProt'$ exactly. Hence,
\begin{align*}
&\mi{\rProt}{\rtheta
\mid \rrho,\Rpublictwo,\ralpha,\eventeq} \\
&\leq \mi{\rProt',\ind[\eventfail]}{\rtheta
\mid \rrho,\Rpublictwo,\ralpha,\eventeq} \tag{by data processing \itfacts{data-processing}}\\
&\leq
\mi{\rProt'}{\rtheta
\mid \rrho,\Rpublictwo,\ralpha,\eventeq} + \mi{\ind[\eventfail]}{\rtheta \mid \rProt', \rrho, \Rpublictwo, \ralpha, \eventeq} \tag{by chain rule \itfacts{mi-chain}} \\
&\leq H(\ind[\eventfail]) + \mi{\rProt'}{\rtheta
\mid \rrho,\Rpublictwo,\ralpha,\eventeq},
\end{align*}
where the last inequality follows by definition of mutual information, and the non-negativity of entropy.
Finally, by \Cref{clm:sequence-construct}, we know that event $\eventfail$ happens with probability at most $1/n^2$. This completes the proof. 
\end{proof}
The following lemma is the heart of the argument.

\begin{lemma}\label{clm:direct-sum-information}
The auxiliary transcript $\rProt'$ of $\prot_{\two}$ satisfies
\[
\mi{\rProt'}{\rtheta
\mid \rrho,\Rpublictwo,\ralpha,\eventeq}
\leq \frac{1}{nd} \cdot \ic_{\cD_{\foplus}}(P) + S/d.
\]
\end{lemma}
\begin{proof}
We first apply
\Cref{clm:msg-cost-2bit-simple} to $\rProt'$. This gives,
\begin{align*}
\mi{\rProt'}{\rtheta
\mid \rrho,\Rpublictwo,\ralpha,\eventeq} 
\leq \sum_{a =1}^k \sum_{c=1}^a
\mi{\rProt'_a}{\rtheta_c \mid \rProt'_{<c}, \rrho,
\Rpublictwo, \ralpha, \eventeq}.
\end{align*}

Let us look at all the variables on the right side of the conditioning. We also use parts $(i)$ and $(ii)$ of \Cref{obs:var-in-conditioning} to see what random variables of the branching program are conditioned on. 

The public randomness samples $\rM$, which along with $\rrho$ and $\rh$ fixes $\rLall$. Similarly, the variables in public randomness along with $\ralpha$ and $\rbeta$ fix $\ry$ sampled from distribution $\cD_{\foplus}$. 
The rest is $\Rpublic$ to sample the deterministic branching program and the state $\specialstate{0} = \rP_{\specialtime{1}-1}$. 
Therefore, the mutual information term becomes,
\begin{align*}
    \sum_{a =1}^k \sum_{c=1}^a
\mi{\rProt'_a}{\rtheta_c \mid \rProt'_{<c}, \rrho,
\Rpublictwo, \ralpha, \eventeq} =  \sum_{a =1}^k \sum_{c=1}^a
\mi{\rProt'_a}{\rtheta_c \mid \rProt'_{<c}, \rLall, \rh, \ry, \Rpublic, \specialstate{0}}.
\end{align*}

We also know $\rProt'_a$ is the same as $\specialstate{a} = \rP_{\specialtime{a+1}-1}$, and $\rtheta_c$ is the same as $\rb_{\specialtime{c}}$. Similarly, $\rProt'_{<c}$ comprises of all $\rP_{\specialtime{2}-1},\rP_{\specialtime{3}-1}, \ldots,  \rP_{\specialtime{c}-1}$.
Substituting these correspondences, 
\begin{align*}
   & \sum_{a =1}^k \sum_{c=1}^a
\mi{\rProt'_a}{\rtheta_c \mid \rProt'_{<c}, \rLall, \rh, \ry, \Rpublic, \specialstate{0}} \\
&=   \sum_{a =1}^k \sum_{c=1}^a
\mi{\rP_{\specialtime{a+1}-1}}{\rb_{\specialtime{c}} \mid \rP_{\specialtime{1}-1},\rP_{\specialtime{2}-1},\rP_{\specialtime{3}-1}, \ldots,  \rP_{\specialtime{c}-1}, \rLall, \rh, \ry, \Rpublic}
\end{align*}

We can remove the conditioning on all prior labels of the branching program barring $\rP_{\specialtime{c}-1}$ using the following observation.
\begin{observation}\label{obs:interim-independence}The state of branching program at $\specialtime{a+1}-1$ is independent of all prior states when conditioned on the state $\rP_{\specialtime{c}-1}$.
\[
\mi{\rP_{\specialtime{a+1}-1}}{\rP_{\specialtime{1}-1},\rP_{\specialtime{2}-1},\ldots,  \rP_{\specialtime{c-1}+1} \mid \rP_{\specialtime{c}-1}, \rLall, \rh, \ry, \Rpublic} = 0.
\]
\end{observation}
\begin{proof}
Conditioned on $\rLall,\rh,\ry$ and $\Rpublic$, the input, the sequence of
indices and the deterministic branching program in $P$ are fixed and only the $\rb$ values are left which are random. Hence, conditioned on
$\rP_{\specialtime{c}-1}$, the state
$\rP_{\specialtime{a+1}-1}$ is fixed by $\rb_{\specialtime{c}}, \rb_{\specialtime{c}+1}, \ldots, \rb_{\specialtime{a+1}-1}$. 
By data processing in \itfacts{data-processing},
\begin{align*}
    &\mi{\rP_{\specialtime{a+1}-1}}{\rP_{\specialtime{1}-1},\rP_{\specialtime{2}-1},\ldots,  \rP_{\specialtime{c-1}+1} \mid \rP_{\specialtime{c}-1}, \rLall, \rh, \ry, \Rpublic} \\
    &\leq \mi{\rb_{\specialtime{c}}, \rb_{\specialtime{c}+1}, \ldots, \rb_{\specialtime{a+1}-1}}{\rP_{\specialtime{1}-1},\rP_{\specialtime{2}-1},\ldots,  \rP_{\specialtime{c-1}+1} \mid \rP_{\specialtime{c}-1}, \rLall, \rh, \ry, \Rpublic}\\
    &\leq  \mi{\rb_{\specialtime{c}}, \rb_{\specialtime{c}+1}, \ldots, \rb_{\specialtime{a+1}-1}}{\rP_{\specialtime{1}-1},\rP_{\specialtime{2}-1},\ldots,  \rP_{\specialtime{c-1}+1},\rP_{\specialtime{c}-1} \mid \rLall, \rh, \ry, \Rpublic} \tag{by \itfacts{mi-chain} and \itfacts{mi-nonnegative}} \\
    &\leq \mi{\rb_{\specialtime{c}}, \rb_{\specialtime{c}+1}, \ldots, \rb_{\specialtime{a+1}-1}}{\rb_1, \rb_2, \ldots, \rb_{\specialtime{c}-1} \mid \rLall, \rh, \ry, \Rpublic} \tag{by \itfacts{data-processing}, data processing} \\
    &= 0,
\end{align*}
where the last line follows because the variables $\rb_i$ are chosen uniformly at random from $\set{0,1}$ irrespective of the other input. \qedhere{\Cref{obs:interim-independence}}
\end{proof}

Using \Cref{obs:interim-independence} and \itfacts{both-heavy-less}, 
\begin{align*}
    &\sum_{a =1}^k \sum_{c=1}^a
\mi{\rP_{\specialtime{a+1}-1}}{\rb_{\specialtime{c}} \mid \rP_{\specialtime{1}-1},\rP_{\specialtime{2}-1},\rP_{\specialtime{3}-1}, \ldots,  \rP_{\specialtime{c}-1}, \rLall, \rh, \ry, \Rpublic} \\
&\leq \sum_{a =1}^k \sum_{c=1}^a
\mi{\rP_{\specialtime{a+1}-1}}{\rb_{\specialtime{c}} \mid\rP_{\specialtime{c}-1}, \rLall, \rh, \ry, \Rpublic}
\end{align*}
The transcript $\rProt'_{<c}$ is comprised of all the $S$-bit labels of the branching programs sent by all the players before $\player_c$. The player $\player_{c-1}$ sends $\rP_{\specialtime{c}-1} = \specialstate{c-1}$ to the blackboard. Conditioned on $\rP_{\specialtime{c}-1}, \rh, \rtheta_c, \ry, \rLall$ and $\Rpublic$ the transcript of $\player_a$ for $a > c$, which is $\rP_{\specialtime{a+1}-1}$ is independent of the transcripts of $\rProt'_{<(c-1)}$. It depends only on the private inputs and private randomness of $\player_{c+1}, \player_{c+2}, \ldots, \player_a$. Hence, by \itfacts{both-heavy-less}, we get,
\begin{align*}
    \sum_{a =1}^k \sum_{c=1}^a
\mi{\rProt'_a}{\rtheta_c \mid \rProt'_{<c}, \rLall, \rh, \ry, \Rpublic} &\leq   \sum_{a =1}^k \sum_{c=1}^a
\mi{\rP_{\specialtime{a+1}-1}}{\rb_{\specialtime{c}} \mid \rP_{\specialtime{c}-1}, \rLall, \rh, \ry, \Rpublic}.
\end{align*}
Now, we move away from mutual information term indexed by players $a, c$ towards the time steps $i, j$ in the cost of $P$. Using definition of conditional mutual information,
\begin{align}\label{eq:interim-eq2}
     &\sum_{a =1}^k \sum_{c=1}^a
\mi{\rP_{\specialtime{a+1}-1}}{\rb_{\specialtime{c}} \mid \rP_{\specialtime{c}-1}, \rLall, \rh, \ry, \Rpublic}  \notag\\
&=  \Exp_{\rM = M, \rh = h}\sum_{a =1}^k \sum_{c=1}^a
\mi{\rP_{\specialtime{a+1}-1}}{\rb_{\specialtime{c}} \mid \rP_{\specialtime{c}-1}, \rLall, \ry, \Rpublic, \rM = M, \rh = h}\notag  \\
&\leq \Exp_{\rM = M, \rh = h}\Big(\sum_{i =1}^{T-1} \sum_{j=1}^i \ind[m_{i+1} = m_j = h] \cdot
\mi{\rP_i}{\rb_{j} \mid \rP_{j-1}, \rLall, \ry, \Rpublic, \rM = M, \rh = h} \notag\\
&\qquad + \sum_{j=1}^T \ind[m_j = h] \cdot \mi{\rP_T}{\rb_{j} \mid \rP_{j-1}, \rLall, \ry, \Rpublic, \rM = M, \rh = h}\Big),
\end{align}
where we have used that the value of $\ell_{\specialtime{c}}$ and $\ell_{\specialtime{a+1}}$ lies in $\set{2h-1, 2h}$ at $\specialtime{c}$ and at $\specialtime{a+1}$.  Otherwise, this would not be a message that $\player_a$ or $\player_c$ writes to the blackboard. Observe that the transcript may also contain mutual information terms beyond the first $k$ messages written on the blackboard, hence, we get an inequality. We have also used mutual information is non-negative \itfacts{mi-nonnegative} to add the extra terms.

Also, the last state $\rP_T$ is written by the last active player irrespective of what the value of sequence $\rM$ is if the number of $t \in [T]$ with $m_t = h$ is at most $k$. 
However, it still contains only information about inputs of prior players, when $m_j = h$. If number of $t \in [T]$ with $m_t = h$ is more than $k$, the last state $\rP_T$ is not written by $\rProt'$. Our upper bound is still valid as mutual information is non-negative, \itfacts{mi-nonnegative}.

We bound the two terms in \Cref{eq:interim-eq2} separately. 
\[
\textnormal{\textbf{Term 1:}} \Exp_{\rM = M, \rh = h}\sum_{i =1}^{T-1} \sum_{j=1}^i \ind[m_{i+1} = m_j = h] \cdot
\mi{\rP_i}{\rb_{j} \mid \rP_{j-1}, \rLall, \ry, \Rpublic, \rM = M, \rh = h}
\]

For this term, 
\begin{align*}
   & \Exp_{\rM = M, \rh = h}\sum_{i =1}^{T-1} \sum_{j=1}^i \ind[m_{i+1} = m_j = h] \cdot
\mi{\rP_i}{\rb_{j} \mid \rP_{j-1}, \rLall, \ry, \Rpublic, \rM = M, \rh = h} \\
&=  \Exp_{\rM = M, \rh = h}\sum_{i =1}^{T-1} \sum_{j=1}^i \ind[m_{i+1} = m_j = h] \cdot
\mi{\rP_i}{\rb_{j} \mid \rP_{j-1}, \rLall, \ry, \Rpublic, \rM = M} \tag{by part $(iii)$ of \Cref{obs:var-in-conditioning}} \\
&= \Exp_{\rh = h} \Exp_{\rM_{\leq i} = M_{\leq i}}\sum_{i =1}^{T-1} \sum_{j=1}^i \ind[ m_j = h] \cdot
\mi{\rP_i}{\rb_{j} \mid \rP_{j-1}, \rLall_{\leq i}, \ry, \Rpublic, \rM_{\leq i} = M_{\leq i}} (\Exp_{\rM_{> i} = M_{>i}} \ind[m_{i+1}=h]),
\end{align*}
where in the last equality, we have used that the joint distribution of all $\rP_i, \rb_j, \rP_{j-1}, \rLall_{\leq i}, \ry, \Rpublic$ is independent of what the value of $\rLall_{> i}, \rM_{>i}$ are.  These random variables do not affect the earlier inputs or the labels of the branching program.  As $m_{i+1}$ is uniform over $[n]$, the expression above simplifies to, 
\begin{align*}
&=\frac1n \cdot\Exp_{\rh = h} \Exp_{\rM_{\leq i} = M_{\leq i}}\sum_{i =1}^{T-1} \sum_{j=1}^i \ind[ m_j = h] \cdot
\mi{\rP_i}{\rb_{j} \mid \rP_{j-1}, \rLall_{\leq i}, \ry, \Rpublic, \rM_{\leq i} = M_{\leq i}}  \\
&= \frac1{nd} \cdot \sum_{i =1}^{T-1} \sum_{j=1}^i  \Exp_{\rM_{\leq i} = M_{\leq i}}\sum_{h = h_1}^{h_d} \ind[ m_j = h] \cdot
\mi{\rP_i}{\rb_{j} \mid \rP_{j-1}, \rLall_{\leq i}, \ry, \Rpublic, \rM_{\leq i} = M_{\leq i}},
\end{align*}
where we sum over the $d = \sens_0[f, \xstar_f]$ many choices (say $h_1, h_2, \ldots, h_d$) for $\rh$ and use the linearity of expectation. By construction, $\rh$ was uniform over these $d$ indices. 

For each choice of $(i,j)$, the mutual information term is:
\[
\Exp_{\rM_{\leq i} = M_{\leq i}} \sum_{h = h_1}^{h_d} \ind[ m_j = h] \cdot
\mi{\rP_i}{\rb_{j} \mid \rP_{j-1}, \rLall_{\leq i}, \ry, \Rpublic, \rM_{\leq i} = M_{\leq i}}.
\]
It is easy to see that if $m_j  \in \set{h_1, \ldots, h_d}$, the term survives, and in fact, is present exactly once. If $m_j$ is not present in $\set{h_1, \ldots, h_d}$, as mutual information is non-negative, we include the term and get an upper bound. 
In either case, we have, 
\begin{align*}
   & \Exp_{\rM_{\leq i} = M_{\leq i}} \sum_{h = h_1}^{h_d} \ind[ m_j = h] \cdot
\mi{\rP_i}{\rb_{j} \mid \rP_{j-1}, \rLall_{\leq i}, \ry, \Rpublic, \rM_{\leq i} = M_{\leq i}} \\
&\leq \Exp_{\rM_{\leq i} = M_{\leq i}}\mi{\rP_i}{\rb_{j} \mid \rP_{j-1}, \rLall_{\leq i}, \ry, \Rpublic, \rM_{\leq i} = M_{\leq i}} \\
&= \mi{\rP_i}{\rb_{j} \mid \rP_{j-1}, \rLall_{\leq i}, \ry, \Rpublic, \rM_{\leq i}}, \tag{by the definition of conditional mutual information} \\
&=  \mi{\rP_i}{\rb_{j} \mid \rP_{j-1}, \rLall, \ry, \Rpublic}, 
\end{align*}
where in the last equality, we have used that $\rM$ is fixed by $\rLall$ and $\rLall_{>i}$ is independent of the joint distribution of all these random variables.
Term 1 in \Cref{eq:interim-eq2} can be upper bounded by:
\begin{align}\label{eq:term1-ub}
     &\Exp_{\rM = M, \rh = h}\Big(\sum_{i =1}^{T-1} \sum_{j=1}^i \ind[m_{i+1} = m_j = h] \cdot
\mi{\rP_i}{\rb_{j} \mid \rP_{j-1}, \rLall, \ry, \Rpublic, \rM = M, \rh = h} \notag\\
&\leq \frac1{nd} \cdot \sum_{i =1}^T \sum_{j=1}^i \mi{\rP_i}{\rb_{j} \mid \rP_{j-1}, \rLall_{\leq i}, \ry, \Rpublic, \rM_{\leq i}} =\frac1{nd} \cdot \ic_{\cD_{\foplus}}(P).
\end{align}

Now, we move on to Term 2 of \Cref{eq:interim-eq2}.
\[
\textbf{Term 2:} \Exp_{\rM = M, \rh = h} \sum_{j=1}^T \ind[m_j = h] \mi{\rP_T}{\rb_{j} \mid \rP_{j-1}, \rLall, \ry, \Rpublic, \rM = M, \rh = h}.
\]
To upper bound this, as a first step, we use part $(iii)$ of \Cref{obs:var-in-conditioning}. As $\rh = h$ is independent of the joint distribution of all the other random variables,
\begin{align*}
    &\Exp_{\rM = M, \rh = h}  \sum_{j=1}^T  \ind[m_j = h] \cdot   \mi{\rP_T}{\rb_{j} \mid \rP_{j-1}, \rLall, \ry, \Rpublic, \rM = M, \rh = h} \\
    &= \Exp_{\rM = M, \rh = h} \sum_{j=1}^T  \ind[m_j = h] \cdot  \mi{\rP_T}{\rb_{j} \mid \rP_{j-1}, \rLall, \ry, \Rpublic, \rM = M}\\
    &= \frac1{d} \cdot\Exp_{\rM = M}  \sum_{j=1}^T  \sum_{h = h_1}^{h_d} \ind[m_j = h] \cdot  \mi{\rP_T}{\rb_{j} \mid \rP_{j-1}, \rLall, \ry, \Rpublic, \rM = M},
\end{align*}
where again we have used that $\rh$ is uniform over $d$ choices, namely, $h_1, \ldots, h_d \in [n]$.
Similar to the upper bound for Term 1, whether $m_j \in \set{h_1, \ldots, h_d}$ or not, we have the following upper bound.
\begin{align*}
&\frac1{d} \cdot\Exp_{\rM = M}  \sum_{j=1}^T   \mi{\rP_T}{\rb_{j} \mid \rP_{j-1}, \rLall, \ry, \Rpublic, \rM = M} \\
    &= \frac1{d} \cdot\sum_{j=1}^T \mi{\rP_T}{\rb_{j} \mid \rP_{j-1}, \rLall, \ry, \Rpublic, \rM} \tag{by definition of conditional mutual information}\\
    &= \frac1{d} \cdot\sum_{j=1}^T \mi{\rP_T}{\rb_{j} \mid \rP_{j-1}, \rLall, \ry, \Rpublic} \tag{as $\rM$ is fixed by $\rLall$}\\
    &\leq \frac1{d} \cdot\sum_{j=1}^T \mi{\rP_T}{\rb_{j} \mid \rb_{<j},\rP_{j-1}, \rP_{<j-1}, \rLall, \ry, \Rpublic} \tag{ as $\rP_{<j-1},\rb_{<j} \perp \rb_j \mid \rP_{j-1}, \rLall, \ry, \Rpublic$ and by \itfacts{both-light-more}}\\
    &\leq  \frac1{d} \cdot\sum_{j=1}^T \mi{\rP_T}{\rP_{j-1}, \rb_{j} \mid \rb_{<j}, \rP_{<j-1}, \rLall, \ry, \Rpublic} \tag{by chain rule \itfacts{mi-chain} and non-negativity \itfacts{mi-nonnegative}} \\
    &= (1/d) \cdot \mi{\rP_T}{\rP_{<T}, \rb_1, \ldots, \rb_T \mid \ry, \rLall, \Rpublic} \tag{by chain rule of mutual information \itfacts{mi-chain}} \\
    &\leq S/d. \tag{as $\en{\rP_T} \leq S$ and by definition of mutual information}
\end{align*}
We have upper bounded Term 2 simply by $S/d$. 
Combining with \Cref{eq:interim-eq2} and \Cref{eq:term1-ub},

\[
\mi{\rProt'}{\rtheta
\mid \rrho,\Rpublictwo,\ralpha,\eventeq}
\leq \frac{1}{nd} \cdot \ic_{\cD_{\foplus}}(P) + S/d,
\]
proving the claim.
\end{proof}

We can now finish the proof of \Cref{thm:equal-lb} under \Cref{ass:sens-0-bit}. 
\begin{corollary}\label{cor:thm-with-assme}
    For any function $f: \set{0,1}^n \rightarrow \set{0,1}$ where \Cref{ass:sens-0-bit} holds, if a randomized branching program $P$ computes $g_{\foplus}$ with probability of success at least $2/3$ with length $T$ and width $2^S$, then, 
    \[
    T \cdot (S + \log T) = \Omega(n \cdot \sens[f]).
    \]
\end{corollary}
\begin{proof}[Proof of \Cref{thm:equal-lb} under \Cref{ass:sens-0-bit}]

First, we change the format of the input and construct a new branching program with length $T' = O(T)$ and width $2^{S'} = O(2^{S} \cdot T)$ that gets input samples $(\ell_i, b_i, w_i)$ as in \Cref{clm:change-bp-input}. We construct program $P'$ such that it succeeds with probability at least $2/3$, and
we know,
\begin{equation}\label{eq:interim-4}
T' = O(T) \qquad S' = O(S + \log T).
\end{equation}

We construct $\prot_{\two}$ using branching program $P'$.
This protocol declares FAIL with
probability at most $1/n^2$ by \Cref{clm:sequence-construct}. Otherwise, by
\Cref{clm:newprot-success}, its success probability is at least:
\[
\textnormal{Probability of success of $\prot_{\two}$} = \frac{2}{3}-\frac1{n^2} \geq \frac58,
\]
for sufficiently large $n$. 
Applying \Cref{lem:two-bit-game},
\begin{align*}
    \Omega(1) &\leq \mi{\rPrott}{\rtheta
\mid \rrho,\Rpublictwo,\ralpha,\eventeq} \\
& \leq  \mi{\rProt'}{\rtheta
\mid \rrho,\Rpublictwo,\ralpha,\eventeq} +H_2(1/n^2) \tag{by \Cref{clm:aux-transcript} for the auxiliary transcript} \\
&\leq \frac1{n \cdot \sens_0[f,\xstar_f]} \cdot \ic_{\cD_{\foplus}}(P') + S' /\sens_0[f,\xstar_f]+ H_2(1/n^2) \tag{by \Cref{clm:direct-sum-information}} \\
&\leq  \frac1{n \cdot \sens_0[f,\xstar_f]} \cdot T' \cdot S' + \frac{S'}{\sens_0[f, \xstar_f]} + H_2(1/n^2). \tag{by \Cref{clm:information-cost-upper-bound}}
\end{align*}
For large $n$, the binary entropy $H_2(1/n^2) \leq (2/n^2) \cdot \log(n^2) \leq 1/n$. We can simplify our bound as, 
\[
T' \cdot S' + n \cdot S'= \Omega(n \cdot \sens_0[f, \xstar_f]).
\]
By \Cref{eq:interim-4}, using upper bounds on $T'$ and $S'$,
\[
T \cdot (S + \log T) + n \cdot (S + \log T) = \Omega(n\cdot \sens_0[f, \xstar_f]). 
\]
As $T = \Omega(n)$ length is required to see the sensitive bits for any non-constant function, 
\[
T \cdot (S + \log T) = \Omega(n \cdot \sens_0[f, \xstar_f]),
\]
finishing the proof.
\end{proof}

\subsubsection*{Removing \Cref{ass:sens-0-bit}}

Under \Cref{ass:sens-0-bit}, we had that $\sens_0[f, \xstar_f] = \Omega(\sens[f])$. However, the sensitive bits of $\xstar$ may be the indices with value $1$ in it. We show that if the inner function is parity, even the bits of $\xstar_f$ with value $1$ can be used to embed a smaller two-bit parity game. 

We define a new function $\fbar:\set{0,1}^n \rightarrow \set{0,1}$ with negated input as follows:
\[
\fbar(x) = f(x \oplus e_1 \oplus e_2 \ldots \oplus e_n) = f(\neg x),
\]
i.e., the output of $\fbar$ is the output of $f$ on the negation of the given input. 

It is easy to see that for any input $x \in \set{0,1}^n$, $\sens_1[f, x]  = \sens_0[\fbar, \neg x]$. 
Similarly, function $\gbar:\set{0,1}^{2n} \rightarrow \set{0,1}$ is defined as:
\begin{align*}
\gbar(y) &= \fbar(y(1) \oplus y(2), y(3) \oplus y(4), \ldots, y(2n-1) \oplus y(2n))\\
&= f(y(1) \oplus \neg y(2), y(3) \oplus \neg y(4), \ldots, y(2n-1) \oplus \neg y(2n)).
\end{align*}

We can give the full proof of \Cref{thm:equal-lb}.

\begin{proof}[Final Proof of \Cref{thm:equal-lb}]
    Given a function $f$, let $\xstar_f$ be some input of $f$ with maximum number of sensitive bits, i.e.,
    \[
    \sens_0[f, \xstar_f] + \sens_1[f, \xstar_f] = \sens[f].
    \]
    If $\sens_0[f, \xstar_f] \geq \sens[f]/2$, \Cref{ass:sens-0-bit} holds, so for any branching program $P$ that solves $g_{\foplus}$ with probability at least $2/3$, we get $T(S+\log T) = \Omega(n \cdot \sens_0[f, \xstar_f]) = \Omega(n \cdot \sens[f])$.
 Otherwise, \Cref{ass:sens-0-bit} holds for $\fbar$ as $\sens_0[\fbar,\neg \xstar_f] = \sens_1[f, \xstar_f] \geq \sens[f]/2$. For any branching program $\Pbar$ that computes $\gbar$ with probability of  success at least $2/3$, we get $T \cdot (S +\log T) = \Omega(n \cdot \sens[\fbar])$.

    To get a lower bound directly for a branching program that computes $g$, it is easy to see that computing $\gbar$ also solves $g$ by simply negating any input $y(2i)$ to $\neg y(2i)$ for $i \in [n]$ and feeding it to a program that computes $\gbar$.
\end{proof}

\subsection{Extension to Other Inner Functions}\label{sec:extend-other-fns}
In this section we give a reduction where the inner function is not Parity, but we can still prove non-trivial lower bounds.

We use the Tribes function as an example to illustrate other lower bounds we can get. 
Given input $y \in \set{0,1}^{2n}$,\[
\tribes(y):= \bigvee_{i \in [n]} (y(2i-1) \wedge y(2i)).
\]

\begin{corollary}\label{cor:tribes}
    For $n\geq 1$, any branching program $P$ that computes $\tribes:\set{0,1}^{2n} \rightarrow \set{0,1}$ with length $T$ and width $2^S$ and probability of success at least $9/10$ must satisfy,
    \[
    T \cdot (S + \log T) = \Omega(n^2).
    \]
\end{corollary}
\newcommand{\ineq}{\textnormal{\textsc{ineq}}}
\begin{proof}
    We show that $P$ can be used to solve the inequality function, which is the OR of many XORs. 
    We define:
    \[
    \ineq(y) = \bigvee_{i \in [n]} (y(2i-1) \oplus y(2i)).
    \]

    Given input $y \in\set{0,1}^{2n}$, we construct a program $P'$ for $\ineq$ as follows:
    \begin{itemize}
        \item Run program $P$ on $y_1 = (y \oplus e_1 \oplus e_3 \oplus \ldots \oplus e_{2n-1})$, i.e., the bits in the odd positions of $y$ are negated. If $P$ returns $1$ output $1$. 
        \item Otherwise, run program $P$ on $y_2 = (y \oplus e_2 \oplus e_4 \oplus \ldots \oplus e_{2n})$, where the even bits are negated. If $P$ returns $1$, output $1$.
        \item If neither of the runs of $P$ return one, we output zero.
    \end{itemize}

It is easy to see that $P'$ has length $T' = O(T)$ and width $2^{S'} = O(2^S)$. The first run of $P$ in $P'$ returns $1$ if there exists some $i \in [n]$ with $y(2i-1)=0$ and $y(2i) = 1$, and returns $0$ in all other cases by the definition of $\tribes$. The second run of $P$ returns $1$ if there exists some $i \in [n]$ with $y(2i-1)=1$ and $y(2i) = 0$, and returns $0$ in all other cases.

Therefore, if $\ineq(y) = 1$, at least one of the two runs of $P$ in $P'$ return $1$. Otherwise both return $0$. When the output of both runs of $P$ in $P'$ are correct, our program $P'$ succeeds as well. It has probability of success at least $2/3$.

By \Cref{thm:equal-lb}, branching program $P'$ must have,
\[
T' (S'+\log(T')) = \Omega(n^2), \qquad \text{ which in turn gives } \qquad T(S + \log T) = \Omega(n^2). \qedhere
\]
\end{proof}

$\tribes_w$ is an extension of the tribes function where there are $w$-indices whose AND is taken inside each term:
For any input $y \in \set{0,1}^{n}$,
\[
\tribes_w(y) = \bigvee_{i \in [n/w]} \paren{\bigwedge_{j \in [w]} y((i-1)w + j)}.
\]

There is a simple branching program that has $T \cdot S = \Ot(n^2 / w)$ to evaluate $\tribes_w$ function: we evaluate an AND function using $\Ot(n)$ samples and $\tilde{O}(1)$ width, and repeat for the $(n/w)$-different AND functions that $\tribes_w$ needs. If width is $2^S$, the $(n/w)$-different AND functions can be batched into $\Ot(n/wS)$ groups of size $\Omgt(S)$ each, and each batch can be evaluated simultaneously. Our techniques also prove a lower bound of $\Omgt(n^2/w)$ for $\tribes_w$ function via another simple reduction. 

\begin{corollary}\label{cor:tribes-w}
     For $n,w\geq 2$, any branching program $P$ that computes $\tribes_w:\set{0,1}^{n} \rightarrow \set{0,1}$ with length $T$ and width $2^S$ and probability of success at least $15/16$ must satisfy,
    \[
    T \cdot S = \Omgt (n^2 / w).
    \]
\end{corollary}
\begin{proof}
    Assume that there exists a branching program for $\tribes_w$ with length $T$ and width $2^S$. We use it to construct a branching program for $\tribes$ where the AND is only over two indices of the input. Given input $x \in \set{0,1}^{2n/w}$, we fix $y \in \set{0,1}^{n}$ as:
    \[
    y(i) = 
    \begin{cases}&x(2j-1) \qquad \text{if $i = (j-1)w + 1$} \\
    &x(2j) \qquad \text{if $i = (j-1)w + 2$} \\
    &1 \qquad \text{otherwise.}
    \end{cases}
    \]
    For this input $y$, we get:
    \[
    \tribes_w(y) = \bigvee_{i \in [n/w]} \paren{\bigwedge_{j \in [w]} y((i-1)w + j)} =  \bigvee_{i \in [n/w]} \paren{x(2i-1) \wedge x(2i)} = \tribes(x).
    \]
    Solving $\tribes_w(y)$ also gives $\tribes(x)$. We construct a branching program $P'$ for $\tribes$ using $P$. At each node, we run $P$ by simulating random indices from $[n]$, till some index requiring a position from $x$ is sampled. At this point, we get the value from the input to $P'$. 
    Only $(2/w)$-fraction of the input indices are from $x$.
    By \Cref{prop:chernoff}, the number of samples used by $P'$ is at most $(2T/w) \cdot \log n$ with high probability.

    Program $P'$ has width $2^{S'}$ where $S' = S + O(\log T)$, as we use $O(\log T)$ bits to keep track of the number of samples given to $P$ so far. It has length $T' = \Ot(T/w)$ and solves $\tribes$ with probability of success at least $9/10$ for large enough $n$.
    By \Cref{cor:tribes}, we get $T \cdot S = \Omgt(n^2 / w)$.
\end{proof}

The choice of outer function as OR in \Cref{cor:tribes} can be changed to any function $f$ if we focus on the input $\xstar_0$ to $f$ with the maximum number of sensitive zeroes. We can compute Inequality using a branching program for $f \circ \wedge_2$ also, to get a lower bound of $\Omgt(n \cdot \sens_0[f, \xstar_0])$.

We are not restricted to inner function as AND: we can also directly get a lower bound for the AND of ORs by a reduction from $\tribes$ using negations. Similarly, with inner function as OR, we can prove a lower bound for any outer function $f$ if we focus on the input to $f$ with maximum number of sensitive ones. We state these extensions in the next corollary, but omit the details as they follow via a reduction from Inequality.
\begin{corollary}\label{cor:others}
The following statements follow from \Cref{thm:equal-lb}.
    \begin{enumerate}
        \item For any function $f:\set{0,1}^n \rightarrow \set{0,1}$, let $\xstar_1$ be the input with the maximum number of sensitive bits with value $1$. That is, $\xstar_1$ is the input with maximum $\sens_1[f, x]$. Let $g_{f \vee}:\set{0,1}^{2n} \rightarrow \set{0,1}$ be the following function.
        \[
        g_{f \vee}(y):= f(y(1) \vee y(2), y(3) \vee y(4), \ldots, y(2n-1) \vee y(2n)).
        \]
        Then any branching program that computes $g_{f\vee}$ with probability of success at least $99/100$, length $T$ and width $2^S$ satisfies $T \cdot S = \Omgt(n \cdot \sens_1[f, \xstar_1])$.
        \item For any function $f:\set{0,1}^n \rightarrow \set{0,1}$, let $\xstar_0$ be the input with the maximum number of sensitive bits with value $0$. That is, $\xstar_0$ is the input with maximum $\sens_0[f, x]$. Let $g_{f \wedge}:\set{0,1}^{2n} \rightarrow \set{0,1}$ be the following function.
        \[
        g_{f \wedge}(y):= f(y(1) \wedge y(2), y(3) \wedge y(4), \ldots, y(2n-1) \wedge y(2n)).
        \]
        Then any branching program that computes $g_{f\wedge}$ with probability of success at least $99/100$, length $T$ and width $2^S$ satisfies $T \cdot S = \Omgt(n \cdot \sens_0[f, \xstar_0])$.
    \end{enumerate}
\end{corollary}

\noindent
\textbf{Element Distinctness}

Element Distinctness asks whether $n$ elements from a list of $m = \poly(n)$ entries are all distinct:
\[
\text{ED}(z(1), \ldots, z(n)) = 1\textnormal{ if and only if $z(i) \neq z(j)$ for all $i \neq j$},
\]
where each $z(i) \in [m]$, and can be represented with $O(\log n)$ bits.

It is a well-studied problem in classical time--space complexity. For list size $m$ polynomial in $n$, ~\cite{BeameCM13} gave a randomized algorithm with $T=\widetilde{O}(n^{3/2}/\sqrt{S})$ under a random-oracle assumption and later this assumption was removed by \cite{LyuZ23}. We also know that the number of queries required by any standard branching program of width $2^S$ is at least $T = \Omega(n\sqrt{\log(n/S)/\log\log(n/S)})$ \cite{Ajtai05,BeameSSV03}.

In the random query model, an index $i \in [n]$ is picked uniformly at random, and $(i, z(i))$ is given as the query. It is easy to see that there is a trivial upper bound of $T\cdot S = \Ot(n^2)$ for this problem, by storing $O(S/\log n)$ elements at a time, and checking for collisions using $O(n \log n)$ samples. 
A reduction from Equality gives the following tight tradeoff in the random-query model for Element Distinctness.\footnote{The input to Element Distinctness can be represented by $O(n \log n)$ bits in total, and
we also get a lower bound of $\Omgt(n^2)$ if each sample is $(j, \text{bit at position }j)$ as opposed to entire elements for index $j$ chosen uniformly at random from the $O(n \log n)$ bits.  However, the upper bound of $\Ot(n^2)$ does not translate directly if bits arrive instead, and this model might be of separate interest. }

\begin{corollary}
Any random-query algorithm for Element Distinctness on $n$ elements from $[m]$ for any $m \geq n$ with probability of success at least $2/3$, using $T$ samples and $S$ bits of space, satisfies
$$
T(S+\log n)=\widetilde{\Omega}(n^2).
$$
\end{corollary}

\begin{proof}
Again, we define the $\ineq: \set{0,1}^{2n} \rightarrow \set{0,1}$ as:
\[
\ineq(y) = \bigvee_{i \in [n]} \paren{y(2i-1) \oplus y(2i)}.
\]
 Given an inequality instance $y\in\{0,1\}^{2n}$, we define $z\in[2n]^{2n}$ as:
$$
z(2i-1)=i+n\cdot y(2i-1),
\qquad
z(2i)=i+n(1-y(2i))
\qquad \text{ for }i\in[n].
$$
First, it is easy to see that $z(2i-1), z(2i) \in \set{i, n+i}$ irrespective of the value of $y$. The two elements $z(2i-1)$ and $z(2i)$ are unequal if and only if $y(2i-1) = y(2i)$. Hence, the elements of $z$ are pairwise distinct if and only if $\ineq(y) = 0$.

Each uniformly random bit sample from $y$ at position $i$ determines the element of $z$ at the same position $i \in [2n]$. Given bit $y(i)$, all the $O(\log n)$ bits that encode $z(i)$ can be computed, and given as queries. We can simulate the Element Distinctness algorithm by sending $(i, z(i))$ every time we receive a uniformly random index $i \in [2n]$ for solving $\ineq(y)$. An Element Distinctness algorithm yields an Inequality algorithm with the same number of samples and error probability, using $S+O(\log n)$ space. The statement follows from \Cref{thm:equal-lb}.
\end{proof}
 
 \section{Two Bit Parity Game}\label{sec:2bitgame}

To prove a lower bound on the information cost, we prove a lower bound on another intermediate distributional communication complexity problem called the two bit parity game.

We know that a successful protocol reveals information about $\ralpha$ and $\rbeta$. However, we want to prove that in order to reveal information about these two input bits, any successful protocol must necessarily send information about whether it received any input bit through the vector $\rtheta$. Moreover, we want to say that information about $\rtheta$ is revealed even conditioned on the event $\eventeq$, and the answer $\ralpha \oplus \rbeta $ is fixed to be $0$. 

The proof of this lemma proceeds in three parts:
\begin{itemize}
    \item Focus on the messages sent by each player individually, and how they incrementally change the distribution of $\alpha, \beta$ reflects on how much information the transcript reveals about their $\theta_i$ value. 
    \item Argue that if some randomness is left in $\alpha$ and $\beta$, and an individual player reveals information about them, then, they reveal information about $\rtheta$ even conditioned on $\eventeq$. This part relies crucially on no player knowing both $\alpha$ and $\beta$.
    \item Lastly, we put the two steps together. This part is mainly concerned with analyzing the transcripts where a lot of information about $\alpha, \beta$ is revealed before some players even send any messages to the board.
\end{itemize}

We start with the first part. This is a fairly simple step in the proof. 

\paragraph{Separation into individual players.}

\begin{claim}[Bounding information in the transcript by information revealed by the individual players]\label{clm:2bit-separate}
For any protocol $\prot$,
\[
\mi{\rProt}{\rtheta
\mid \rrho,\Rpublic,\ralpha,\eventeq}
\geq \sum_{i \in [k]} \mi{\rProt_i}{\rtheta_i \mid \rrho_i, \Rpublic, \ralpha, \rProt_{<i}, \eventeq}.
\]
\end{claim}
\begin{proof}
   Using chain rule of mutual information, \itfacts{mi-chain}
   \begin{align*}
   \mi{\rProt}{\rtheta
\mid \rrho,\Rpublic,\ralpha,\eventeq} &= \sum_{i \in [k]}\mi{\rProt_i}{\rtheta
\mid \rProt_{<i},\rrho,\Rpublic,\ralpha,\eventeq} \\
&= \sum_{i \in [k]} \sum_{j \in [k]}\mi{\rProt_i}{\rtheta_j
\mid\rtheta_{<j}, \rProt_{<i},\rrho,\Rpublic,\ralpha,\eventeq}.
   \end{align*}

We know $\rProt_i$ is independent of $\rtheta_{<i}, \rrho_{<i}$ conditioned on $\rProt_{<i}$, $\rtheta_i$, values $\ralpha, \rbeta$, $\rrho_i$ and public randomness $\Rpublic$. Hence for $j < i$, we have,
\begin{equation}\label{eq:interim-1}
    \mi{\rProt_i}{\rtheta_j
\mid\rtheta_{<j}, \rProt_{<i},\rrho,\Rpublic,\ralpha,\eventeq} = 0.
\end{equation}
For $j = i$, again as $\rtheta_{<i}$ is independent of $\rProt_i$ conditioned on $\rProt_{<i}, \ralpha, \rbeta, \rrho$ and $\Rpublic$, by \itfacts{both-light-more}, we get,
\begin{equation}\label{eq:interim-2}
    \mi{\rProt_i}{\rtheta_i
\mid\rtheta_{<i}, \rProt_{<i},\rrho,\Rpublic,\ralpha,\eventeq} \geq  \mi{\rProt_i}{\rtheta_i
\mid\rProt_{<i},\rrho_i,\Rpublic,\ralpha,\eventeq}.
\end{equation}
Lastly for $j > i$, we know $\rProt_i$ is independent of all the values $\rtheta_j$ that come after $i$ even conditioned on $\rProt_{<i}$, $\rrho, \ralpha, \rbeta$ and $\Rpublic$. Hence all terms with $j > i$ become zero by \itfacts{mi-nonnegative}.

Putting together \Cref{eq:interim-1} and \Cref{eq:interim-2}, we finish the proof of this claim. 
\end{proof}

Using \Cref{clm:2bit-separate}, we can separate the analysis into the transcripts of each player.

\subsection{Learning theta-values through the two input bits}
All we know about the final protocol is that information about $\ralpha$ and $\rbeta$ are revealed. We leverage this information to argue that information about $\rtheta$ is revealed also. We fix the bits of public randomness, a player $\player_i$ for some $i \in [k]$ and a transcript prefix $\Prot_{<i}$ sampled from the distribution of $\rProt_{<i}$, and analyze the distribution of $\rProt_i$ conditioned on these random variables.

Given $\Prot_{<i}$ and the public randomness $\Rpublicn$, for $u, v \in \set{0,1}$, we define:
\begin{align*}
    \delta_{uv}:=\text{Probability that $\ralpha=u$ and $\rbeta = v$ conditioned on $\rProt_{<i} = \Prot_{<i}$, $\Rpublic = \Rpublicn$}
\end{align*}

Initially, for $i = 1$, all $\delta_{uv} = 1/4$ as $\alpha$ and $\beta$ are chosen uniformly at random. The transcript reveals some information about these bits to change the probability distributions. We argue that any \emph{additional} information revealed by $\rProt_i$ also reveals information about $\rtheta_i$. 

\begin{lemma}[Information about $\alpha, \beta$ should reveal information about $\rtheta_i$ even conditioned on $\eventeq$]\label{lem:thru-theta}
For any fixed prior transcript $\Prot_{<i}$ and public randomness $\Rpublicn$, let $\delta_{uv} \geq 0.1$ for all $u,v \in \set{0,1}$. Then,
    \[
    \mi{\rProt_i}{\rtheta_i \mid \rrho_i, \ralpha, \Rpublicn, \Prot_{<i}, \eventeq}
    \geq \frac1{64} \cdot\mi{\rProt_i}{\ralpha, \rbeta \mid \rrho_i, \Rpublicn, \Prot_{<i}}.
    \]
\end{lemma}

\Cref{lem:thru-theta} proves that as long as there is some randomness left in $\ralpha$ and $\rbeta$, information about $\rtheta_i$ will be revealed. We show how to remove the assumption in the next subsection. 
We define multiple distributions of $\rProt_i$ based on the input that player $\player_i$ gets:
\begin{align*}
    \cD^{(0)}_{\alpha}&:= \text{Distribution of $\rProt_{i}$ when $\player_i$ gets $\rtheta = 0$, $\rrho_i = $`$\alpha$'} \\
    \cD^{(0)}_{\beta}&:= \text{Distribution of $\rProt_{i}$ when $\player_i$ gets $\rtheta = 0$, $\rrho_i = $`$\beta$'} \\
    \cD^{(1)}_{\alpha u}&:= \text{Distribution of $\rProt_{i}$ when $\player_i$ gets $\rtheta = 1$, $\rrho_i = $`$\alpha$', $\alpha = u$} \\
    \cD^{(1)}_{\beta u}&:= \text{Distribution of $\rProt_{i}$ when $\player_i$ gets $\rtheta = 1$, $\rrho_i = $`$\beta$', $\beta = u$} 
\end{align*}
These distributions are conditioned on $\rProt_{<i} = \Prot_{<i}$ and $\Rpublic = \Rpublicn$, but not the event $\eventeq$. We argue that this conditioning does not change things either.

\begin{claim}\label{clm:remove-alphabeta}
The distribution of $\rProt_i$ is independent of event $\eventeq$ after:
\begin{enumerate}[label=$(\roman*)$]
    \item conditioning on $\rtheta = 0$, $\rrho_i = \charalpha$, $\Prot_{<i}$ and $\Rpublicn$.
    \item conditioning on $\rtheta = 1$, $\rrho_i = \charalpha$, $\Prot_{<i}$, $\Rpublicn$ and value of $\ralpha$.
    \item similarly, conditioning on $\rtheta = 0$, $\rrho_i = \charbeta$, $\Prot_{<i}$ and $\Rpublicn$.
    \item conditioning on $\rtheta = 1$, $\rrho_i = \charbeta$, $\Prot_{<i}$, $\Rpublicn$ and value of $\rbeta$.
\end{enumerate}
\end{claim}
\begin{proof}
Let $\rv{R}_i$ denote the private randomness of player $\player_i$, and let $\ind[\eventeq]$ be the indicator random variable of the event $\ralpha=\rbeta$. We argue that the mutual information between $\rProt_i$ and $\ind[\eventeq]$ is zero after all the conditioning, and this proves the claim by \itfacts{mi-nonnegative}.

For case $(i)$, when $\rtheta_i=0$, player $\player_i$ receives neither $\ralpha$ nor $\rbeta$. After conditioning on $\rrho_i=\charalpha$, $\Prot_{<i}$ and $\Rpublicn$, its message $\rProt_i$ depends only on $\rv{R}_i$. This private randomness is independent of $\ralpha,\rbeta$ and the transcript prefix.
\begin{align*}
&\mi{\rProt_i}{\ind[\eventeq] \mid \rtheta_i=0, \rrho_i=\charalpha, \Prot_{<i}, \Rpublicn} \\
&\leq \mi{\rProt_i}{\ralpha,\rbeta \mid \rtheta_i=0, \rrho_i=\charalpha, \Prot_{<i}, \Rpublicn} \tag{by \itfacts{data-processing}, as $\ind[\eventeq]$ is fixed by $\ralpha,\rbeta$} \\
&\leq \mi{\rv{R}_i}{\ralpha,\rbeta \mid \rtheta_i=0, \rrho_i=\charalpha, \Prot_{<i}, \Rpublicn} \tag{by \itfacts{data-processing}, as $\rProt_i$ depends only on $\rv{R}_i$} \\
&=0. \tag{by independence of the private randomness}
\end{align*}

For case $(ii)$, we fix the value $\ralpha=u$ for any $u \in \set{0,1}$. When $\rtheta_i=1$ and $\rrho_i=\charalpha$, player $\player_i$ receives $u$ but does not receive $\rbeta$. After additionally conditioning on $\Prot_{<i}$ and $\Rpublicn$,  $\rProt_i$ depends only on $\rv{R}_i$, which is independent of $\rbeta$. Hence,
\begin{align*}
&\mi{\rProt_i}{\ind[\eventeq] \mid \rtheta_i=1, \rrho_i=\charalpha, \ralpha=u, \Prot_{<i}, \Rpublicn} \\
&\leq \mi{\rProt_i}{\rbeta \mid \rtheta_i=1, \rrho_i=\charalpha, \ralpha=u, \Prot_{<i}, \Rpublicn} \tag{by \itfacts{data-processing}, as $\ind[\eventeq]=\ind[\rbeta=u]$} \\
&\leq \mi{\rv{R}_i}{\rbeta \mid \rtheta_i=1, \rrho_i=\charalpha, \ralpha=u, \Prot_{<i}, \Rpublicn} \tag{by \itfacts{data-processing}, as $\rProt_i$ depends only on $\rv{R}_i$} \\
&=0. \tag{by independence of private randomness, \itfacts{mi-nonnegative}}
\end{align*}

 Cases $(iii)$ and $(iv)$ follow similarly.
\end{proof}

We are ready to give the proof of \Cref{lem:thru-theta}, which is the main lemma of this subsection. 

 \begin{proof}[Proof of \Cref{lem:thru-theta}]
Using \Cref{fact:weighted-js-info}, we expand both the LHS and RHS of \Cref{lem:thru-theta} using the appropriately weighted JS-divergences of these distributions.

\noindent
\textbf{RHS:}
Let $p_{\alpha}, p_{\beta}$ be the probabilities with which $\alpha$ and $\beta$ become zero respectively. 
We know:
\begin{align*}
p_\alpha&=\delta_{00}+\delta_{01} \tag{the probability with which $\alpha = 0$} \\
p_\beta&=\delta_{00}+\delta_{10} \tag{the probability with which $\beta = 0$}
\end{align*}
For $u\in\set{0,1}$, we define two distributions:
\[
\cG_u:=\frac12\cD^{(0)}_\alpha+\frac12\cD^{(1)}_{\alpha u},
\qquad
\cH_u:=\frac12\cD^{(0)}_\beta+\frac12\cD^{(1)}_{\beta u}.
\]
 $\cG_u$ is the distribution of $\rProt_i$ conditioned additionally on
$\rrho_i = `\alpha$' and $\ralpha=u$ along with $\Prot_{<i}, \Rpublicn$. Distribution $\cH_u$ is defined similarly for $\beta$.   Using \Cref{fact:weighted-js-info}, we can now write the RHS
of \Cref{lem:thru-theta} exactly as
\begin{align*}
&\mi{\rProt_i}{\ralpha,\rbeta\mid\rrho_i,\Rpublicn,\Prot_{<i}} \\
&=\frac12 \cdot (\js{(p_\alpha,1-p_\alpha)}{\cG_0}{\cG_1}
+\js{(p_\beta,1-p_\beta)}{\cH_0}{\cH_1}) \\
&\leq \js{(1/2,1/2)}{\cG_0}{\cG_1}
+
\js{(1/2,1/2)}{\cH_0}{\cH_1}.\tag{by \Cref{fact:js-weight-change}}
\end{align*}
Hence, at least one of the two pairs $\cG_0,\cG_1$ and $\cH_0,\cH_1$ has
unweighted JS-divergence at least one half of the RHS of the lemma statement.  Without loss of generality, we assume: 
\begin{equation}
\js{(1/2,1/2)}{\cG_0}{\cG_1}
\geq
\frac12\cdot\mi{\rProt_i}{\ralpha,\rbeta
\mid\rrho_i,\Rpublicn,\Prot_{<i}}.
\label{eq:far-alpha-pair}
\end{equation}
The case when $\cH_0$ and $\cH_1$ is the pair with high JS-divergence follows similarly.

The two distributions $\cG_0$ and $\cG_1$ in \Cref{eq:far-alpha-pair} are far apart in JS-divergence. We show that at least one of these two distributions is far from $\cD^{(0)}_{\alpha}$ using  JS-distance. Using \Cref{fact:js-distance} and applying the triangle inequality to the square
root of JS gives:
\begin{align*}
\sqrt{\js{(1/2,1/2)}{\cG_0}{\cG_1}}
&\leq
\sqrt{\js{(1/2,1/2)}{\cG_0}{\cD^{(0)}_\alpha}}
+
\sqrt{\js{(1/2,1/2)}{\cD^{(0)}_\alpha}{\cG_1}} \\
\js{(1/2,1/2)}{\cG_0}{\cG_1}
&\leq
\js{(1/2,1/2)}{\cG_0}{\cD^{(0)}_\alpha}
+
\js{(1/2,1/2)}{\cD^{(0)}_\alpha}{\cG_1}\\
&+ 2 \paren{\sqrt{\js{(1/2,1/2)}{\cG_0}{\cD^{(0)}_\alpha}}}\paren{\sqrt{\js{(1/2,1/2)}{\cD^{(0)}_\alpha}{\cG_1}}}\\
&\leq
2(\js{(1/2,1/2)}{\cG_0}{\cD^{(0)}_\alpha}
+
\js{(1/2,1/2)}{\cD^{(0)}_\alpha}{\cG_1}) \tag{using AM $\geq$ GM inequality}
\end{align*}
Hence, for at least one  of $u\in\set{0,1}$, we have:
\begin{equation}
\js{(1/2,1/2)}{\cD^{(0)}_\alpha}{\cG_u}
\geq
\frac18 \cdot \mi{\rProt_i}{\ralpha,\rbeta
\mid\rrho_i,\Rpublicn,\Prot_{<i}},
\label{eq:far-from-no-input}
\end{equation}
where for the last inequality we have used \Cref{eq:far-alpha-pair}.
By joint convexity of unweighted JS in \Cref{fact:js-convex} and the definition of $\cG_u$,
\[
\js{(1/2,1/2)}{\cD^{(0)}_\alpha}{\cG_u}
\leq
\frac12 \cdot \js{(1/2,1/2)}
{\cD^{(0)}_\alpha}{\cD^{(1)}_{\alpha u}}.
\]
Combining with \Cref{eq:far-from-no-input}, we get that there exists $u \in \set{0,1}$ with
\begin{equation}
\js{(1/2,1/2)}
{\cD^{(0)}_\alpha}{\cD^{(1)}_{\alpha u}}
\geq
\frac14 \cdot \mi{\rProt_i}{\ralpha,\rbeta
\mid\rrho_i,\Rpublicn,\Prot_{<i}}.
\label{eq:theta-js-lower-bound}
\end{equation}

Therefore, we have argued that the RHS being large directly relates to the differences in whether $\rtheta$ was 0 or 1 in at least one of the possible input cases.
We compare this quantity with the LHS. 

\noindent
\textbf{LHS:}

We know $\rtheta_i$ is uniform over $\set{0,1}$ even after conditioning
on $\rrho_i$, $\Rpublicn$, $\eventeq$, $\ralpha$, and the transcript $\Prot_{<i}$. Thus the weighted-JS identity from \Cref{fact:weighted-js-info} expresses this information as the average
of the unweighted JS-divergences between the $\theta_i=0$ and
$\theta_i=1$ message distributions.  

Let $q_v$ be the probability that $\ralpha = \rbeta  = v$ for $v \in \set{0,1}$, conditioned on
$\eventeq$, and $\Prot_{<i}, \Rpublicn$.
From the statement of the lemma we know that each of the four possible choices of $(\alpha, \beta)$ values happen with probability at least $0.1$, after conditioning on $\Prot_{<i}, \Rpublicn$. Hence,
\begin{align*}
    q_v &= \Pr[\ralpha = \rbeta = v \mid \eventeq, \Prot_{<i}, \Rpublicn] \\
    &= \frac{\text{$\Pr[\ralpha = \rbeta = v \mid \Prot_{<i}, \Rpublicn]$}}{\text{$\Pr[\eventeq \mid \Prot_{<i}, \Rpublicn]$}} \tag{by Bayes rule}\\
    &= \delta_{vv} \cdot \frac1{\delta_{00} + \delta_{11}} \\
    &= \frac{\delta_{vv}}{1-(\delta_{01}+\delta_{10})} \geq \frac{0.1}{1-0.2} = \frac18.
\end{align*}

Now, we can simplify the LHS of the statement. 
\begin{align*}
&\mi{\rProt_i}{\rtheta_i
\mid\rrho_i,\ralpha,\Rpublicn,\Prot_{<i},\eventeq}\\
&= \sum_{v \in \set{0,1}} q_v \cdot \mi{\rProt_i}{\rtheta_i \mid \rrho_i, \ralpha = \rbeta = v, \Prot_{<i}, \Rpublicn}\tag{by definition $q_v$, and expanding conditional mutual information} \\
&=
\Pr[\rrho_i = \charalpha] \cdot \paren{\sum_{v\in\set{0,1}}q_v\cdot 
\mi{\rProt_i}{\rtheta_i \mid \rrho_i=\charalpha, \ralpha = \rbeta = v, \Prot_{<i}, \Rpublicn}}
\\
&\qquad\qquad+
\Pr[\rrho_i = \charbeta]\cdot \paren{\sum_{v\in\set{0,1}}q_v \cdot
\mi{\rProt_i}{\rtheta_i \mid \rrho_i=\charbeta, \ralpha = \rbeta = v, \Prot_{<i}, \Rpublicn}}\tag{as $\rrho_i$ is independent of $\ralpha, \rbeta, \Prot_{<i}, \Rpublicn$}\\
&=
\frac12 \cdot \paren{\sum_{v\in\set{0,1}}q_v\cdot 
\mi{\rProt_i}{\rtheta_i \mid \rrho_i=\charalpha, \ralpha = \rbeta = v, \Prot_{<i}, \Rpublicn}}
\\
&\qquad\qquad+
\frac12\cdot \paren{\sum_{v\in\set{0,1}}q_v \cdot
\mi{\rProt_i}{\rtheta_i \mid \rrho_i=\charbeta, \ralpha = \rbeta = v, \Prot_{<i}, \Rpublicn}}\tag{as $\rrho_i$ is uniform over $\charalpha,\charbeta$}\\
&=
\frac12 \cdot \paren{\sum_{v\in\set{0,1}}q_v\cdot 
\js{(1/2,1/2)}
{\cD^{(0)}_\alpha}{\cD^{(1)}_{\alpha v}}}
+\frac12\cdot \paren{\sum_{v\in\set{0,1}}q_v \cdot
\js{(1/2,1/2)}
{\cD^{(0)}_\beta}{\cD^{(1)}_{\beta v}}},
\end{align*}
where in the last line, we have used \Cref{fact:weighted-js-info}, as $\rtheta_i$ is uniform over $\set{0,1}$. These distributions should be conditioned on $\eventeq$, but by \Cref{clm:remove-alphabeta}, the distribution of $\rProt_i$ is independent of this event, conditioned on the input of $\player_i$, the transcript prefix $\Prot_{<i}$ and $\Rpublicn$.
Continuing, as JS-divergence is positive from \Cref{fact:kl-positive},
\begin{align*}
\textnormal{\textbf{LHS}} \geq \frac12\cdot \paren{\sum_{v\in\set{0,1}}q_v \cdot
\js{(1/2,1/2)}
{\cD^{(0)}_\alpha}{\cD^{(1)}_{\alpha v}}}
&\geq
\frac1{16} \cdot
\js{(1/2,1/2)}
{\cD^{(0)}_\alpha}{\cD^{(1)}_{\alpha u}}\tag{again, as JS-divergence is positive, \Cref{fact:kl-positive}, and $q_v \geq 1/8$}\\
&\geq
\frac1{64} \cdot \mi{\rProt_i}{\ralpha,\rbeta
\mid\rrho_i,\Rpublicn,\Prot_{<i}}, \tag{by \Cref{eq:theta-js-lower-bound}}
\end{align*}
proving the lemma.
\end{proof}

\subsection{Finishing the two-bit game lower bound}
We now make the last step towards the proof of \Cref{lem:two-bit-game}. \Cref{lem:thru-theta} only gives a lower bound on the information revealed about $\rtheta_i$ if the previous transcript has some randomness left to exploit in $\ralpha$ and $\rbeta$. However, this transcript has already revealed information about $\rtheta_{<i}$, and we argue that this is already $\Omega_{\epsilon}(1)$ information. 

We call a transcript prefix $\Prot_{<i}$ along with public randomness $\Rpublicn$ \textbf{good prefix} if the probability that $\alpha = u$ and $\beta = v$, conditioned on $\Prot_{<i}$ and $\Rpublicn$, is at least $0.1$ for all $u, v \in \set{0,1}$. That is, there is enough randomness left in $\alpha, \beta$ to apply \Cref{lem:thru-theta}.

We define the protocol $\protgood$ as follows. For any $i \in [k]$, as long as the current transcript prefix $\Prot_{<i}$ is good, the players send the same messages as in $\prot$. If the transcript prefix $\Prot_{<i}$ is not good, all the remaining players starting from $\player_i$ send $\bot$ or the null message containing no information. We use $\rProtgood$ to denote its transcript random variable. Notice that $\rProtgood$ can be obtained from $\rProt$ and the public randomness $\Rpublic$, as this reveals whether the partial transcript is  good or not.

We can prove that even though $\protgood$ is obtained by players sending dummy messages without completing $\prot$, it still reveals $\Omega_{\varepsilon}(1)$ information about $\alpha, \beta$.

\begin{claim}\label{clm:good-protocol-information}
For any $\eps < 1/2$, the protocol $\protgood$ reveals constant information about $\ralpha,\rbeta$. 
\[
\mi{\rProtgood}{\ralpha,\rbeta\mid\Rpublic}=\Omega_{\eps}(1).
\]
\end{claim}
\begin{proof}
Let $\eventgood$ be the event that $\prot$ reaches a transcript prefix which is not good, and let $\pgood:=\Pr[\eventgood ]$. This is also the probability with which some transcript in $\protgood$ is shortened by dummy messages. 

We begin by proving that the KL-divergence of $\ralpha$ and $\rbeta$ conditioned on a message of $\prot$ if $\eventgood$ happens is already $\Omega(1)$. 

Fix any transcript $\Prot$ of $\prot$ such that there is some prefix $\Prot_{<i}$ for $i \in [k]$ where $\delta_{uv} < 0.1$ for some $u, v \in \set{0,1}$. That is, the probability that $\alpha = u, \beta = v$ is at most $0.1$. 
Then, 
\begin{align*}
    \kl{(\ralpha,\rbeta \mid \Rpublicn,\Prot_{<i})}{(\ralpha,\rbeta)} &\geq 2 (\tvd{(\ralpha,\rbeta \mid \Rpublicn,\Prot_{<i})}{(\ralpha,\rbeta)})^2 \tag{by Pinsker's inequality \Cref{fact:pinsker-js-tvd}} \notag\\
    &\geq 2 (\frac12 \cdot (0.1-0.25)^2) \tag{by definition of TVD} \notag \\
    &= \Omega(1).
\end{align*}
Hence, for transcripts which are shortened in $\protgood$, even the prefix that is present (not shortened or replaced with dummy messages) already has:
\begin{equation}\label{eq:interim-3}
    \kl{(\ralpha,\rbeta \mid \Rpublicn,\Prot_{<i})}{(\ralpha,\rbeta)} = \Omega(1).
\end{equation}

 We can directly write the mutual information as,
\begin{align*}
\mi{\rProtgood}{\ralpha,\rbeta\mid\Rpublic}
&= \Exp_{\Protgood,\Rpublicn} \paren{\kl{(\ralpha,\rbeta\mid\Protgood,\Rpublicn)}
{(\ralpha,\rbeta)}} \tag{by \Cref{fact:KL-info}} \\
&\geq \Pr[\eventgood] \cdot \Exp_{\Protgood,\Rpublicn \mid \eventgood} \paren{\kl{(\ralpha,\rbeta\mid\Protgood,\Rpublicn)} 
{(\ralpha,\rbeta)}} \tag{as KL-divergence is non-negative, \Cref{fact:kl-positive}}\\
&\geq \Omega(1)\cdot \pgood. \tag{by \Cref{eq:interim-3}}
\end{align*}
 Therefore, if $\pgood = \Omega_{\epsilon}(1)$, we are already done. 
 
 Otherwise, we assume $\pgood<\eps/2$ to get a contradiction. In this case, we can prove that protocols $\protgood$ and $\prot$ are not too different from each other. 
Conditioned on $\neg \eventgood$, the transcripts of $\protgood$ and $\prot$ are the same. We can use the output of $\prot$ in case the messages are not truncated, and if truncated output a random bit in $\protgood$. By \tvdfacts{event-prob-tvd}, $\protgood$ has probability of success at least
 \[
 \Pr[\text{Success of $\prot$}] - \pgood \geq \frac12 + \frac{\eps}2.
 \]

 Using Fano's inequality \itfacts{fanos} on the transcript $\rProtgood$, \[
 \en{\ralpha \oplus \rbeta \mid \rProtgood, \Rpublic} \leq H_2(1/2+\eps/2).
 \]
 By definition of mutual information, 
\begin{align*}
\mi{\rProtgood}{\ralpha,\rbeta\mid\Rpublic}
&= \mi{\rProtgood}{\ralpha\oplus\rbeta, \beta\mid\Rpublic}\tag{ as $\ralpha \oplus \beta , \beta$ is fixed by $\ralpha, \rbeta$ and vice-versa}\\
&\geq \mi{\rProtgood}{\ralpha\oplus\rbeta\mid\Rpublic} \tag{as mutual information is non-negative, and by \itfacts{mi-chain}}
&= \en{\ralpha \oplus \rbeta \mid \Rpublic} - \en{\ralpha \oplus \rbeta \mid \rProtgood, \Rpublic} \\
&\geq 1-H_2(1/2+\eps/2)
=\Omega_{\eps}(1),
\end{align*}
which completes the proof. 
\end{proof}

We can prove \Cref{lem:two-bit-game} by focusing on $\protgood$.
\begin{proof}[Proof of \Cref{lem:two-bit-game}]
We start the proof by showing that the event $\eventeq$ only reduces the probability of a good prefix happening by a constant factor. 
 For any $i$, and any partial transcript $\Prot_{<i}$ along with public randomness $\Rpublicn$, 
\begin{align*}
    \Pr[\Prot_{<i}, \Rpublicn \mid \eventeq] &= \frac{\Pr[\Prot_{<i}, \Rpublicn \textnormal{ and } \eventeq]}{\Pr[\eventeq]} \tag{by Bayes' rule} \\
    &= \frac{\Pr[\eventeq \mid \Prot_{<i}, \Rpublicn] \cdot \Pr[\Prot_{<i}, \Rpublicn]}{\Pr[\eventeq]} \tag{again by Bayes' rule}\\
    &\geq \frac{0.2 \cdot \Pr[\Prot_{<i}, \Rpublicn]}{\Pr[\eventeq]}  \tag{as $\Prot_{<i}, \Rpublicn$ is a good prefix} \\
    &= 0.4 \cdot \Pr[\Prot_{<i}, \Rpublicn]. \tag{as $\ralpha, \rbeta$ are uniform random bits}
\end{align*}

We have concluded that, for any good prefix $\Prot_{<i}, \Rpublicn$,
\begin{equation}\label{eq:interim-pref}
     \Pr[\Prot_{<i}, \Rpublicn \mid \eventeq] \geq 0.4 \cdot \Pr[\Prot_{<i}, \Rpublicn].
\end{equation}

The next step of the proof is to analyse how much information about $\rtheta$ is revealed in protocol $\protgood$. We use a similar notation where $\rProtgood_i$ denotes the random variable for messages sent by $\player_i$ and $\rProtgood_{<i}$ denotes the random variable for the prefix. 
\begin{align*}
&\mi{\rProtgood}{\rtheta
\mid \rrho,\Rpublic,\ralpha,\eventeq}\notag\\ 
&\geq \sum_{i \in [k]}\mi{\rProtgood_i}{\rtheta_i \mid \rrho_i, \Rpublic, \ralpha, \rProtgood_{<i}, \eventeq} \tag{as \Cref{clm:2bit-separate} can be applied to $\protgood$}\\
&= \sum_{i \in [k]} \Exp_{\Protgood_{<i}, \Rpublicn \mid \eventeq}\mi{\rProtgood_i}{\rtheta_i \mid \rrho_i, \ralpha, \Protgood_{<i},\Rpublicn, \eventeq} \tag{by definition of conditional mutual information}\\
&\geq \frac1{64} \cdot \sum_{i\in[k]} \Exp_{\Protgood_{<i}, \Rpublicn \mid \eventeq} 
\mi{\rProtgood_i}{\ralpha,\rbeta
\mid \rrho_i,\Rpublicn,\Protgood_{<i}},
\end{align*}
where in the last inequality, we have applied \Cref{lem:thru-theta}. This is because, if $\Rpublicn, \Protgood_{<i}$ form a good prefix, the claim applies. If they do not, $\rProtgood_i$ does not reveal any information by construction. 

Continuing, using \Cref{eq:interim-pref}, we get, 
\begin{align*}
\mi{\rProtgood}{\rtheta
\mid \rrho,\Rpublic,\ralpha,\eventeq} 
&\geq \frac1{64} \cdot \sum_{i\in[k]} \Exp_{\Protgood_{<i}, \Rpublicn \mid \eventeq} 
\mi{\rProtgood_i}{\ralpha,\rbeta
\mid \rrho_i,\Rpublicn,\Protgood_{<i}},\\
&\geq \frac1{160} \cdot \sum_{i\in[k]} \Exp_{\Protgood_{<i}, \Rpublicn} 
\mi{\rProtgood_i}{\ralpha,\rbeta
\mid \rrho_i,\Rpublicn,\Protgood_{<i}},
\end{align*}
again, as any $\Protgood_{<i}, \Rpublicn$ pair that appears in the conditioning is a good prefix. For the prefixes that are not good, the mutual information term is zero. 
For the next step,
\begin{align*}
\mi{\rProtgood}{\rtheta
\mid \rrho,\Rpublic,\ralpha,\eventeq} 
&\geq \frac1{160} \cdot \sum_{i\in[k]} \Exp_{\Protgood_{<i}, \Rpublicn} 
\mi{\rProtgood_i}{\ralpha,\rbeta
\mid \rrho_i,\Rpublicn,\Protgood_{<i}},\\
&=  \frac1{160} \cdot \sum_{i\in[k]} 
\mi{\rProtgood_i}{\ralpha,\rbeta
\mid \rrho_i,\Rpublic,\rProtgood_{<i}} \tag{by definition of mutual information} \\
&\geq  \frac1{160} \cdot \sum_{i\in[k]} 
\mi{\rProtgood_i}{\ralpha,\rbeta
\mid \Rpublic,\rProtgood_{<i}} \tag{as $\rho_i \perp \ralpha, \rbeta \mid \Rpublic,\rProtgood_{<i}$ and by \itfacts{both-light-more}} \\
&= \frac1{160} \cdot \mi{\rProtgood}{\ralpha, \rbeta \mid \Rpublic} \tag{by chain rule of mutual information \itfacts{mi-chain}} \\
&= \Omega_{\eps}(1).  \tag{by \Cref{clm:good-protocol-information}}
\end{align*}
Now we lower bound the mutual information of the original transcript. The transcript $\rProtgood$ is fixed by $\rProt$ and public randomness $\Rpublic$, therefore, 
\begin{align*}
    \mi{\rProt}{\rtheta \mid \rrho, \Rpublic, \ralpha, \eventeq} \geq \mi{\rProtgood}{\rtheta
\mid \rrho,\Rpublic,\ralpha,\eventeq}  = \Omega_{\eps}(1). \qedhere
\end{align*}
\end{proof}

This proves \Cref{lem:two-bit-game} and concludes this section.

\section{Bounds based on Degeneracy}\label{sec:dgn-lb}

In this section, we prove the lower bound for any function parameterized by its degeneracy.

ChatGPT 5.6 Sol was used in identifying that the notion here was degeneracy in a back-and-forth conversation with the authors. The notion for symmetric functions from \Cref{def:sided-sens} was provided by the authors without using AI and ChatGPT identified the measure for general functions. The proof using degeneracy was done by the authors. 

The main result for this subsection follows.

\begin{lemma}\label{lem:dgn-lb-1}
For any function $f:\{0,1\}^n \rightarrow \{0,1\}$, any distribution over branching programs that computes $f$ with probability of success at least $2/3$, with width $2^S$ and length $T$ must satisfy:
\[
T \cdot S = \Omgt(n \cdot \dgn[f]).
\]
\end{lemma}

The work of \cite{Dinur24} uses a reduction from the Missing Element problem on the uniform distribution of inputs, whereas we choose a different distribution particular to the function $f$. We begin by stating the missing element problem, and the associated impossibility result. Then, we describe our input distribution specific to function $f$.

\subsection{Missing Element Problem}
We use the communication complexity lower bound of \cite{Dinur24} to prove our general lower bound based on the degeneracy. We state the lower bound here. 

\begin{definition}[Missing Element, $\MEproblem_{k,n} $ cf.~\cite{Dinur24}]\label{def:missing-element}
There are $k$ players each with input $A_i \subset [n]$ for $i \in [k]$. There is a bit $b \in \{0,1\}$ chosen uniformly at random, and an element $\jstar \in [n]$ chosen uniformly at random. 
\begin{itemize}
    \item When $b = 0$, $A_i$ does not contain $\jstar$ for all $i \in [k]$. For any $i \in [k]$ and $\ell \neq \jstar$, element $\ell$ is added to $A_i$ with probability $1/2$ independently. 
    \item When $b = 1$, each $\ell \in [n]$ is added with probability $1/2$ to $A_i$ independently for all $i \in [k]$.
\end{itemize}
The players are allowed to communicate on a blackboard that is visible to everyone. At the end of all the communication, they have to output the value of $b$. 
\end{definition}
The following theorem was proved by \cite{Dinur24} for $\MEproblem_{k,n}$. 

\begin{proposition}[cf.~\cite{Dinur24}]\label{thm:miss-element-lb}
    Let $\zeta = \zeta(n)$ such that $0 < \zeta(n) < 1/2$ for sufficiently large $n$. Assume that there is a public-coin protocol $\prot$ for $\MEproblem_{k, n}$ such that for all large $n$, 
    \begin{align*}
       \max_{j \in [n]} \Pr\paren{\prot(A) \neq 0 \mid \rb = 0, \rjstar = j} &\leq 1-\zeta \\
        \max_{j \in [n]} \Pr\paren{\prot(A) \neq 1 \mid \rb = 1, \rjstar = j} &\leq \zeta/200. 
    \end{align*}
    Then the total communication sent by protocol $\prot$ is $\zeta \cdot \Omega(n)$. 
\end{proposition}

We begin by defining a hard distribution over the inputs, and this hard distribution crucially depends on degeneracy. Then, we show that a protocol that solves the input over this hard distribution must require a lot of space via a reduction from the lower bound on missing element.

\subsection{A Hard Distribution}

Let $G$ be the sensitivity graph of $f$ (recall from \Cref{def:sens-graph}). This is a bipartite graph $G_f = (V_1 \cup V_0, E_f)$ with:
    \begin{align*}
    V_0 &= \set{x \in \set{0,1}^n \mid f(x) = 0} \qquad \text{(this is the set of inputs with $f(x) = 0$)} \\
    V_1 &= \set{x \in \set{0,1}^n \mid f(x) = 1} \qquad \text{(this is the set of inputs with $f(x) = 1$)} \\
    E_f &= \set{(x,y) \mid x \in V_0, y \in V_1, \text{$x$ and $y$ differ in exactly one bit}}.
    \end{align*}

The degeneracy of $G_f$ is the minimum value $\ell$ such that every subgraph of $G$ contains at least one vertex with degree at most $\ell$. Let $H$ be a subgraph of $G$ that maximizes minimum degree. 
\begin{tbox}
	\textbf{Distribution $\cD_f$ over vertices of $H$:}
	\begin{enumerate}[label=$(\roman*)$]
		\item Choose an edge $e = (u, v) \in H$ uniformly at random from all the edges of $E(H)$.
		\item Choose one of the end-points of $e$ uniformly at random.
	\end{enumerate}
\end{tbox}

Distribution $\cD_f$ is the well-known stationary distribution of a random walk on the vertices of $H$. We prove some simple properties about this distribution.

\commented{
\begin{claim}\label{clm:miss-coordinate-error}
For any $x \sim \cD_f$, and index $\eta$ chosen uniformly at random, any algorithm (deterministic or randomized), given access only to $\eta$ and $x(-\eta)$ that outputs $f(x)$ must have error at least $\ell/2n$. 
\end{claim}
\begin{proof}
 Fix an edge \(\{u,v\}\in E(H)\) whose endpoints differ at coordinate \(i\). When \(\eta=i\), the algorithm receives identical observations on \(u\) and \(v\). Since \(f(u)\ne f(v)\), its error probabilities on these inputs sum to one. Their weighted contribution to the error is therefore at least \(\min\{D_f(u),D_f(v)\}\).

For each coordinate, the corresponding edges form a matching. Summing these contributions and averaging over \(\eta\) gives

$$
\begin{aligned}
\Pr[A(\eta,X(-\eta))\ne f(X)]
&\ge \frac1n\sum_{\{u,v\}\in E(H)}
       \min\{D_f(u),D_f(v)\}\\
&=\frac{1}{2nm}\sum_{\{u,v\}\in E(H)}
       \min\{\deg_H(u),\deg_H(v)\}\\
&\ge \frac{d}{2n}.
\end{aligned}
$$

\(\square\)
\end{proof}}

\begin{observation}\label{obs:out-random}
	For any $x \in \{0,1\}^n$ sampled according to distribution $\cD_f$, $f(x)$ is uniform over $\set{0,1}$.
\end{observation}
\begin{proof}
	Irrespective of what edge is sampled in step $(i)$ of $\cD_f$, the end points of the edge have different outputs in $f$. Thus, the output of $f$ on $x \sim \cD_f$ is uniformly random. 
\end{proof}

For any index $\eta \in [n]$, let $\Inp(\eta) \subset \{0,1\}^{n-1}$ be the set of possible values of $x(-\eta)$ \footnote{For any $i \in [n]$, $x(-i)$ is the $(n-1)$-length string containing the $(n-1)$ values of $x$ at every position in $[n] \setminus \set{i}$.} such that there is an edge in $H$ where both end points have values $x(-\eta)$ on the indices in $[n] \setminus \set{\eta}$. That is, $\Inp(\eta)$ is the set of all $x(-\eta)$ such that, there exist vertices $x, x'$ in $H$ satisfying,
\[
f(x) \neq f(x'), \qquad \textnormal{where $x$ (resp. $x'$) has $0$ (resp. $1$) at index $\eta$ and $x(-\eta)$ everywhere else.}
\]
\begin{observation}\label{obs:output-j}
	In distribution $\cD_f$, we have, 
	\[
	\sum_{\eta \in [n]} \card{\Inp(\eta)} = \card{E(H)}.
	\]
\end{observation}
\begin{proof}
	For each $\eta \in [n]$, $\card{\Inp(\eta)}$ is the number of edges in $H$ whose endpoints differ at coordinate $\eta$, as each such edge has a unique $x(-\eta)$. These edge classes form a disjoint partition of $E(H)$, which gives the result.
\end{proof}

\begin{observation}\label{obs:vert-deg}
	Any vertex $u \in \cD_f$ is sampled with probability $(1/2) \cdot \deg(u)/\card{E(H)}$.
\end{observation}
\begin{proof}
	There are $\deg(u)$ many edges that can give the output as $u$ among $\card{E(H)}$ many edges, and for each edge, the end point $u$ is chosen with probability half.
\end{proof}

Our aim is to solve the Missing Element problem using an algorithm that receives random samples and solves $f$. If such an algorithm \emph{does not} get random samples from the input, and instead never gets some index $\eta \in [n]$, the value of the input $x$ at index $\eta$ is unknown to the algorithm, conditioned on the rest of the input $x(-\eta)$. For such an $\eta$ and $x(-\eta)$, any output of the algorithm has to be incorrect with some probability. We analyse this probability in the next claim. 

\begin{claim}\label{clm:error-prob-alpha}
	For any $\eta \in [n]$ and $z \in \Inp(\eta)$, let $u, v$ be the two vertices in the support of $\cD_f$ whose restrictions to $[n]\setminus\set{\eta}$ equal $z$. Then, any algorithm that outputs $f(x)$ based only on $x(-\eta)=z$ and the index $\eta$, for an input $x \sim \cD_f$, must make error at least:
	\[
	\frac{\min\set{\deg(u), \deg(v)}}{\deg(u) + \deg(v)}.
	\]
\end{claim}
\begin{proof}
	The algorithm can either output $0$ or $1$ or a mixture of both. We prove that whatever the output is, the error is at least the quantity in the statement. Without loss of generality, let $u$ be the vertex with $f(x) = 0$.
	
	For any $x \sim \cD_f$,
	\begin{align*}
		\Pr[f(x) = 0 \mid x(-\eta) = z] &= \frac{\Pr[f(x) = 0 \text{ and } x(-\eta) = z]}{\Pr[x(-\eta) = z]} \tag{by Bayes rule} \\
		&= \frac{\Pr[\cD_f = u]}{\Pr[\cD_f = u] + \Pr[\cD_f = v]} \tag{as these are the only inputs with $x(-\eta)=z$}\\
		&= \frac{\deg(u)}{\deg(u) + \deg(v)}. \tag{by \Cref{obs:vert-deg}} 
	\end{align*}
	Similarly, we have, 
	\begin{align*}
		\Pr[f(x) = 1 \mid x(-\eta) = z] = \frac{\deg(v)}{\deg(u) + \deg(v)}.
	\end{align*}
	Hence, whatever value the algorithm outputs, the error is at least:
	\[
	\min\set{\deg(u), \deg(v)} \cdot \frac1{\deg(u) + \deg(v)}. \qedhere
	\]
\end{proof}

We have understood the error for a fixed value of $x(-\eta)$ and index $\eta$. We use \begin{equation}\label{eq:err-alpha}
\err(x, \eta):= \begin{cases}  \frac{\min\set{\deg(u), \deg(v)}}{\deg(u) + \deg(v)} &\text{ if $x(-\eta) \in \Inp(\eta)$ }, \\
0 &\text{ otherwise.}
\end{cases}
\end{equation}
where $u, v$ are the two vertices with $x(-\eta)$ as the input in indices $[n]\setminus \set{\eta}$, if they exist.  Now, we can analyze the error based on the randomness of input $x$ sampled from $\cD_f$ and a random missing index. 

\begin{claim}\label{clm:output-min-deg}
	Let $\eta \in [n]$ be chosen uniformly at random, and $x \sim \cD_f$ independently of $\eta$. Then,
	\[
	\Exp_{\eta} \Exp_{x \sim \cD_f}  \err(x, \eta) \geq \frac1{2n} \cdot \dgn[f].
	\]
\end{claim}
\begin{proof}
	For any vertex $x$, we use $N_H(x)$ to denote the neighborhood of $x$ in $H$.
	To continue, we have:
	\begin{align*}
		&\Exp_{\eta} \Exp_{x \sim \cD_f} \ind[x(-\eta) \in \Inp(\eta)] \cdot \err(x, \eta) \\
		&= \Exp_{x \sim \cD_f} \frac1n \cdot \sum_{\eta \in [n]} \ind[x(-\eta) \in \Inp(\eta)] \cdot \err(x, \eta) \\
		&= \Exp_{x \sim \cD_f} \frac1n \cdot \sum_{y \in N_H(x)} \min\set{\deg(x), \deg(y)} \cdot \frac1{\deg(x)+\deg(y)},
	\end{align*}
	where, in the last line, we have used that $\ind[x(-\eta) \in \Inp(\eta)]$ is $1$ for all $\eta \in [n]$ where there is an edge leading out of $x$ with the value at index $\eta$ flipped. Hence, it is sufficient to iterate over the neighbors of $x$. We have also used the definition of $\err(x, \eta)$. Next, using the definition of degeneracy, we get,
	\begin{align*}
		&\Exp_{x \sim \cD_f} \frac1n \cdot \sum_{y \in N_H(x)} \min\set{\deg(x), \deg(y)} \cdot \frac1{\deg(x)+\deg(y)} \\
		&\geq \dgn[f] \cdot \frac1n \cdot \Exp_{x \sim \cD_f} \sum_{y \in N_H(x)} \frac1{\deg(x)+\deg(y)} \\
		&=\frac{\dgn[f]}{2n} \cdot \frac1{\card{E(H)}}\sum_{x} \deg(x) \sum_{y \in N_H(x)} \frac1{\deg(x)+\deg(y)} \tag{by \Cref{obs:vert-deg}} \\
		&= \frac{\dgn[f]}{2n} \cdot \frac1{\card{E(H)}} \sum_{e=(x,y)} \frac{\deg(x) + \deg(y)}{\deg(x)+\deg(y)} \\
		&=  \frac{\dgn[f]}{2n},
	\end{align*}
	which proves the claim.
\end{proof}

We are ready to prove the main theorem in the next subsection. 

\subsection{Proof of \Cref{thm:dgn-lb}}

In this subsection, we prove \Cref{thm:dgn-lb} via a reduction from \Cref{thm:miss-element-lb}. This part is largely similar to the reduction in \cite{Dinur24}.

We run the algorithm that solves $f:\set{0,1}^n$ to solve the \MEproblem$_{k,n}$.
First, we show how the players of $\MEproblem$ can generate a uniformly random stream of $\nprime$ indices given $\nprime$ elements of $[n]$ for any $\nprime \leq n$.
\begin{claim}\label{clm:stochastic-queries}
    For any $1 \leq \nprime \leq n$, given $\nprime$ distinct elements from $[n]$ chosen uniformly at random, there exists a way to generate a uniformly random stream of $\nprime$ indices from $[n]$.
\end{claim}
\begin{proof}
    Let $W$ be the set of $\nprime$ elements chosen uniformly at random from $[n]$ from which we want to generate the $\nprime$ indices.
    
   We use $W'$ to keep track of the indices that have been generated so far. Initially $W'$ is the empty set.
    To generate index $i$ for $i \geq 1$, we do:
    \begin{itemize}
        \item With probability $\card{W'}/n$, choose a uniformly random element from set $W'$.
        \item Otherwise, choose a uniformly random element from $W \setminus W'$, and add this element to $W'$.
    \end{itemize}
    
    We argue that in this process, for each index $i$, every element from $[n]$ is equally likely to be chosen irrespective of the other indices. For the elements in $W'$, each of them is chosen with probability 
    \[
        \underset{\substack{\text{probability of the }\\\text{ element being from $W'$}}}{\underbrace{\card{W'} \cdot \frac1n}} \cdot \underset{\substack{\text{uniformly random}\\\text{ element from $W'$}}}{\underbrace{\frac{1}{\card{W'}}}} = \frac1n.
    \]
    For elements outside of $W'$, we have:
     \[
        \underset{\substack{\text{probability of the }\\\text{ element not being from $W'$}}}{\underbrace{(n-\card{W'}) \cdot \frac1n}} \cdot \underset{\substack{\text{uniformly random}\\\text{ element from $[n]\setminus W'$}}}{\underbrace{\frac{1}{n-\card{W'}}}} = \frac1n,
    \]
where in the second part we have used that $W$ is a uniformly random subset. The fresh element chosen out of $W'$ will be uniformly random from $[n] \setminus W'$.

Irrespective of all the previous indices in the sequence, the next index remains uniformly random, as our analysis works for any choice of subset $W'$. Therefore, our process generates a uniformly random sequence from $[n]$.
\end{proof}

We are ready to describe the final protocol that players in $\MEproblem$ use.  We assume that the players have access to an algorithm $\alg$ that succeeds with probability at least $1-\delta$ over the randomness of choosing the branching program and the distribution of the input indices.

\begin{tbox}
    \textbf{Protocol $\protME$ for $\MEproblem_{k,n}$ given $\cDalg_n$ on branching programs with length $T = \floor{nk/100}$ and width $2^S$ that solves $f$:}
    \begin{enumerate}[label=$(\roman*)$]
        \item \label{item:branch} The players use public randomness to sample a branching program  from $\cDalg$. They also use public randomness to sample a random $z \sim \cD_f$. 
     \item Each player checks if their input $\card{A_i} \geq n/100$. If not, the protocol fails. 
\item Using public randomness, a permutation of $[n]$, labeled $\sigma:[n] \rightarrow [n]$ is chosen uniformly at random. Each player relabels their input according to $\sigma$, i.e., for each $i \in [k]$,
\[
    A_i' = \set{\sigma(a) \mid a \in A_i}.
\]
\item \label{item:indices} Each player generates $\nprime = n/100$ uniformly random elements of $[n]$ using the sets $A'_i$ for $i \in [k]$ (see \Cref{clm:stochastic-queries}). Hence, they also have $\nprime$ samples of the form $(j, z(j))$ for $j \in [n]$ uniformly at random as $z$ is sampled with public randomness in step \ref{item:branch}.
 \item All the players collectively run the branching program sampled in step \ref{item:branch} on input $z$. Player $i \in [k]$ runs the branching program from $(i-1)\nprime + 1$ to $i \cdot \nprime$ layers with the $\nprime$ samples they generate in step \ref{item:indices}. Player $i$  writes the identity of the node of the branching program on the blackboard after layer $i \cdot \nprime$.
\item Player $k$ looks at the node of the branching program after all the queries, finds the output $\alg(z)$. If $\alg(z)$ is equal to $f(z)$, they output $b = 1$ and otherwise they output $b = 0$.
    \end{enumerate}
\end{tbox}

We prove some simple properties about this protocol. Let $\eventbad$ be the event that there is some $i \in [k]$ with $\card{A_i} < n/100$. 

\begin{claim}\label{clm:size-input}
The probability that $\eventbad$ happens is at most $k \cdot e^{-n/10}$.
\end{claim}
\begin{proof}
    In $\MEproblem_{k, n}$, when $b = 0$, there are $n-1$ elements, and each of them is added to $A_i$ with probability $1/2$ independently. When $b = 1$, there are $n$ elements and they are chosen to be added to $A_i$ with probability $1/2$ as well. In either case, we have, 
    \[
    \Exp[\card{A_i}] \geq (n-1)/2,
    \]
    by the linearity of expectation. We use \Cref{prop:chernoff} for independent random variables. 
    \begin{align*}
        \Pr[\card{A_i} \leq n/100] \leq \exp(-(\frac{24}{25})^2 \cdot (n/4) \cdot \frac12) \leq \exp(-n/10).
    \end{align*}
    A union bound over all $i \in [k]$ proves the claim.
\end{proof}

Hence, $\eventbad$ is a very low probability event for small values of $k$.
Next, we analyse the probability that $\protME$ succeeds based on the value of $b$.

\begin{claim}\label{clm:suc-b-one}
   When $b = 1$, protocol $\protME$ outputs the right answer with  probability at least $1-\delta-k\cdot \exp(-n/10)$.
\end{claim}

\begin{proof}
    We condition on $\eventbad$ not happening. Now, the players of $\MEproblem_{k,n}$ run $\alg$ on the correct input distribution for $k \cdot n/100  = T$ samples accurately. This is because, conditioned on $b = 1$, the sets given to all the players of $\MEproblem_{k,n}$ are uniformly random. By \Cref{clm:stochastic-queries}, the input distribution given to $\alg$ is uniformly random over indices of $[n]$.  

    The output of $\alg$ is incorrect with probability at most $\delta$, and event $\eventbad$ happens with probability at most $k \cdot \exp(-n/10)$ by \Cref{clm:size-input}. Hence, barring the output being incorrect and event $\eventbad$, the algorithm succeeds and $\protME$ outputs $b = 1$ with probability at least $1-\delta - k \cdot \exp(-n/10)$. 
\end{proof}
We perform a similar analysis when $b = 0$.

\begin{claim}\label{clm:suc-bzero}
When $b = 0$, protocol $\protME$ outputs that $b = 1$ with probability at most $k \cdot \exp(-n/10) + 1-\dgn[f]/2n$. 
\end{claim}
\begin{proof}
    
When $b = 0$ for $\MEproblem_{k,n}$, the stream given to $\alg$ is not uniformly random over $[n]$ because the element $\sigma(\rjstar)$ never appears as the index. Let $b'$ be the bit output by $\alg$ in this case. Note that $\sigma(\jstar)$ is chosen uniformly at random from $[n]$ as $\sigma$ was a uniformly random permutation in protocol $\protME$. Algorithm does not get access to the value of $z(\sigma(\rjstar))$, and hence $\alg(z)$ remains independent of $z(\sigma(\rjstar))$, when conditioned on the value of $z(-\sigma(\rjstar))$.

When outputting $b'$, algorithm only has access to $z(-\sigma(\rjstar))$. This algorithm must make an error if there is some edge leading out of $z$ that is labeled by $\sigma(\rjstar)$, since $z \oplus e_{\sigma(\rjstar)}$ has a different output compared to $z$. The algorithm cannot distinguish  between these two inputs. 

The quantity $\err(z, \sigma(\rjstar))$ from \Cref{eq:err-alpha} quantifies this error for our needs by \Cref{clm:error-prob-alpha}. Therefore,
\begin{equation}\label{eq:err-dgn-lb}
    \Pr[\alg(z) \neq f(z) \mid \neg \eventbad, b=0] \geq \Exp_{\sigma(\rjstar)} \Exp_{z \sim \cD_f} \err(z, \sigma(\rjstar)) 
    \geq \frac{\dgn[f]}{2n},
\end{equation}
where the last inequality follows by \Cref{clm:output-min-deg}.
We are ready to upper bound the error of $\protME$.
\begin{align*}
    \Pr[\text{Protocol }\protME\text{ outputs $b = 1$} \mid b = 0] 
    &\leq \Pr[\eventbad] + \Pr[\alg(z)=f(z) \mid \neg \eventbad, b=0] \\
    &\leq k \cdot e^{-n/10} + 1-\frac{\dgn[f]}{2n}, \tag{by \Cref{clm:size-input} and \Cref{eq:err-dgn-lb}}
\end{align*}
which finishes this proof. 
\end{proof}

The rest of the proof of \Cref{thm:dgn-lb} is fairly straightforward, similar to the proof of Theorem 2 in \cite{Dinur24}. We first show how to amplify success probability of branching programs. 

\begin{claim}\label{clm:bp-amplify}
    For any $0< \delta <1$, given a randomized branching program $P$ with length $T$ and width $2^S$ and error probability at most $\epsilon < 1/2$ solving function $f$, there is another branching program $P'$ of length $O(T \log(1/\delta))$ and width $2^S \cdot O(\log (1/\delta))$ that computes $f$ with error at most $\delta$.
\end{claim}
\begin{proof}
In $P'$, we can run $\ell = O(\log(1/\delta))$ copies of $P$ and output the majority value. In addition to running the copies, we also have to keep track of the number of copies, and the number of ones seen so far, which incurs an additional $O(\log(1/\delta))$ factor in the width. 

As the copies are independent, the probability of success follows directly from \Cref{prop:chernoff}.
\end{proof}

We are ready to give the proof of the lower bound in \Cref{thm:dgn-lb}.
 
\begin{proof}[Proof of \Cref{lem:dgn-lb-1}]
We use \Cref{clm:bp-amplify} on the branching program from the statement to get another branching program $P'$ that has length $T' = T \cdot O(\log(1/\delta))$ and width $2^{S'}$ with $S' = S + O(\log \log(1/\delta))$, and error probability at most $\delta = \dgn[f]/500n$.

We want to use the impossibility results of  missing element problem with $k$ people where $k  = \ceil{100T'/n}$. ($T' = \Omega(n)$, because otherwise, the function uses only $o(n)$ samples, and can only compute constant functions.)
Observe that $k = O(n)$, as otherwise $T' = \Omgt(n^2)$ and we are done. 

We use protocol $\protME$ to solve this problem using branching program $P'$. By \Cref{clm:suc-b-one} and \Cref{clm:suc-bzero} when $b=1$ and $b = 0$ respectively, the error of the protocol is at most :
\begin{align*}
\Pr\paren{\prot(A) \neq 0 \mid \rb = 0, \rjstar = j} &\leq k \cdot \exp(-n/10) + 1-\dgn[f]/2n \\
      \Pr\paren{\prot(A) \neq 1 \mid \rb = 1, \rjstar = j} &\leq \dgn[f]/500n + k \cdot \exp(-n/10). 
    \end{align*}

Let $\zeta$ be a parameter related to the error probability:
\[
\zeta=\frac{\dgn[f]}{2n}-k \cdot \exp(-n/10).
\]
For sufficiently large $n$, we have $0 < \zeta < 1/2$ and
\begin{align*}
\Pr\paren{\prot(A) \neq 0 \mid \rb = 0, \rjstar = j} &\leq k \cdot \exp(-n/10) + 1-\dgn[f]/2n = 1-\zeta \\
      \Pr\paren{\prot(A) \neq 1 \mid \rb = 1, \rjstar = j} &\leq \dgn[f]/500n + k \cdot \exp(-n/10) \leq \zeta/200.
    \end{align*}

Thus, using \Cref{thm:miss-element-lb}, the communication of $\protME$ is at least
\[
\zeta\cdot\Omega(n)=\Omega(\dgn[f]).
\]
We also have an upper bound on the communication of at most,
\[
(k-1)S'+1=O\paren{\frac{T'S'}{n}} = O(TS) \cdot (1/n) \cdot (\log(500n/\dgn[f])) \cdot O(\log \log (500n/\dgn[f])),
\]
as each of the first $(k-1)$ players writes an $S'$-bit state of the branching program, and the last player writes one bit corresponding to the answer. 
Combining the two bounds:
\[
TS=\Omega(n\cdot \dgn[f]) \cdot \frac{1}{\log^2(n)},
\]
proving the statement.
\end{proof}


\subsection{Algorithm for Symmetric Functions}

In this subsection, we give an algorithm for symmetric functions showing that the lower bound in \Cref{thm:dgn-lb} is tight up to logarithmic factors. More specifically, for every symmetric function $f$ and space parameter $w$, our algorithm uses $S = \Ot(w)$ bits, and the width of the corresponding branching program is $2^S = 2^{\widetilde{O}(w)}$. The length is at most
\[
    \widetilde{O}\paren{\frac{n\cdot \dgn[f]}{w}}.
\]
We begin with a simpler characterization of degeneracy for symmetric functions and then give the algorithm.


A Boolean function $f:\{0,1\}^n\rightarrow\{0,1\}$ is symmetric if $f(x)=f(y)$ whenever $\card{x}=\card{y}$. For each $0\leq k\leq n$, let $f_k$ denote the value of $f$ on input $x$ with $\card{x} = k$. We describe the complexity of $f$ by the locations where the sequence $f_0,f_1,\ldots,f_n$ changes value, similar to the characterization of approximate degree by \cite{Paturi92}.

\begin{definition}\label{def:sided-sens}
For a symmetric function $f$, define
\begin{align*}
    \sleft[f]
    &:= \max\left(\set{k+1:0\leq k<n/2 \text{ and } f_k\neq f_{k+1}}\cup\set{0}\right),\\
    \sright[f]
    &:= \max\left(\set{n-k:n/2\leq k<n \text{ and } f_k\neq f_{k+1}}\cup\set{0}\right).
\end{align*}
Thus, $\sleft[f]$ measures the last change in $f$ when moving from the all-zero input towards the middle layers, and $\sright[f]$ is the analogous quantity when moving from the all-one input towards the middle layers.

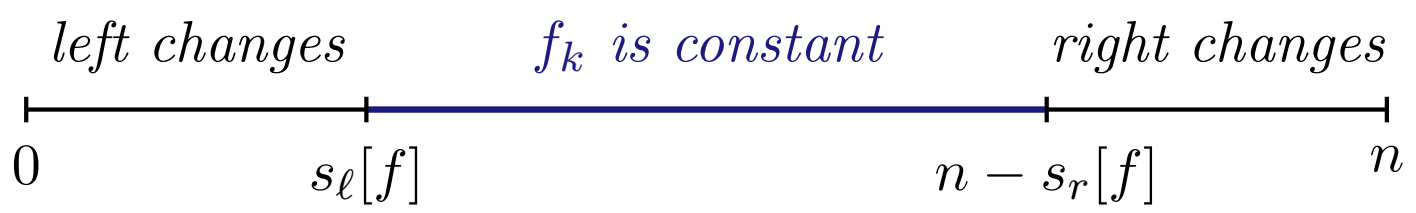
\begin{center}
\begin{tikzpicture}[x=0.9cm,y=0.65cm]
    \draw[thick] (0,0) -- (10,0);
    \draw[very thick,DarkBlue] (2.5,0) -- (7.5,0);
    \foreach \x/\label in {
        0/{0},
        2.5/{\sleft[f]},
        7.5/{n-\sright[f]},
        10/{n}
    } {
        \draw[thick] (\x,-0.13) -- (\x,0.13);
        \node[below=3pt] at (\x,0) {$\label$};
    }
    \node[above=3pt,DarkBlue] at (5,0) {$f_k$ is constant};
    \node[above=3pt] at (1.25,0) {left changes};
    \node[above=3pt] at (8.75,0) {right changes};
\end{tikzpicture}
\end{center}

\end{definition}

The following claim relates these two quantities to degeneracy and shows that the upper bound proved in this section is $\widetilde{O}(n\cdot\dgn[f])$.

\begin{claim}\label{clm:symmetric-degeneracy}
For every symmetric Boolean function $f$,
$\sleft[f]+\sright[f]=\Theta(\dgn[f])$.
\end{claim}

\begin{proof}
The edges in the sensitivity graph $G_f$ between the inputs of Hamming weight $k$ and $k+1$ are present exactly when $f_k\neq f_{k+1}$. 
For any such $k$, we consider the subgraph induced by the inputs of Hamming weight both $k$ and $k+1$.
Vertices with Hamming weight $k$ have degree $n-k$, and vertices with Hamming weight $k+1$ have degree $k+1$. The minimum degree of this subgraph is $\min\set{k+1,n-k}$, and $\dgn[f]$ is at least this value for all $k$ by definition. 

We consider two cases:
\begin{itemize}
    \item If $k < n/2$, the minimum of $\set{k+1, n-k}$ is $k+1$ and the largest such $k+1$ equals $\sleft[f]$.
    \item If $k \geq n/2$, the minimum of $\set{k+1, n-k}$ is $n-k$. For the smallest such $k$, $\sright[f] = n-k$. 
\end{itemize}

In either case, we have: $\dgn[f] \geq \max\set{\sleft[f], \sright[f]} = \Omega(\sleft[f] + \sright[f])$.

Now we prove that $\dgn[f]$ is upper bounded by $O(\max\set{\sleft[f], \sright[f]})$.

Let $H_f$ be a subgraph of $G_f$ that has maximum minimum degree. Let $D$ be the set of $i \in ([n] \cup \set{0})$ such that at least one input of $\card{x} = i$ is present in $H_f$. We look at a continuous sequence of values that occur in $D$ such that the value of function $f$ alternates across these indices. Let $p$ be the least value of this sequence, followed by $p+1, p+2, \ldots, q$ where $q$ is the largest value. We know:
\begin{itemize}
    \item Vertex with Hamming weight $p$ has degree at most $n-p$, as it only has edges to vertices with higher Hamming weight. We also know if $p \geq n/2$, this value is at most $\sright[f]$.
    \item Vertex with Hamming weight $q$ has degree at most $q$, as it has edges to vertices with Hamming weight $q-1$. If $q \leq n/2$, this degree is at most $\sleft[f]$. 
\end{itemize}
So if either $q \leq n/2$ or $p \geq n/2$, we are done, as the degree is at most $\sleft[f] + \sright[f]$. Otherwise we have $p < n/2 < q$, which implies $\sleft[f] \geq n/2-2$ as the value of $f$ alternates across indices $n/2-1$ and $n/2$. The function $f$ in this case has $\sleft[f] + \sright[f] = \Omega(n) = \Omega(\dgn[f])$, which concludes the proof. 
\end{proof}

The definition also immediately gives the following useful property.

\begin{observation}\label{clm:value-f-fixed}
The function value $f_k$ is the same for every $\sleft[f]\leq k\leq n-\sright[f]$.
\end{observation}

\begin{proof}
Assume towards a contradiction that there is some $\ell$ with $\sleft[f] \leq \ell < n-\sright[f]$ such that $f_{\ell} \neq f_{\ell+1}$. 
 If $\ell<n/2$, then by definition $\sleft[f]\geq \ell+1$. If $\ell\geq n/2$, then by definition of $\sright[f]$, we have $\sright[f]\geq n-\ell$. In both cases we get a contradiction. 
\end{proof}

We are ready to give the algorithm, formalized in the following lemma. 

\begin{lemma}\label{lem:symmetric-ub}
For every symmetric Boolean function $f$ and for every $w$ with $1\leq w\leq \sleft[f]+\sright[f]$, there is an algorithm that computes $f$ with high probability using $S = \widetilde{O}(w)$ space and number of queries at most
\[
    \widetilde{O}\paren{\frac{n}{w}\cdot \left(\sleft[f]+\sright[f]\right)}
    =\widetilde{O}\paren{\frac{n\cdot\dgn[f]}{w}}.
\]
\end{lemma}

It is sufficient to accurately determine one of three cases to compute the value of $f(x)$:
 \begin{itemize}
     \item If $\card{x} \leq \sleft[f]$, we want to find $\card{x}$ exactly. 
     \item If $\card{x} \geq n-\sright[f]$, we also want to find $\card{x}$ exactly.
     \item Otherwise, we know the output $f(x)$ by \Cref{clm:value-f-fixed}.
 \end{itemize}
We first give an algorithm that exactly counts the number of ones as long as it is at most $\sleft[f]$.
 
 We assume that the algorithm has access to a separate tape of randomness that is not counted as part of its memory.\footnote{This corresponds to picking a random branching program among a set of deterministic branching programs.}

\begin{tbox}
\textbf{Algorithm $\textsc{CountOnes}$:} Output the number of ones in $x$ as long as it is at most $\sleft[f]$.
\begin{enumerate}[label=$(\roman*)$]
    \item Pick a uniformly random permutation $\sigma$ of $[n]$ using the separate tape of random bits.
    \item Let $g=\max\set{1,\ceil{\sleft[f]/w}}$ and divide $[n]$ into $g$ groups of size at most $\ceil{n/g}$ according to $\sigma$. Initialize a counter $c=0$.
    \item Process the groups sequentially. For each group, get $100n\log n$ input samples and maintain a list $O$ of the distinct indices $i \in [n]$ where $i$ belongs to this group and $x(i) = 1$.
    \item At any point, if $\card{O}>10w\log n$, stop and declare that $\card{x}>\sleft[f]$. Otherwise, add $\card{O}$ to $c$ and continue to the next group.
    \item Output $c$ after all groups have been processed.
\end{enumerate}
\end{tbox}

We prove that this algorithm works as intended.

\begin{claim}\label{clm:num-ones-less}
For every input $x$, with high probability, either:
\begin{itemize}
\item $\textsc{CountOnes}$ declares $\card{x} > \sleft[f]$ and the actual input $x$ has more than $\sleft[f]$ ones, or, 
\item \textsc{CountOnes} outputs $\card{x}$ exactly.
\end{itemize} 
The algorithm uses $\widetilde{O}(w)$ space and $\widetilde{O}(n \cdot\max\set{1,\ceil{\sleft[f]/w}})$ queries totally.
\end{claim}

\begin{proof}
The space used by the algorithm is $O(w \log^2 n)$, to maintain $O(w \log n)$ indices in $O$, and each index takes $O(\log n)$ bits. The number of input samples is
\[
    100n\log n\cdot\max\set{1,\ceil{\sleft[f]/w}}
    =\widetilde{O}\paren{n\max\set{1,\ceil{\sleft[f]/w}}}.
\]
By the coupon collector bound, all indices in a fixed group are seen among the input samples allotted for this group with probability at least $1-1/\poly(n)$. It remains to show that the list $O$ does not overflow.

Fix a group of indices $J$. Since $J$ is a uniformly random set of size at most $\ceil{n/g}$, the expected number of ones in $J$ is at most
\[
     \card{x}\cdot\frac{\ceil{n/g}}{n}=O(w),
\]
by the linearity of expectation. The random variables that make up this expectation are negatively correlated.
By \Cref{prop:chernoff},
\[
    \Pr\paren{\text{The number of ones exceeds }10w\log n}\leq \frac{1}{\poly(n)}.
\]
A union bound over at most $n$ groups shows that with high probability, every list has size at most $10w\log n$ and every index in every group is queried. Conditioned on these events, the algorithm counts the total number of ones exactly.
\end{proof}

By similarly dividing into random groups and instead keeping track of the distinct number of zeroes, we get an algorithm $\textsc{CountZeroes}$ that, for every input with at most $\sright[f]$ zeroes, counts the number of zeroes exactly with high probability using $\widetilde{O}(w)$ space and $\widetilde{O}(n\max\set{1,\ceil{\sright[f]/w}})$ queries also. 
We run these two algorithms sequentially. The proof of \Cref{lem:symmetric-ub} is now trivial.

\begin{proof}[Proof of \Cref{lem:symmetric-ub}]
If $\sleft[f]+\sright[f]=0$, then function $f$ is constant and there is nothing to prove. Otherwise, we run $\textsc{CountOnes}$ first and then run $\textsc{CountZeroes}$, reusing the same memory.

If either algorithm returns an exact count, we compute $f(x)$ directly from this count. Otherwise, $\textsc{CountOnes}$ certifies that $\card{x}>\sleft[f]$, and $\textsc{CountZeroes}$ certifies that $n-\card{x}>\sright[f]$. Hence,
\[
    \sleft[f]<\card{x}<n-\sright[f].
\]
By \Cref{clm:value-f-fixed}, the function is constant in this range, and we output this constant.

The two algorithms use $\widetilde{O}(w)$ space in total, since they are run sequentially, and their total number of queries is
\begin{align*}  \widetilde{O}\paren{n\max\set{1,\ceil{\sleft[f]/w}}+n\max\set{1,\ceil{\sright[f]/w}}}
    &=\widetilde{O}\paren{\frac{n}{w}\left(\sleft[f]+\sright[f]\right)}\\
    &=\widetilde{O}\paren{\frac{n\cdot\dgn[f]}{w}},
\end{align*}
where the last equality follows from \Cref{clm:symmetric-degeneracy}. This completes the proof.
\end{proof}

\Cref{lem:symmetric-ub} and \Cref{lem:dgn-lb-1} together give the proof of \Cref{thm:dgn-lb}.

\paragraph{AI Methodology.} For \Cref{thm:equal-lb}, AI was not used except for proof reading. For the upper bound for symmetric functions in \Cref{thm:dgn-lb}, AI was not used.

AI was used in identifying the degeneracy notion for \Cref{thm:dgn-lb}. More specifically,
the authors had the lower bound argument in \Cref{thm:dgn-lb} for symmetric functions without AI use with the characterization by \cite{Paturi92} (see \Cref{def:sided-sens}). Further connections to degeneracy for all Boolean functions were given by ChatGPT 5.6 Sol. 
The authors have verified the results and take full responsibility for the correctness of this paper.


\bibliographystyle{alpha}
\bibliography{general}

\clearpage
\appendix

\part*{Appendix}

\section{Background on Information Theory}\label{sec:information-theory}

In this section, we recall the standard information-theoretic definitions and facts used throughout our proofs. More details can be found in \cite{CoverT03}.

Let $\rv{X}$ be a random variable with distribution $P(x)=\Pr[\rv{X}=x]$.
The entropy of $\rv{X}$ is:
 $$\en{\rv{X}}
    := -\sum_{x} P(x)\log P(x),$$
with the convention that $0\log 0=0$. For two random variables $\rv{X}$ and $\rv{Y}$, the conditional entropy of $\rv{X}$ given $\rv{Y}$ is
\[
    \en{\rv{X}\mid \rv{Y}}
    := \sum_y \Pr[\rv{Y}=y] \cdot \en{\rv{X}\mid \rv{Y}=y}.
\]
For an event $E$ of positive probability, $\en{\rv{X}\mid E}$ denotes the entropy of $\rv{X}$ under the distribution conditioned on $E$.
The mutual information between $\rv{X}$ and $\rv{Y}$ is defined as:
\[
    \mi{\rv{X}}{\rv{Y}}
    := \en{\rv{X}}-\en{\rv{X}\mid \rv{Y}}
    = \en{\rv{Y}}-\en{\rv{Y}\mid \rv{X}}.
\]
For three random variables $\rX, \rY$ and $\rZ$, we define:
\[
    \mi{\rv{X}}{\rv{Y}\mid \rv{Z}}
    := \en{\rv{X}\mid \rv{Z}}-\en{\rv{X}\mid \rv{Y},\rv{Z}}.
\]
In words, $\mi{\rv{X}}{\rv{Y}\mid \rv{Z}}$ is the expected mutual information between $\rv{X}$ and $\rv{Y}$ after fixing $\rv{Z}$.

We state below the basic facts that we use without proof.

\begin{fact}[Standard information-theoretic facts]\label{fact:it-facts}
Let $\rv{W}, \rv{X},\rv{Y},\rv{Z}$ be four jointly distributed random variables. Then:
\begin{enumerate}[label=(\roman*)]
    \item \label{part:entropy-nonnegative}
    Entropy and conditional entropy are nonnegative. Moreover, conditioning cannot increase entropy:
    \[
        0\leq \en{\rv{X}\mid \rv{Y}}\leq \en{\rv{X}}.
    \]

    \item \label{part:entropy-support}
    If $\rv{X}$ takes values in a set of size at most $N$, then
    \[
        \en{\rv{X}}\leq \log N,
    \]
    where equality holds for the uniform random variable over all the elements of the set.
    \item \label{part:entropy-chain}
    Chain rule of entropy:
    $
    \en{\rv{X}, \rv{Y}}  = \en{\rv{X}} + \en{\rv{Y} \mid \rv{X}}$.

    \item \label{part:mi-nonnegative}
    Mutual information is symmetric and nonnegative:
    \[
        \mi{\rv{X}}{\rv{Y}}=\mi{\rv{Y}}{\rv{X}}\geq 0,
    \]
    where equality holds if and only if $\rv{X}$ and $\rv{Y}$ are independent.
    \item \label{part:mi-chain}
    Chain rule of mutual information: 
    \[
        \mi{\rv{X},\rv{Y}}{\rv{W}\mid \rv{Z}}
        = 
        \mi{\rv{X}}{\rv{W}\mid \rv{Z}} + \mi{\rv{Y}}{\rv{W} \mid \rv{X}, \rv{Z}}.
    \]
    \item \label{part:data-processing} Data Processing for Mutual Information: 
    If $\rv{X} \perp \rv{Z} \mid \rv{Y}$, then $\mi{\rv{X}}{\rv{Z}}\leq \mi{\rv{X}}{\rv{Y}}$.
    \item \label{part:both-heavy-less}
    If $\rv{W} \perp \rv{X} \mid \rv{Y},\rv{Z}$, then, 
    $\mi{\rv{X}}{\rv{Y} \mid \rv{Z}}\geq  \mi{\rv{X}}{\rv{Y} \mid \rv{W},\rv{Z}}$.
    \item \label{part:both-light-more}
    If $\rv{W} \perp \rv{X}\mid \rv{Z}$, then, 
    $
    \mi{\rv{X}}{\rv{Y} \mid \rv{Z}}\leq \mi{\rv{X}}{\rv{Y} \mid \rv{W},\rv{Z}}$.
    \item \label{part:fanos} Fano's inequality: Let $\rX, \rY$ be two random variables with support $\set{0,1}$. If random variable $\rY$ can be used to estimate another random variable $\rX$ with probability of error at most $1/2 - \epsilon$, then, 
    \[
    \en{\rX \mid \rY} \leq H_2(1/2 - \epsilon), 
    \]
    where $H_2(p) = -p \log (p) - (1-p) \log(1-p)$ is the binary entropy function. 
\end{enumerate}
\end{fact}

\paragraph{Distances and Divergences of Distributions.}

We describe the standard types of distances and divergences we use in our result.  

We start with the KL-divergence of two distributions. 
\begin{definition}[Kullback--Leibler divergence]\label{def:kl-divergence}
For distributions $P$ and $Q$ on the same finite set $\Omega$, their Kullback--Leibler divergence is defined as:
\[
    \kl{P}{Q}
    := \sum_{\omega \in \Omega} P(\omega)\log\frac{P(\omega)}{Q(\omega)}.
\]
\end{definition}

KL-divergence is useful because of its relation to mutual information.
We use $\dist(\rA)$ to denote the distribution of random variable $\rA$.

\begin{fact}\label{fact:KL-info}
    Let $\rX, \rY$ and $\rZ$ be three random variables. Then, 
    \[
\mi{\rX}{\rY \mid \rZ} = \Exp_{(Y, Z) \sim \dist(\rY, \rZ)} \kl{\dist(\rX \mid \rY = Y, \rZ = Z)}{\dist(\rX \mid \rZ = Z)}.
    \]
\end{fact}

We know KL-divergence is always non-negative. 
\begin{fact}\label{fact:kl-positive}
    For any two distributions $P$ and $Q$, $\kl{P}{Q} \geq 0$.
\end{fact}
The proof is a simple consequence of Jensen's inequality. KL-divergence can, however, be unbounded.
We also need the notion of Jensen-Shannon divergence, which is a generalised version of KL-divergence applicable to many distributions.

\begin{definition}[Jensen-Shannon divergence]\label{def:js-divergence}
    For any $\tau:= (\tau_1, \ldots, \tau_n)$ with each $\tau_i \in [0,1]$ for $i \in [n]$ such that $\sum_i \tau_i = 1$ and $n$ distributions $P_1, \ldots, P_n$ on the same finite set $\Omega$, their weighted Jensen-Shannon divergence parametrized by $\tau$ is defined as:
\[
    \js{\tau}{P_1}{P_2 || \ldots ||P_n}
    :=\sum_{i \in [n]} \tau_i \kl{P_i}{M},
\]
where $M = \sum_i \tau_i \cdot P_i $ is the mixture distribution weighted by $\tau$ values.

\end{definition}

The weighted JS-divergence is precisely the information carried by a
random variable about a finite-valued random variable.
\begin{fact}[Weighted JS and information]\label{fact:weighted-js-info}
Let $\rB$ take values in $[n]$, with
$\Pr[\rB=b]=\tau_b$, and let
$P_b:=\dist(\rX\mid\rB=b)$. Then
\[
    \mi{\rX}{\rB}
    =
    \mathbb{J}_{(\tau_b)_{b\in\mathcal{B}}}
    \bigl( P_1 || P_2 || \ldots ||P_n\bigr).
\]
\end{fact}
\begin{proof}
Apply \Cref{fact:KL-info}. The marginal distribution of $\rX$ is the
mixture $\sum_b\tau_bP_b$, so the resulting expectation of KL
divergences is exactly the definition of weighted JS-divergence.
\end{proof}

We state some simple properties of the JS-divergence measure. 

\begin{fact}\label{fact:js-distance}
For two distributions $P, Q$, we define the Jensen-Shannon distance as 
\[
\mathbb{JSD}(P, Q) = \sqrt{\js{(1/2, 1/2)}{P}{Q}}.
\]
It is known that this forms a metric over distributions and obeys the triangle inequality.  
\end{fact}

We can also relate JS-divergence parametrized by different weights. Intuitively, when focussing on only two distributions, if the JS-divergence when parametrized by some value $\tau$ is large, so should the divergence when parametrized by $1/2$. 
\begin{fact}\label{fact:js-weight-change}
    For any two distributions $P, Q$ over support $\Omega$, and any parameter $\tau \in [0,1]$, we have,\[
    \js{(\tau, 1-\tau)}{P}{Q} \leq 2 \cdot \js{(1/2, 1/2)}{P}{Q}.
    \]
\end{fact}
\begin{proof}
Let $M_1 = \tau \cdot P + (1-\tau) \cdot Q$ and $M_2 = 1/2 (P+Q)$ be the  two mixture distributions.
    \begin{align*}
        \js{(\tau, 1-\tau)}{P}{Q} &= \tau \kl{P}{M_1} + (1-\tau) \kl{Q}{M_1} \\
        &=\sum_{\omega \in \Omega} P(\omega) \tau \log (\frac{P(\omega)}{M_1(\omega)}) + Q(\omega) (1-\tau) \log (\frac{Q(\omega)}{M_1(\omega)}) \tag{by definition} \\
        &= \sum_{\omega \in \Omega} P(\omega) \tau \log (\frac{P(\omega)}{M_2(\omega)}) + Q(\omega) (1-\tau) \log (\frac{Q(\omega)}{M_2(\omega)}) + M_1(w) \log(\frac{M_2(w)}{M_1(w)}) \\
        &=  \tau \kl{P}{M_2} +(1-\tau) \kl{Q}{M_2} - \kl{M_1}{M_2} \\
        &\leq \tau \kl{P}{M_2} +(1-\tau) \kl{Q}{M_2}  \tag{by \Cref{fact:kl-positive}} \\
        &\leq 2 \js{(1/2,1/2)}{P}{Q}, \tag{as $\tau, 1-\tau \leq 1$}
    \end{align*}
    which completes the proof.
\end{proof}

Lastly, we also use the convexity of JS-divergence when parametrized by $(1/2,1/2)$, which leads to the following fact. 
\begin{fact}\label{fact:js-convex}
    For any two sets of distributions $P_1, \ldots, P_n$ and $Q_1, \ldots, Q_n$ and parameters $\tau_1, \tau_2, \ldots, \tau_n$ such that $\sum_i \tau_i = 1$, we have, \[
    \js{(1/2, 1/2)}{\sum_{i \in [n]} \tau_i P_i}{\sum_{i \in [n]} \tau_i Q_i} \leq \sum_{i \in [n]}\tau_i \cdot \js{(1/2,1/2)}{P_i}{Q_i}.
    \]
\end{fact}

This concludes our properties about JS-divergence. 
We use one last notion of distance called the total variation distance. 
\begin{definition}[Total Variation Distance]\label{def:tvd}
    For any two distributions $P, Q$ over the finite set $\Omega$, their total variation distance is defined as:
    \[
        \tvd{P}{Q} = \frac12 \cdot \sum_{\omega \in \Omega} \card{P(\omega)-Q(\omega)}.
    \]
\end{definition}

Total variation distance is useful because it relates well to the probability of success of any randomized protocol. We give some simple properties of this distance.
\begin{fact}\label{fact:tvd-facts}
Facts about total variation distance:
\begin{enumerate}[label=(\roman*)]
\item \label{fact:tvd-success}
    For any two distributions $P$ and $Q$ with the same support $\Omega$, given a sample $\omega$ from $P$ or $Q$, the best probability of success of deciding whether $\omega$ came from $P$ or $Q$ is at most $1/2 \cdot (1+\tvd{P}{Q})$.

    \item \label{fact:tvd-cond-low}
For any distribution $\mu$ and event $\mathcal{E}$, $\tvd{\mu}{(\mu \mid \neg \mathcal{E})} \leq \Pr[\mathcal{E}]$.

    \item \label{fact:event-prob-tvd} For any two distributions
     $\mu$ and $\nu$ and event $\mathcal{E}$, $\mu(\mathcal{E}) \leq \nu(\mathcal{E}) + \tvd{\mu}{\nu}$.
\end{enumerate}
\end{fact}

We can relate JS-divergence and the total variation between two distributions using the well-known Pinsker's inequality. 
\begin{fact}[Pinsker's inequality]\label{fact:pinsker-js-tvd}
    For any two distributions $P$ and $Q$, \[
\kl{P}{Q} \geq  2(\tvd{P}{Q})^2 \qquad \js{1/2}{P}{Q} \geq (\tvd{P}{Q}^2)/2.    \]
\end{fact}

\end{document}